\documentclass{article}

\usepackage{authblk}
\usepackage[square]{natbib}
\usepackage{yhmath}
\usepackage{mathtools, amsmath, amsthm, amssymb, amsbsy, amsfonts, amscd}
\usepackage{euscript}
\usepackage{bm}
\usepackage{breqn}
\usepackage{xcolor, soul}
\usepackage{empheq}
\usepackage{rotating}
\usepackage{enumerate}
\usepackage{enumitem}
\usepackage{bbold}

\usepackage{ftnxtra}
\usepackage{fnpos}
\usepackage{appendix}

\usepackage{graphicx}
\usepackage{pstool}
\usepackage{psfrag}

\usepackage{comment}

\usepackage{empheq}

\numberwithin{equation}{section}

\theoremstyle{plain}

\theoremstyle{definition}	
 
 \newtheorem{remark}{Remark}[section]
 \newtheorem{example}{Example}[section]

\usepackage{caption2}

\usepackage{hyperref}
\hypersetup{colorlinks=true, linkcolor=blue}
\hypersetup{colorlinks=true,citecolor=blue}

\DeclareMathAlphabet{\mathpzc}{OT1}{pzc}{m}{it}

\usepackage{mathrsfs}

\definecolor{lighter_purple_mathematica}{rgb}{0.6666666666,0.33333333333,0.666666666666}

\makeatletter
\newsavebox{\@brx}
\newcommand{\llangle}[1][]{\savebox{\@brx}{\(\m@th{#1\langle}\)}%
  \mathopen{\copy\@brx\mkern2mu\kern-0.9\wd\@brx\usebox{\@brx}}}
\newcommand{\rrangle}[1][]{\savebox{\@brx}{\(\m@th{#1\rangle}\)}%
  \mathclose{\copy\@brx\mkern2mu\kern-0.9\wd\@brx\usebox{\@brx}}}%
\let\oldabs\abs
\def\abs{\@ifstar{\oldabs}{\oldabs*}}
\makeatother

\usepackage{accents}
\newcommand{\Fe}{\accentset{e}{\mathbf{F}}}
\newcommand{\Fa}{\accentset{a}{\mathbf{F}}}
\newcommand{\Fp}{\accentset{p}{\mathbf{F}}}
\newcommand{\cFa}{\accentset{a}{\mathrm F}}
\newcommand{\cFe}{\accentset{e}{\mathrm F}}
\newcommand{\Fv}{\accentset{v}{\mathbf{F}}}
\newcommand{\cFv}{\accentset{v}{\mathrm F}}

\newcommand{\phie}{\accentset{e}{\varphi}}
\newcommand{\phiv}{\accentset{v}{\varphi}}
\newcommand{\Gv}{\accentset{v}{\mathbf{G}}}
\newcommand{\Ce}{\accentset{e}{\mathbf{C}}}
\newcommand{\cCe}{\accentset{e}{\mathrm C}}
\newcommand{\Cv}{\accentset{v}{\mathbf{C}}}

\newcommand{\Ie}{\accentset{e}{I}}

\newcommand{\ce}{\accentset{e}{\mathbf{c}}}
\newcommand{\be}{\accentset{e}{\mathbf{b}}}
\newcommand{\Bv}{\accentset{v}{\mathbf{B}}}
\newcommand{\Jv}{\accentset{v}{J}}

\newcommand{\Ne}{\accentset{e}{\mathbf{n}}}
\newcommand{\Lambdav}{\accentset{v}{\boldsymbol{\Lambda}}}

\usepackage{dsfont}
\newcommand{\FFe}{\accentset{e}{\boldsymbol{\mathds{F}}}}
\newcommand{\FFv}{\accentset{v}{\boldsymbol{\mathds{F}}}}

\DeclareMathOperator\erf{erf}

\begin{document}

\title{\textbf{Nonlinear Anisotropic Viscoelasticity
}}

\author[1]{Souhayl Sadik}
\author[2,3]{Arash Yavari\thanks{Corresponding author, e-mail: arash.yavari@ce.gatech.edu}}
\affil[1]{\small \textit{Department of Mechanical and Production Engineering, Aarhus University, 8000~Aarhus~C, Denmark}}
\affil[2]{\small \textit{School of Civil and Environmental Engineering, Georgia Institute of Technology, Atlanta, GA 30332, USA}}
\affil[3]{\small \textit{The George W. Woodruff School of Mechanical Engineering, Georgia Institute of Technology, Atlanta, GA 30332, USA}}

\date{October 27, 2023}

\maketitle

%-----------------------------
%-----------------------------
\begin{abstract}
In this paper, we revisit the mathematical foundations of nonlinear viscoelasticity. We study the underlying geometry of viscoelastic deformations, and in particular, the intermediate configuration. Starting from the direct multiplicative decomposition of the deformation gradient $\mathbf{F}=\Fe\Fv\, \,$, into elastic and viscous distortions $\Fe$ and $\Fv\,$, respectively, we point out that $\Fv$ can be either a material tensor ($\Fe$ is a two-point tensor) or a two-point tensor ($\Fe$ is a spatial tensor). We show, based on physical grounds, that the second choice is unacceptable. It is assumed that the free energy density is the sum of an equilibrium and a non-equilibrium part. The symmetry transformations and their action on the total, elastic, and viscous deformation gradients are carefully discussed. Following a two-potential approach, the governing equations of nonlinear viscoelasticity are derived using the Lagrange--d'Alembert principle. We discuss the constitutive and kinetic equations for compressible and incompressible isotropic, transversely isotropic, orthotropic, and monoclinic viscoelastic solids. We finally semi-analytically study creep and relaxation in three examples of universal deformations. 
\end{abstract}

\begin{description}
\item[Keywords:] Nonlinear viscoelasticity, multiplicative decomposition, intermediate configuration, anisotropic solids.
\end{description}

\tableofcontents

%-----------------------------
%-----------------------------
\section{Introduction}
The linear theory of viscoelasticity was formulated $150$ years ago by~\citet{Boltzmann1874} for isotropic solids (and by~\citealp{Volterra1909} for anisotropic solids), see~\citet{GurtinSternberg1962} and \citet{ColemanNoll1961}. The nonlinear theories of viscoelasticity appeared much later. \citet{RivlinEricksen1955} formulated a theory of viscoelasticity for isotropic solids in which stress at a material point depends on the deformation gradient and gradients of velocity, acceleration, and higher order accelerations up to some finite order at that point.\footnote{There are some recent works on ``nonlinear viscoelasticity of strain rate type''~\citep{Csengul2021, Mielke2020, Badal2023}, which is a very special case of the Rivlin--Ericksen theory. Apparently, the authors of these recent papers were not aware of the seminal paper of~\citet{RivlinEricksen1955}.} Most of the early studies of finite viscoelasticity~\citep{Pipkin1964,Rivlin1965,PipkinRogers1968} were based on
the theory of fading memory~\citep{GreenRivlin1957,GreenRivlinSpencer1959,GreenRivlin1959,Wang1965}. 

Much of the recent developments in the literature of viscoelasticity stem from the pioneering work of~\citet{green1946new} on rubber-like viscoelastic relaxation and its subsequent extension by~\citet{Lubliner1985} to finite rubber-like viscoelasticity using the Bilby--Kr\"oner--Lee decomposition following~\citep{sidoroff1974}.  As detailed in~\citep{Sadik2017},  although largely credited to~\citet{leeliu1967} and~\citet{lee1969ElastoPlast}, the multiplicative decomposition of the deformation gradient was first formally introduced by~\citet{bilby1955} and~\citet{kroner1959}.  In the context of nonlinear viscoelasticity, it was first introduced by~\citet{sidoroff1974} inspired by its use in elasto-plasticity~\citep{BerdiSedov1967,leeliu1967,lee1969ElastoPlast,sidoroff1973geometrical}.\footnote{Note, however, that the difference in the underlying conceptual rational for the use of the multiplicative decomposition of the deformation gradient in viscoelasticity as opposed to anelasticity has been to the best of our knowledge so far ignored\textemdash or at best not explicitly discussed\textemdash in the literature. As we later point out, this does have important consequences on the nature of the so-called intermediate configuration and the validity of the model otherwise.} \citet{sidoroff1974} formulated a finite deformation viscoelastic model by assuming a free energy that depends explicitly on both the total deformation gradient and its elastic (or viscous) part. Constraining the free energy by the Clausius--Duhem inequality, he found the constitutive relations\textemdash including an additive split of the stress into elastic and viscoelastic parts without assuming any such split of the free energy. He then introduced a quadratic dissipation potential following the Casimir--Onsager's reciprocity principle to obtain evolution equations.  Following~\citet{green1946new} and~\citet{sidoroff1974}, \citet{Lubliner1985} considered the nonelastic part of the deformation gradient as an additional internal variable governed by a linear rate equation. Without assuming isotropy, he formulated a constitutive model such that the free energy is additively split into an elastic part uncoupling the volumetric and deviatoric contributions of the deformation, and a viscous part that depends on the additional internal variable. \citet{Latorre2016} explicitly acknowledged that the intermediate  configuration in viscoelasticity is not stress free (after their Eq.~(20) they write ``(i.e.,~the intermediate configuration is not strictly speaking a ``stress-free'' configuration)'').  \citet{Latorre2016} suggested using a reverse decomposition of deformation gradient in viscoelasticity, i.e.,~$\mathbf{F}=\Fe\Fv=\FFv\FFe$ (see also~\citealp{Bahreman2022} who compared the direct and reverse decompositions for viscoelasticity. It should be noted that these authors assumed the elastic and viscous distortions to be compatible (see Eqs.~(8)~\&~(12)), which is incorrect.). Similar to anelasticity~\citep{YavariSozio2023}, the reverse decomposition is expected to result in an equivalent theory.

Based on the generalized one-dimensional linear Maxwell rheological model,~\citet{simo1987fully} first sketches an alternative formulation of the standard linear solid that he subsequently generalizes to a nonlinear formulation of viscoelastic solids. In his formulation, he assumes an additive split of the free energy into an initial (elastic) and a non-equilibrium contribution. Similarly to~\citet{Lubliner1985}, he uncouples the bulk and deviatoric components of the deformation gradient to forgo the isotropy assumption in his constitutive model. The viscous response is introduced by considering a strain-like tensor as an internal variable in the non-equilibrium part of the free energy; a variable whose evolution is governed by a linear rate constitutive equation. When specialized to the particular case of neo-Hookean solids, he shows that his theory is consistent with the Bilby--Kr\"oner--Lee decomposition with the internal variable related to the non-elastic contribution of the deformation gradient as
in~\citep{Lubliner1985}. It is worth mentioning that Simo's finite linear convolution model has never been demonstrated to conform to the second law. For a recent relevant study, see~\citet{Liu2021}.

\citet{LeTallec1993} assumed the multiplicative decomposition of the deformation gradient into elastic and viscous parts. For incompressible viscoelastic solids, they assumed that both the total deformation gradient and the viscous deformation gradient (or equivalently both the elastic and viscous deformation gradients) are volume preserving (the same assumption had been made earlier by~\citealp{Leonov1976}). They assumed an additive split of the free energy density into equilibrium and non-equilibrium parts that depend on $\mathbf{C}$ and {$\Ce$}, respectively. Finally, they assumed a dissipation potential that explicitly depends on {$\dot{\Cv}$}, i.e.,~{$\phi=\phi(\dot{\Cv})$}. For fiber-reinforced viscoelastic composites,~\citet{Nguyen2007} used an isotropic dissipation potential written in terms of {$\Cv$} and identical to that used by~\citet{Reese1998}. When written in terms of {$\Cv$}, the quadratic dissipation potential is expressed in terms of two positive viscosities (deviatoric and volumetric).  

Without any mention of the Bilby--Kr\"oner--Lee decomposition,~\citet{holzapfel1996a} introduced an additive decomposition of the free energy into a purely thermoelastic contribution and a non-equilibrium contribution; where, following~\citet{ColeGurt1967internal}, the latter is described as a configurational free energy dependent on a set of additional internal variables, akin to strain, characterizing the irreversible viscoelastic response of the material. Starting from the generalized one-dimensional linear Maxwell rheological model~\citep{Holzapfel1996b}, they introduced an evolution equation for the conjugate internal non-equilibrium stresses following~\citep{valanis1972irreversible}.

Starting with the generalized linear Maxwell rheological model and generalizing the constitutive model of~\citet{Lubliner1985},~\citet{ReeseGovindjee1998,Reese1998} proposed an additive split of the free energy into an equilibrium and a non-equilibrium part. The equilibrium part depends on the total deformation gradient and gives the free energy in the thermodynamic equilibrium state at infinite time. The non-equilibrium part of the free energy depends however solely on the elastic part of the Bilby--Kr\"oner--Lee deformation gradient decomposition and eventually vanishes as the body relaxes in the thermodynamic equilibrium state. In the framework of~\citet{holzapfel1996a}, they effectively took the inelastic part of the deformation gradient to be the internal variable of interest, and instead assumed its evolution to be given by a positive semi-definite quadratic dissipation potential; a sufficient condition to fulfill the second law of thermodynamics. More recently,~\citet{Kumar2016} formulated nonlinear viscoelasticity using a two-potential approach. They critically reviewed some of the previous works in the literature on the kinetic equations and pointed out some inconsistencies regarding objectivity (material-frame-indifference) of some of the proposed kinetic equations.

There have been attempts in the literature to model anisotropic nonlinear viscoelastic solids.~\citet{biot1954theory} presented a Lagrangian treatment of anisotropic viscoelasticity based on Onsager's reciprocal relations~\citep{onsager1931reciprocal} using potential energy and dissipation function and introduced operational tensors to relate stress and strain. For a viscoelastic solid reinforced by one family of fibers (a transversely isotropic viscoelastic solid),~\citet{Merodio2006} assumed that the Cauchy stress depends on the fiber orientation, the right Cauchy--Green strain, and its time derivative, i.e.,~$\boldsymbol{\sigma}=\boldsymbol{\sigma}(\mathbf{N},\mathbf{C},\dot{\mathbf{C}})\,$,  where $\mathbf{N}=\mathbf{N}(X)$ is the unit tangent vector to the fiber at the material point $X\,$. There have been several efforts in the literature in modeling nonlinear viscoelasticity of fiber-reinforced viscoelastic solids using the multiplicative decomposition of the deformation gradient~\citep{Nedjar2007,Nguyen2007,Liu2019}. They assumed separate multiplicative decompositions of the deformation gradient for the matrix and the fibers. \citet{Nguyen2007} assumed that the equilibrium and non-equilibrium free energies have the same symmetry group. There have also been recent efforts in modeling viscoelasticity of nematic liquid crystal elastomers~\citep{Wang2022}. We should also mention that there are several reviews of viscoelasticity in the literature~\citep{Schapery2000,Drapaca2007,Banks2011,Wineman2020,Csengul2021}.

This paper is organized as follows. In  \S\ref{Sec:Kinematics}, kinematics of viscoelasticity is discussed. In particular, we assume the multiplicative decomposition {$\mathbf{F}=\Fe\Fv$}, and the tensorial characters of {$\Fe$} and {$\Fv$} are carefully investigated. Additive decomposition of the free energy density into an equilibrium and a non-equilibrium part is assumed. In  \S\ref{Sec:BalanceLaws}, balance of mass, balance of linear and angular momenta, and the kinetic equation for {$\Fv$} are derived, and the constitutive relations are discussed. The balance of linear and angular momenta, and the kinetic equations for {$\Fv$} are derived using a two-potential approach and the Lagrange--d'Alembert principle. The first and second laws of thermodynamics are discussed and used to find the constitutive relations for viscoelastic solids. Material symmetry in viscoelasticity is studied in  \S\ref{Sec:MaterialSymmetry}. In particular, it is seen that the symmetry group acts on both the equilibrium and non-equilibrium parts of the free energy, as well as the dissipation potential. The representation of the Cauchy stress in terms of the material integrity basis is derived for transversely isotropic, orthotropic, and monoclinic solids both in the compressible and incompressible cases. The dissipation potential and its functional form for both isotropic and anisotropic solids is discussed. Three examples of universal deformations of isotropic and anisotropic viscoelastic solids are analyzed in detail in  \S\ref{Sec:Examples}. Concluding remarks are given in  \S\ref{Sec:Conclusions}.

%-----------------------------
%-----------------------------
\section{Kinematics, Free Energy, and Dissipation Potential} \label{Sec:Kinematics}

%-----------------------------
%-----------------------------
\subsection{Kinematics}
Let us consider a body that is made of a viscoelastic solid. We identify the body with an embedded $3$-submanifold $\mathcal{B}$ of the Euclidean ambient space $\mathcal{S}=\mathbb R^3\,$.
We adopt the standard convention to denote objects and indices by uppercase characters in the material manifold $\mathcal{B}$ (e.g.,~$X\in\mathcal{B}$) and by lowercase characters in the spatial manifold $\mathcal{S}$ (e.g.,~$x\in\mathcal{S}$).
We denote by $\{X^A\}\,$,  and $\{x^a\}\,$,  the local coordinate charts on $\mathcal{B}$ and $\mathcal{S}\,$,  respectively; by $\left\{\partial_A=\frac{\partial}{\partial X^A}\right\}$ and $\left\{\partial_a=\frac{\partial}{\partial x^a}\right\}\,$,  we denote the corresponding local coordinate bases, respectively; and by $\left\{dX^A\right\}$ and $\left\{dx^a\right\}\,$,  we denote the corresponding dual bases. We also adopt Einstein's repeated index summation convention, e.g.,~$u^i v_i\coloneq \sum_i u^i v_i\,$.

Motion is represented by a one-parameter family of maps $\varphi_t:\mathcal{B}\to\mathcal{C}_t\subset\mathcal{S}\,$,  where $\mathcal{C}_t=\varphi_t(\mathcal{B})$ is the current configuration of the body. A material point $X\in\mathcal{B}$ is mapped to ${x=x(X,t)=\varphi_t(X)=\varphi_X(t)}\,$. The Euclidean ambient space has the flat metric $\mathbf{g}\,$,  which has the representation $\mathbf{g}=\mathrm{g}_{ab}\,dx^a\otimes dx^b\,$. For example, if $\{x^a\}$ are Cartesian coordinates, the metric reads off $\mathbf{g}=\delta_{ab}\,dx^a\otimes dx^b\,$. 
Given two vectors $\mathbf{u}\,,\mathbf{w}\in T_x\mathcal{S}$\textemdash the tangent space of $\mathcal S$ at $x\,$,  their dot product is denoted by $\llangle \mathbf{u},\mathbf{w} \rrangle_{\mathbf{g}}=\mathrm{u}^a\,\mathrm{w}^b\,\mathrm{g}_{ab}\,$. Given a vector $\mathbf{u}\in T_x\mathcal{S}$ and a 1-form $\boldsymbol{\omega}\in T^{*}_x\mathcal{S}$\textemdash the cotangent space of $\mathcal S$ at $x\,$,  their natural pairing is denoted by $\langle \boldsymbol{\omega},\mathbf{u} \rangle=\boldsymbol{\omega}(\mathbf{u})=\omega_a\,\mathrm{u}^a\,$. The spatial volume form reads ${dv = \sqrt{\det\mathbf g}\,dx^1 \wedge dx^2 \wedge dx^3}\,$. Let $\nabla^{\mathbf g}$ be the Levi-Civita connection of $(\mathcal{S},\mathbf{g})\,$. We denote its Christoffel symbols by ${\gamma^a}_{bc}$ in the local coordinate chart $\{x^a\}\,$.
The Euclidean metric $\mathbf{g}$ induces the Euclidean metric $\mathbf{G}$ on $\mathcal{B}\,$. The natural distances in the body before deformation are calculated using the metric $\mathbf{G}$; this is the material metric and has the representation ${\mathbf{G}=\mathrm{G}_{AB}\,dX^A\otimes dX^B}\,$. For example, if $\{X^A\}$ are Cartesian coordinates, the metric reads off ${\mathbf{G}=\delta_{AB}\,dX^A\otimes dX^B}$; in cylindrical coordinates $\{R,\Theta,Z\}\,$,  it reads off ${\mathbf{G}=dR\otimes dR+R^2\,d\Theta\otimes d\Theta+dZ\otimes dZ}\,$. Given two vectors $\mathbf{U}\,,\mathbf{W}\in T_X\mathcal{B}$\textemdash the tangent space of $\mathcal B$ at $X\,$,  their dot product is denoted by $\llangle \mathbf{U},\mathbf{W} \rrangle_{\mathbf{G}}=\mathrm{U}^A\,\mathrm{W}^B\,\mathrm{G}_{AB}\,$. Given a vector $\mathbf{U}\in T_X\mathcal{B}$ and a 1-form $\boldsymbol{\Omega}\in T^{*}_X\mathcal{B}$\textemdash the cotangent space of $\mathcal B$ at $X\,$,  their natural pairing is denoted by $\langle \boldsymbol{\Omega},\mathbf{U} \rangle=\boldsymbol{\Omega}(\mathbf{U})=\Omega_A\,\mathrm{U}^A\,$. The material volume form reads $dV = \sqrt{\det\mathbf G} \,dX^1 \wedge dX^2 \wedge dX^3\,$.
Let $\nabla^{\mathbf G}$ be the Levi-Civita connection of $(\mathcal{B},\mathbf{G})\,$. We denote its Christoffel symbols by ${\Gamma^A}_{BC}$ in the local coordinate chart $\{X^A\}\,$.

A local elastic deformation is measured with respect to a local stress-free state and induces change of distances. It may be quantified by the derivative of the deformation mapping\textemdash the so-called deformation gradient, denoted by $\mathbf{F}(X,t)=T\varphi_t(X):T_X\mathcal{B}\to T_{\varphi_t(X)}\mathcal{C}_t\,$,  which has components ${\mathrm{F}^a}_A=\partial \varphi^a/\partial X^A\,$.
The dual $\mathbf F^\star$ of $\mathbf F$ is defined as ${\mathbf F^\star(X,t):T_{\varphi_t(X)}\mathcal{C}_t \to T_X\mathcal{B}}\,$,  ${\langle\boldsymbol\alpha,\mathbf F \mathbf U\rangle=\langle\mathbf F^\star \boldsymbol\alpha,\mathbf U\rangle}\,$,  $\forall\, \mathbf U \in T_X\mathcal B\,$,  $\forall\, \boldsymbol\alpha \in T^{*}_{\varphi(X)}\mathcal S\,$,  and reads in components ${\left(\mathrm{F}^\star\right)_A}^a={\mathrm{F}^a}_A\,$.
The transpose $\mathbf F^{\mathsf T}$ of $\mathbf F$ is defined as ${\mathbf F^{\mathsf T}(X,t):T_{\varphi_t(X)}\mathcal{C}_t \to T_X\mathcal{B}}\,$,  ${\llangle \mathbf F\mathbf{U},\mathbf{u} \rrangle_{\mathbf{g}}=\llangle \mathbf{U},\mathbf F^{\mathsf T}\mathbf{u} \rrangle_{\mathbf G}}\,$,  $\forall\, \mathbf U \in T_X\mathcal B\,$,  $\forall\, \mathbf u \in T_{\varphi(X)}\mathcal S\,$,  and has components ${\left(\mathrm{F}^{\mathsf T}\right)^A}_a = \mathrm{G}^{AB}\, {\mathrm{F}^b}_B\, \mathrm{g}_{ba}\,$. Note that $\mathbf F^{\mathsf T}=\mathbf G^\sharp \mathbf F^\star \mathbf g\,$,  where $(.)^\sharp$ denotes the musical isomorphism for raising indices.
The right Cauchy--Green\textemdash also known as Green\textemdash deformation tensor is defined as  $\mathbf{C}\coloneq \mathbf{F}^{\mathsf{T}}\mathbf{F}\,$,  and reads in components ${\mathrm{C}^A}_B=\mathrm{G}^{AK}\,{\mathrm{F}^a}_K\,\mathrm{g}_{ab}\,{\mathrm{F}^b}_B\,$. Note that $\mathbf{C}^\flat$ agrees with the pull-back of the spatial metric $\mathbf{g}$ by $\varphi\,$,  i.e.,~$\mathbf{C}^\flat=\varphi{}^{*}\mathbf{g}=\mathbf F^\star \mathbf g \mathbf F\,$,  where $(.)^\flat$ denotes the musical isomorphism for lowering indices.
The Piola deformation tensor is defined as $\mathbf{B}\coloneq \mathbf C^{-1}=\mathbf F^{-1} \mathbf F^{-\mathsf{T}}\,$,  and has components ${\mathrm{B}^A}_B={(\mathrm{F}^{-1})^A}_a\, \mathrm{g}^{ab}{(\mathrm{F}^{-1})^C}_b\, \mathrm{G}_{CB}\,$. Note that $\mathbf{B}^\sharp$ agrees with the pull-back of the inverse material metric $\mathbf{g}^\sharp$ by $\varphi\,$,  i.e.,~$\mathbf{B}^\sharp=\varphi{}^{*}\mathbf{g}^\sharp=\mathbf F^{-1} \mathbf g^\sharp \mathbf F^{-\star}\,$.
The left Cauchy--Green\textemdash also known as Finger\textemdash deformation tensor is defined as  $\mathbf{b}\coloneq \mathbf{F}\mathbf{F}^{\mathsf{T}}\,$,  and reads in components ${b^a}_b={F^a}_A G^{AB} {F^c}_B \,g_{cb}\,$. Note that $\mathbf{b}^{\sharp}$ agrees with the push-forward of the inverse of the material metric $\mathbf{G}^\sharp$ by $\varphi\,$,  i.e.,~$\mathbf{b}^{\sharp}=\varphi{}_*\mathbf G^\sharp=\mathbf F \mathbf G^\sharp \mathbf F^\star\,$.
The inverse Finger deformation tensor is denoted by  $\mathbf{c}\coloneq \mathbf{b}^{-1}=\mathbf{F}^{-\mathsf{T}}\mathbf{F}^{-1}\,$,  and has components ${\mathrm{C}^A}_b=\mathrm{g}^{ac}{(\mathrm{F}^{-1})^A}_c \,\mathrm{G}_{AB}{(\mathrm{F}^{-1})^B}_b\,$. Note that $\mathbf{c}^\flat$ agrees with the push-forward of the material metric $\boldsymbol{G}$ by $\varphi\,$,  i.e.,~$\boldsymbol{c}^{\flat}=\varphi{}_*\mathbf{G} = \mathbf F^{-\star} \mathbf G \mathbf F^{-1}\,$.
The Jacobian of the motion relates the material and spatial volume elements as $ \textrm dv = J \textrm dV\,$,  and it can be shown that $J=\sqrt{\det\mathbf C} = \sqrt{\det\mathbf g/\det\mathbf G}\det\mathbf F\,$.\footnote{Denoting the Riemannian volume $3$-forms corresponding to the Riemannian metrics $\mathbf{g}$ and $\mathbf{G}$ by $\boldsymbol{\mu}_{\mathbf{g}}$ and $\boldsymbol{\mu}_{\mathbf{G}}\,$,  respectively, they are related as $\varphi^{*}\boldsymbol{\mu}_{\mathbf{g}}=J\,\boldsymbol{\mu}_{\mathbf{G}}\,$.}

The material velocity $\mathbf V$ of the motion is defined as ${\mathbf{V}:\mathcal{B}\times\mathbb{R}^+\to T \mathcal{S}\,, \mathbf{V}(X,t)\coloneq \partial \varphi(X,t)/\partial t}\,$,  and in components reads $\mathrm{V}^a = \frac{\partial \varphi^a}{\partial t}\,$. The spatial velocity is defined as ${\mathbf{v}:\varphi_t(\mathcal{B})\times\mathbb{R}^+\to T \mathcal{S}}\,$,  ${\mathbf{v}(x,t)\coloneq \mathbf{V}(\varphi_t^{-1}(x),t)}\,$.
The material acceleration is defined as ${\mathbf{A}:\mathcal{B}\times\mathbb{R}^+\to T \mathcal{S}}\,$,  ${\mathbf{A}(X,t)\coloneq D_t^{\mathbf g}\mathbf{V}(X,t)}\,$,  where $D_t^{\mathbf g}$ denotes the covariant derivative along $\varphi_X:t\mapsto \varphi(X,t)\,$. In components, $\mathrm{A}^a=\frac{\partial \mathrm{V}^a}{\partial t}+{\gamma^a}_{bc}\,\mathrm{V}^b\,\mathrm{V}^c\,$.
The spatial acceleration is defined as $\mathbf{a}:\varphi_t(\mathcal{B})\times\mathbb{R}^+\to T \mathcal{S}\,$,  $\mathbf{a}(x,t)\coloneq \mathbf{A}(\varphi_t^{-1}(x),t)\in T_x\mathcal{S}\,$,  and in components reads, $\mathrm{a}^a=\frac{\partial\mathrm{v}^a}{\partial t}+\frac{\partial\mathrm{v}^a}{\partial x^b}\,\mathrm{v}^b+{\gamma^a}_{bc}\,\mathrm{v}^b\,\mathrm{v}^c\,$.

%-----------------------------
%-----------------------------
\subsection{Multiplicative decomposition of the deformation gradient}
\label{S:FeFv}

Let us consider a viscoelastic body in its loaded deformed state. If we proceed to unload the body, we observe an instantaneous partial relaxation into an intermediate stressed state (that is embedded in the Euclidean ambient space), followed by a slower relaxation back into its initial undeformed state. Note that in this experiment, the intermediate state may, in general, still contain unresolved residual elastic strain. As a mater of fact, while the instantaneous partial relaxation is purely elastic, the slow relaxation is, in general, not purely viscous and may involve some residual elastic deformation that might have been prevented from resolving instantaneously.\footnote{This is essentially an expression of the incompatibility of the elastic strain in the body.}

\begin{figure}[t!]
\centering
\includegraphics[width=0.90\textwidth]{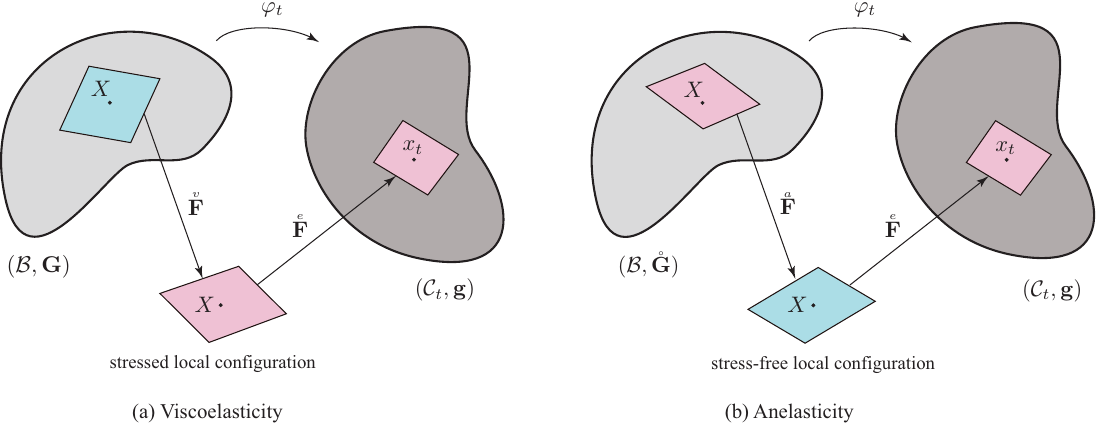}
\vspace*{-0.0in}
\caption[Local intermediate configurations in viscoelasticity.....]{Local intermediate configurations in viscoelasticity and anelasticity. Blue and pink squares indicate locally stress-free and locally stressed configurations, respectively.
(a) In viscoelasticity, the local intermediate configuration is stressed, and the material manifold is the Euclidean manifold $(\mathcal{B},\mathbf{G})\,$. (b) In anelasticity, the local intermediate configuration is stress-free, and the material manifold is $(\mathcal{B},\mathbf{G})\,$, where $\mathbf{G}$ is the non-flat material metric, which is related to the Euclidean metric $\mathring{\mathbf{G}}$ via pull-back.} 
\label{Local-Configurations}
\end{figure}

Instead of the global picture above, let us look at this thought experiment locally by considering a volume element in a viscoelastic body in its loaded state, i.e.,~a small neighborhood of a spatial point\textemdash generally deformed and stressed. We let this volume element be isolated and proceed to unload it independently of the rest of the body. We then observe an instantaneous purely elastic relaxation of the total elastic strain in the isolated volume element into an intermediate stressed state,\footnote{The union of all such partially relaxed volume elements constituting the body does not, in general, result in a body embeddable in the Euclidean ambient space. It can, however, be described as an abstract manifold with a non-trivial metric; and it is, in general, different from the partially relaxed intermediate state embedded in the Euclidean ambient space discussed in the global experiment above.} followed by a slow viscous relaxation into its initial undeformed state. The instantaneous local purely elastic unloading map is denoted by $\Fe^{-1}\,$. The slow final local purely viscous relaxation map is denoted by $\Fv^{-1}\,$. Therefore, one has a local multiplicative decomposition of the deformation gradient {$\mathbf{F}=\Fe\Fv$}. The two maps {$\Fe$} and {$\Fv$} are incompatible, in general, i.e.,~global maps $\phiv_t:\mathcal{B}\to\mathcal{B}$ and $\phie_t:\mathcal{B}\to\mathcal{C}_t$ (or $\phiv_t:\mathcal{B}\to\mathcal{C}_t$ and $\phie_t:\mathcal{C}_t\to\mathcal{C}_t$) such that $\Fv=T\phiv_t$ and $\Fe=T\phie_t$ do not exist, in general.
%\footnote{\citet{Kumar2016} %assumed the existence of %global elastic and viscous deformation maps. This incorrect assumption affected their formulation of material symmetries as will be discussed in  \S\ref{Sec:CriticalDiscussion}.}
Notice that the local configuration that results after an instantaneous local elastic unloading is not stress-free, in general (see \S\ref{InterConfig}). This is in contrast with anelasticity for which a locally unloaded configuration is stress-free.  This is the fundamental difference between viscoelasticity and anelasticity (see Fig.~\ref{Local-Configurations}).

%-----------------------------
%-----------------------------
%-----------------------------
%-----------------------------

It should be noted that in the decomposition $\mathbf{F}=\Fe\Fv$ there are two possibilities:
%---------------------
\begin{subequations} \label{eq:viscoelasticity-choices}
\begin{align}
	\label{eq:viscoelasticity-choice1}
	i) \quad &\Fv(X):T_X\mathcal{B}\to T_X\mathcal{B}\,, \quad 
	\Fe(X):T_X\mathcal{B}\to T_x\mathcal{C}\,,	\\
	\label{eq:viscoelasticity-choice2}
	ii) \quad &\Fv(X):T_X\mathcal{B}\to T_x\mathcal{C}\,, \quad\; 
	\Fe(x):T_x\mathcal{C}\to T_x\mathcal{C}\,,
\end{align}
\end{subequations}
%---------------------
where $x=\varphi(X)\,$. In the next section, we show that the second choice will have physically-inconsistent consequences, and hence, it is not acceptable.

%-----------------------------
%-----------------------------
\subsection{Additive decomposition of the free energy into equilibrium and non-equilibrium parts}
\label{Additive_decomp}

We assume that the constitutive model of a nonlinear viscoelastic solid is described by a pair of functionals $(\Psi,\phi)\,$,  where $\Psi=\Psi(X,\Theta,\mathbf F,\Fe,\mathbf G,\mathbf g)$ is the free energy density functional (per unit undeformed volume) and $\phi=\phi(X,\Theta,\mathbf{F},\Fv,\dot\Fv,\mathbf{G},\mathbf{g})$ is a dissipation potential density (per unit undeformed volume)\textemdash or Rayleigh functional, where $\Theta=\Theta(X,t)$ is the temperature field. In the literature, it has usually been assumed that the free energy can be additively decomposed into an equilibrium part and a non-equilibrium part: $\Psi=\Psi_{\text{EQ}}+\Psi_{\text{NEQ}}\,$,  where the equilibrium free energy depends on the total deformation gradient, and the non-equilibrium free energy depends on the instantaneous elastic contribution of the deformation gradient {$\Fe$}~\citep{Reese1998,Kumar2016}. For either choice in \eqref{eq:viscoelasticity-choices}, one may write $\Psi_{\text{EQ}}=\Psi_{\text{EQ}}(X,\Theta,\mathbf{F},\mathbf{G},\mathbf{g})$; and material frame indifference\footnote{Material frame indifference, objectivity, or invariance under the ambient space rigid body motions of $\Psi_{\text{EQ}}$ are equivalent in the case of a Euclidean ambient space to $\Psi_{\text{EQ}}(X,\mathbf q\mathbf{F},\Theta,\mathbf{G}, \mathbf{g})=\Psi_{\text{EQ}}(X,\mathbf{F},\Theta,\mathbf{G},\mathbf{g})$ for all deformation gradients $\mathbf F$ and any arbitrary $\mathbf g$-orthogonal second-order tensor $\mathbf q:T_x\mathcal S \to T_x\mathcal S\,$,  i.e.,~$\mathbf{q}^{\mathsf T} \mathbf{q} = \mathrm{id}_{\mathcal{S}}\,$,  which is equivalent to writing $\mathbf q^{*} \mathbf g = \mathbf g\,$,  where $\mathbf q^{*} \mathbf g=\mathbf q^\star \mathbf g \mathbf q\,$.} implies that $\Psi_{\text{EQ}}(X,\mathbf{F},\Theta,\mathbf{G},\mathbf{g})=\hat{\Psi}_{\text{EQ}}(X,\Theta,\mathbf{C}^{\flat},\mathbf{G})\,$.
However, the functional form of the non-equilibrium free energy depends on the choices in \eqref{eq:viscoelasticity-choices}:
%---------------------
\begin{subequations} \label{eq:Psi-choices}
\begin{align}
	\label{eq:Psi-choice1}
	i)  & \quad \Psi_{\text{NEQ}}=\Psi^{(1)}_{\text{NEQ}}(X,\Theta,\Fe,\mathbf{G},\mathbf{g})\,,	\\
	\label{eq:Psi-choice2}
	ii) & \quad \Psi_{\text{NEQ}}=\Psi^{(2)}_{\text{NEQ}}(x,\theta,\Fe,\mathbf{g}) \,,
\end{align}
\end{subequations}
%---------------------
where $x=\varphi(X)$ and $\theta=\Theta\circ\varphi_t^{-1}\,$. Considering \eqref{eq:Psi-choice1}, objectivity implies that $\Psi^{(1)}_{\text{NEQ}}=\hat{\Psi}^{(1)}_{\text{NEQ}}(X,\Theta,\Ce^{\flat},\mathbf{G})\,$,  where $\Ce=\Fe^{\mathsf T}\Fe\,$.\footnote{Note that $\Ce^{\flat}=\Fe^{*}\mathbf{g}=\Fe^{\star}\mathbf{g}\,\Fe\,$.}
For the second choice, material frame indifference forces $\Psi^{(2)}_{\text{NEQ}}$ to be isotropic.
More specifically, for \eqref{eq:Psi-choice2}, spatial covariance (an assumption that implies material frame indifference) of the non-equilibrium free energy $\Psi^{(2)}_{\text{NEQ}}$ holds if under a spatial diffeomorphism $\xi:\mathcal{S}\to\mathcal{S}$ (such that $T\xi$ is an isometry in the case of a Euclidean ambient space) one has
%---------------------
\begin{equation}
	\Psi^{(2)}_{\text{NEQ}}(\xi(x),\xi_*\theta,\xi_*\Fe,\xi_*\mathbf{g})
	=\Psi^{(2)}_{\text{NEQ}}(x,\theta,\Fe,\mathbf{g})\,,	
\end{equation}
%---------------------
where ${\xi_*\theta=\theta\circ\xi}\,$, ${\xi_*\Fe=T\xi\cdot\Fe\cdot(T\xi)^{-1}}\,$, and ${\xi_*\mathbf{g}=(T\xi)^{-1}\,\mathbf{g}\,(T\xi)^{-\star}}\,$.
This implies that
%---------------------
\begin{equation}
	\Psi^{(2)}_{\text{NEQ}}(x',\theta,\Fe,\mathbf{g})=\Psi^{(2)}_{\text{NEQ}}(x,\theta,\xi^*\Fe,\xi^*\mathbf{g})\,,
\end{equation}
%---------------------
where ${x'=\xi(x)}\,$,  for all such $\xi\,$,  which hence means that $\Psi^{(2)}_{\text{NEQ}}$ is an isotropic functional of {$\Fe$}.\footnote{Note that material symmetries put constraints on $\Psi^{(1)}_{\text{NEQ}}$ but not on $\Psi^{(2)}_{\text{NEQ}}\,$.} It follows that the non-equilibrium free energy is isotropic for any viscoelastic solid; and consequently, viscoelastic materials experience creep (deformation increase under constant load) and relaxation (stress decrease under constant deformation) in an isotropic fashion. However, experimental evidence contradicts this hypothetical situation since viscoelastic creep and relaxation have experimentally been observed to be
anisotropic in different classes of materials, e.g.,~skin \emph{in vivo}~\citep{khatyr2004model}, some single-crystal superalloys~\citep{segersall2014creep}, and soft
soil~\citep{sivasithamparam2015modelling}. Therefore, we conclude that the physically consistent decomposition is indeed \eqref{eq:viscoelasticity-choice1}. 
From here on, we assume that $\Fv(X):T_X\mathcal{B}\to T_X\mathcal{B}\,$,  $\Fe(X):T_X\mathcal{B}\to T_x\mathcal{C}$ and write
%---------------------
\begin{equation}\label{eq:Psi_F_decomp}
	\Psi=\Psi(X,\Theta,\mathbf F, \Fe, \mathbf{G},\mathbf{g})
	=\Psi_{\text{EQ}}(X,\Theta,\mathbf F, \mathbf{G},\mathbf{g})
	+\Psi_{\text{NEQ}}(X,\Theta,\Fe,\mathbf{G},\mathbf{g})\,,
\end{equation}
%---------------------
or equivalently
%---------------------
\begin{equation}\label{eq:Psi_C_decomp}
	\Psi=\hat\Psi(X,\Theta,\mathbf C^\flat, \Ce^\flat, \mathbf G)
	=\hat\Psi_{\text{EQ}}(X,\Theta,\mathbf C^\flat, \mathbf G)+\hat\Psi_{\text{NEQ}}(X,\Theta,\Ce^\flat, \mathbf G)	\,.
\end{equation}
%---------------------

\begin{remark}\label{rmrk:V_intermediate}
One may argue that \eqref{eq:viscoelasticity-choice1} and \eqref{eq:viscoelasticity-choice2} are not the only possibilities for the direct decomposition $\mathbf F = \Fe \Fv\,$. In the most general case, one may assume that $\Fv : T_X\mathcal{B} \to \mathcal{V}\,$, and  $\Fe: \mathcal{V}\to T_x\mathcal{C}\,$,  for some arbitrary vector space $\mathcal{V}\,$. In such a case, we ought to have $\Psi_{\text{NEQ}}=\Psi_{\text{NEQ}}(X,\Theta,\Fe,\mathbf{m},\mathbf{g})\,$,  for some metric $\mathbf m$ on $\mathcal V$\, as {$\Fe$} is a two-point tensor. This then implies that $\Psi_{\text{NEQ}}$ is independent of the material configuration $\mathcal B$ and its metric $\mathbf G\,$. Hence, unless $\mathcal V$ is isometric to $T_X\mathcal B\,$,  it would follow that one may not enforce any material symmetry constraint on $\Psi_{\text{NEQ}}\,$. This would consequently preclude $\Psi_{\text{NEQ}}$\,\textemdash and subsequently the material's creep and viscous relaxation behaviors\textemdash from reflecting the material symmetry, which is not physical. It follows then that $\mathcal V$ has to be isometric to $T_X\mathcal B\,$. Therefore, one may assume that $\mathcal V=T_X\mathcal B$  without any loss of generality.
\end{remark}

\begin{remark}
Instead of looking at the deformed body and a local unloading, let us start with the stress-free undeformed body. Consider a material volume element (a small neighborhood of a material point\textemdash undeformed and stress-free) and imagine that it is locally loaded, i.e.,~it is isolated and deformed independently from the rest of the body. This element undergoes an instantaneous elastic deformation followed by a slow viscous relaxation. It should be noted that both the elastically deformed intermediate state and the final deformed state are generally stressed. This thought experiment motivates the reverse decomposition of the deformation gradient: $\mathbf{F}=\FFv\FFe\,$. It should also be noted that if the reverse decomposition $\mathbf{F}=\FFv\FFe$ is used, it may be proved\textemdash similarly to \S\ref{Additive_decomp}\textemdash  that $\FFe:T_X\mathcal{B}\to T_x\mathcal{C}\,$,  and $\FFv:T_x\mathcal{C}\to T_x\mathcal{C}\,$. It is observed, in both the direct and reverse decompositions, that the elastic deformation gradient is a two-point tensor. In the direct decomposition, the viscous deformation gradient is a material tensor while it is a spatial tensor in the reverse decomposition. We expect this decomposition to lead to an equivalent theory of viscoelasticity.\footnote{It has been shown that the direct and inverse decompositions are equivalent for anelasticity~\citep{YavariSozio2023}, and we expect a similar result for viscoelasticity.} In this paper, we work with the direct decomposition.
\end{remark}

%-----------------------------
%-----------------------------
\subsection{Dissipation potential for isotropic and anisotropic viscoelastic solids}

The dissipation potential complements the free energy functional $\Psi$ to form the full constitutive model of a viscoelastic solid; it is assumed to have the functional form $\phi=\phi(X,\Theta,\mathbf{F},\Fv,\dot\Fv,\mathbf{G},\mathbf{g})$ and is such that the generalized force driving the evolution of the viscous deformation is given by
%---------------------
\begin{equation}\label{Dissipative_B}
	\Bv= -\frac{\partial \phi}{\partial \dot\Fv}\,.
\end{equation}
%---------------------
Notice that for a general viscoelastic solid, while the dependence of $\phi$ on $\mathbf F$ can be reduced to a dependence on the symmetric tensor $\mathbf C^\flat$ following from material frame indifference, i.e.,~$\phi=\hat{\phi}(X,\Theta,\mathbf{C}^{\flat},\Fv,\dot\Fv,\mathbf{G})\,$,  the dependence of $\phi$ on {$\Fv$} and $\dot{\Fv}$ cannot always be reduced to a dependence on the symmetric tensors $\Cv^\flat$ and $\dot{\Cv}^\flat\,$. This implies that the model used by~\citet{LeTallec1993} is applicable to only a subset of viscoelastic solids. It is assumed that $\phi$ is a convex functional of $\dot\Fv$~\citep{ziegler1958attempt, ziegler1987derivation, Germain1983, Goldstein2002, Kumar2016}, which is equivalent to
%---------------------
\begin{equation}\label{eq:convex}
	\left(\frac{\partial \phi}{\partial \dot{\Fv}_2}-\frac{\partial \phi}{\partial \dot{\Fv}_1}\right)
	:\left(\dot{\Fv}_2-\dot{\Fv}_1\right)
	\geq 0 \,,
\end{equation}
%---------------------
for any $\dot{\Fv}_1$ and $\dot{\Fv}_2\,$.

%-----------------------------
%-----------------------------
\section{Balance Laws} \label{Sec:BalanceLaws}

%-----------------------------
%-----------------------------
\subsection{Conservation of mass}

We denote the material and spatial mass densities by $\rho_o(X)$ and $\rho(x,t)\,$,  respectively. The conservation of mass in local form reads $\rho J = \rho_o \,$,  which yields the continuity equation
%-----------------------------
\begin{equation}\nonumber
\frac{d\rho}{dt} + \rho\, \operatorname{div}\mathbf v = 0 \,,
\end{equation}
%-----------------------------
where $\operatorname{div}=\operatorname{div}_{\mathbf{g}}$ denotes the spatial Levi-Civita divergence operator corresponding to the metric $\mathbf{g}\,$.

%-----------------------------
%-----------------------------
\subsection{The Lagrange-d'Alembert principle}

The configuration of a viscoelastic body is given by a pair $(\varphi,\Fv)\,$,  and we denote the configuration space containing all such pairs by $\mathfrak{C}\,$. The governing equations of the viscoelastic body can be derived as the Euler-Lagrange equations associated with a variational principle defined as follows: Let us fix a time interval $[t_{0} , t_1]\,$,  and look at paths $c:[t_{0},t_1]\to\mathfrak{C}$  in the configuration space such that $c(t_{0})$ and $c(t_1)$ are fixed.
We define an action functional $\mathsf{S}$ on the space of paths as
%---------------------
\begin{equation}
	\mathsf{S} ( c ) = \int_{t_0}^{t_1}\!\int_{\mathcal B} \mathscr{L}  \, dV dt \,,
\end{equation}
%---------------------
where $\mathscr{L}$ is the Lagrangian density per unit undeformed volume, defined as ${\mathscr{L} = \mathscr{T} - \Psi}\,$,  where $\mathscr{T} = \frac{1}{2}\rho_o\Vert\mathbf V\Vert^2_{\mathbf g}$ is the kinetic energy density per unit undeformed volume. One may hence choose the following functional dependence\footnote{One instead may equivalently choose the functional dependence $\mathscr L=\hat{\mathscr L}(X,\Theta,\mathbf V,\mathbf F, \Fe, \mathbf G, \mathbf g)\,$.}
%---------------------
\begin{equation}\label{eq:Lagrangian}
	\mathscr L=\hat{\mathscr L}(X,\Theta,\mathbf V,\mathbf C^\flat, \Ce^\flat, \mathbf G, \mathbf g)
	=\frac{1}{2}\rho_o\Vert\mathbf V\Vert^2_{\mathbf g} 
	- \hat{\Psi}(X,\Theta,\mathbf{C}^{\flat},\Ce^{\flat},\mathbf{G})\,.
\end{equation}
%---------------------
Variations of the generalized configurations $(\varphi,\Fv)$ are represented by a one-parameter family $(\varphi_{t,\epsilon},\Fv_{t,\epsilon})$ of motions $\varphi_{t,\epsilon}$ and viscous deformation gradients $\Fv_{t,\epsilon}$ such that
%---------------------
\begin{subequations}\label{cd_variations}
\begin{align}
	(\varphi_{t,0},\Fv_{t,0})&=(\varphi_t,\Fv_t)\,, ~~\qquad \forall t\in[t_0,t_1]\,,\\
	(\varphi_{t_0,\epsilon},\Fv_{t_0,\epsilon})&=(\varphi_{t_0},\Fv_{t_0})\,, \qquad
	\forall \epsilon >0\,,\label{cd_variations3}\\
	(\varphi_{t_1,\epsilon},\Fv_{t_1,\epsilon})&=(\varphi_{t_1},\Fv_{t_1})\,, 
	\qquad \forall \epsilon >0\,.	
	\label{cd_variations4}
\end{align}
\end{subequations}
%---------------------
Notice that $\delta\varphi=d\varphi_{t,\epsilon}(X)/d\epsilon \vert _{\epsilon=0}$ is a vector in the ambient space, whereas {$\delta\Fv=d\Fv_{t,\epsilon}(X)/d\epsilon \vert _{\epsilon=0}$} is a material tensor, i.e.,~a tensor in the material manifold. The Lagrange--d'Alembert variational principle states that the physical configuration of the body satisfies the following identity~\citep{MarsRat2013MechSym}\footnote{In this work, we assume that the temperature field remains unaltered by perturbations of the deformation mapping, i.e.,~$\delta\Theta=0\,$. Otherwise, in the case of thermoelasticity, we ought to consider variations of the temperature field\textemdash see~\citet{Sadik2017Thermoelasticity}.}
%---------------------
\begin{equation} \label{eq:LD-Principle}
\begin{split}
	\delta \int_{t_0}^{t_1}\!\int_{\mathcal B} \mathscr{L}  \, dV dt
	+ \int_{t_0}^{t_1} \int_{\mathcal B} \Bv\!:\!\delta\Fv   \, dV dt
	+ \int_{t_0}^{t_1} \int_{\mathcal B} \rho_o \llangle \boldsymbol{\mathsf{B}} , 
	\delta\varphi \rrangle_{\mathbf{g}}  \, dV dt
	+ \int_{t_0}^{t_1} \int_{\partial \mathcal B} \llangle \boldsymbol{\mathsf{T}} , 
	\delta\varphi \rrangle_{\mathbf{g}}  \, dA \, dt
	= 0 \,,
\end{split}
\end{equation}
%---------------------
for all variation fields $\delta\varphi$ and $\delta\Fv\,$.
The vector fields $\boldsymbol{\mathsf{B}}=\boldsymbol{\mathsf{B}}(X,t)$ and $\boldsymbol{\mathsf{T}}=\boldsymbol{\mathsf{T}}(X,t)$ are the body force per unit mass and the boundary traction fields per unit undeformed area, respectively.
It follows from the Lagrange--d'Alembert variational principle \eqref{eq:LD-Principle} that
%---------------------
\begin{equation}\label{eq:LD1}
\begin{split}
	 & \int_{t_0}^{t_1}\!\int_{\mathcal B} \left(
	\frac{\partial \hat{\mathscr L}}{\partial \mathbf V} \delta \mathbf V
	+\frac{\partial \hat{\mathscr L}}{\partial \mathbf C^\flat}\!:\!\delta \mathbf C^\flat
	+ \frac{\partial \hat{\mathscr L}}{\partial \Ce^\flat}\!:\!\delta \Ce^\flat
	+ \frac{\partial \hat{\mathscr L}}{\partial \mathbf G}\!:\!\delta \mathbf G
	+ \frac{\partial \hat{\mathscr L}}{\partial \mathbf g}\!:\!\delta \mathbf g 
	\right) \, dV dt  \\
	& \quad + \int_{t_0}^{t_1} \int_{\mathcal B} \Bv\!:\!\delta\Fv   \, dV dt
	+ \int_{t_0}^{t_1} \int_{\mathcal B} \rho_o \llangle \boldsymbol{\mathsf{B}} , 
	\delta\varphi \rrangle_{\mathbf{g}} \, dV\, dt
	+ \int_{t_0}^{t_1} \int_{\partial \mathcal B} \llangle \boldsymbol{\mathsf{T}} , 
	\delta\varphi \rrangle_{\mathbf{g}}  \, dA \, dt
	= 0 \,.
\end{split}
\end{equation}
%---------------------
Using \eqref{Dissipative_B}, \eqref{eq:LD_Vf}, \eqref{eq:LD_Cf}, and \eqref{eq:LD_Cef}-\eqref{eq:LD_g} as detailed in Appendix~\ref{eq:LD_deets}, the above identity is simplified to read
%---------------------
\begin{equation}\label{eq:LD_f}
\begin{split}
	\int_{t_0}^{t_1}\!\int_{\mathcal B} \left[
	\llangle - \rho_o \mathbf A + \operatorname{Div}\left(2\mathbf F \frac{\partial \hat{\Psi}}
	{\partial \mathbf C^\flat} + 2 \Fe\frac{\partial \hat{\Psi}}{\partial \Ce^\flat} \right) 
	+ \rho_o \boldsymbol{\mathsf{B}}, \delta\varphi \rrangle_{\mathbf g}  
	+ \left(2\Ce^\flat\frac{\partial \hat{\Psi}}{\partial \Ce^\flat}\Fv^{-\star} -\frac{\partial \phi}
	{\partial \dot\Fv} \right): \delta\Fv  \right] \, dV dt &\\
	- \int_{t_0}^{t_1}\int_{\partial \mathcal B} \llangle \left(2\mathbf F \frac{\partial \hat{\Psi}}
	{\partial \mathbf C^\flat} +2 \Fe\frac{\partial \hat{\Psi}}{\partial \Ce^\flat}\right)
	\mathbf N - \boldsymbol{\mathsf{T}} , \delta\varphi \rrangle_{\mathbf{g}} \,dA\,dt  
	&= 0\,,
\end{split}
\end{equation}
%---------------------
where $\mathbf N$ is the $\mathbf G$-unit normal to $\partial \mathcal B\,$.
Since \eqref{eq:LD_f} above is valid for all variations $\delta\varphi$ and $\delta\Fv\,$, 
one finds the balance of linear momentum together with its boundary conditions\footnote{The balance of angular momentum will later be discussed in \S\ref{sec:const_rel}.}
%---------------------
\begin{subequations}\label{eq:LD_LinM}
\begin{align}
	\label{eq:LD_LinM1}
	\operatorname{Div}\left[2\mathbf F \frac{\partial \hat{\Psi}}{\partial \mathbf C^\flat} 
	+2 \Fe\frac{\partial \hat{\Psi}}{\partial \Ce^\flat}\right] 
	+ \rho_o \boldsymbol{\mathsf{B}} &= \rho_o \mathbf A\,,\\
	\left.\left[2\mathbf F \frac{\partial \hat{\Psi}}{\partial \mathbf C^\flat} 
	+2 \Fe\frac{\partial \hat{\Psi}}{\partial \Ce^\flat}\right]\right|_{\partial\mathcal B} \mathbf N 
	&= \boldsymbol{\mathsf{T}}\,,
\end{align}
\end{subequations}
%---------------------
where $\operatorname{Div}$ denotes the two-point Levi-Civita divergence operator.\footnote{
For a two-point tensor with components $Q^{aA}\,$, $\operatorname{Div}\mathbf{Q}$ has components, $Q^{aA}{}_{|A}=Q^{aA}{}_{,A}+\Gamma^A{}_{AB}\,Q^{aB}+\gamma^a{}_{bc}F^b{}_A\,Q^{cA}\,$.}
One also finds the kinetic equation governing the evolution of the internal variable $\Fv$: 
%---------------------
\begin{equation}\label{eq:LD_kinetic}
	\frac{\partial \phi}{\partial \dot\Fv} - 2\Ce^\flat\frac{\partial \hat{\Psi}}{\partial \Ce^\flat}\Fv^{-\star} = \mathbf 0\,.
\end{equation}
%---------------------
The initial condition for the kinetic equation is in physical components $\widehat{\Fv}(X,0)=\mathbf{I}\,$.
For the deformation map, $\varphi(X,0)=\iota_{\mathcal{B}}(X)\,$, and $\partial \varphi(X,0)/\partial t=\mathbf{V}_0\,$, where $\iota_{\mathcal{B}}$ is the inclusion map and $\mathbf{V}_0$ is the initial velocity field of the viscoelastic body.

\paragraph{Incompressible viscoelastic solids.}
For an incompressible viscoelastic solid, one assumes both the total deformation and its purely viscous part to be volume preserving \citep{Leonov1976,LeTallec1993}, i.e.,
%---------------------
\begin{equation}\label{eq:incompressibility}
	J=\det\mathbf{F}\sqrt{\frac{\det\mathbf{g}}{\det\mathbf{G}}}=1\,,\qquad \Jv=\det\Fv=1\,.
\end{equation}
%---------------------
The free energy is hence augmented by the above constraints and their corresponding Lagrange multipliers, i.e., the free energy for the incompressible solid is modified to read $\Psi_{\text{inc}}=\Psi -p(J-1) -q(\Jv-1)\,$, where $p=p(X,t)$ and $q=q(X,t)$ are the Lagrange multipliers corresponding to the constraints given in \eqref{eq:incompressibility}. Therefore, the Lagrangian \eqref{eq:Lagrangian} is modified to read $\mathscr L_{\text{inc}} = \mathscr L+p(J-1) +q(\Jv-1)\,$. Now, the Lagrange-d'Alembert principle reads
%---------------------
\begin{equation} \label{eq:inc_LD-Principle}
\begin{split}
	\int_{t_0}^{t_1}\!\int_{\mathcal B} \left(\delta \mathscr{L} 
	+ p\,\delta J |_{\accentset{{}^{}}J=1} + q\,\delta \Jv |_{\Jv=1}
	+   \Bv\!:\!\delta\Fv 
	+  \rho_o \llangle \boldsymbol{\mathsf{B}} , \delta\varphi \rrangle_{\mathbf{g}}  \right) \, 
	dV dt
	+ \int_{t_0}^{t_1} \int_{\partial \mathcal B} \llangle \boldsymbol{\mathsf{T}} , 
	\delta\varphi \rrangle_{\mathbf{g}}  \, dA \, dt
	= 0 \,.
\end{split}
\end{equation}
%---------------------
Consequently, by using \eqref{Dissipative_B}, \eqref{eq:LD_Vf}, \eqref{eq:LD_Cf}, \eqref{eq:LD_Cef}-\eqref{eq:LD_g}, \eqref{eq:LD_J}, and \eqref{eq:LD_Jv} as detailed in Appendix~\ref{eq:LD_deets}, \eqref{eq:inc_LD-Principle} yields
%---------------------
\begin{subequations}\label{eq:inc_LD_LinM}
\begin{align}
	\operatorname{Div}\left(2\mathbf F \frac{\partial \hat{\Psi}}{\partial \mathbf C^\flat} +2 \Fe\frac{\partial \hat{\Psi}}{\partial \Ce^\flat}-p \mathbf g^\sharp \mathbf F^{-\star}\right) + \rho_o \boldsymbol{\mathsf{B}} &= \rho_o \mathbf A\,,\\
	\left.\left(2\mathbf F \frac{\partial \hat{\Psi}}{\partial \mathbf C^\flat} +2 \Fe\frac{\partial \hat{\Psi}}{\partial \Ce^\flat}-p \mathbf g^\sharp \mathbf F^{-\star}\right)\right|_{\partial \mathcal B} \mathbf N &= \boldsymbol{\mathsf{T}}\,,
\end{align}
\end{subequations}
%---------------------
and
%---------------------
\begin{equation}\label{eq:LD_kinetic_inc}
	\frac{\partial \phi}{\partial \dot\Fv} 
	- 2\Ce^\flat\frac{\partial \hat{\Psi}}{\partial \Ce^\flat}\Fv^{-\star} = q \Fv^{-\star}\,.
\end{equation}
%---------------------

%---------------------
\begin{remark}
Note that if one considers the equivalent functional dependence for $\Psi$ in terms of $\Fv\,$, i.e., ${\Psi=\check\Psi(X,\mathbf F,\Fv,\mathbf G,\mathbf g)=\Psi(X,\mathbf F,\mathbf F \Fv^{-1},\mathbf{G},\mathbf{g})}\,$, it may be seen that
%---------------------
\begin{equation}\label{eq:der_Ce->Fv}
	2\Ce^\flat\frac{\partial \hat\Psi}{\partial\Ce^\flat}\Fv^{-\star} = -\frac{\partial \check\Psi}{\partial \Fv}\,.
\end{equation}
%---------------------
This consequently transforms the kinetic equations \eqref{eq:LD_kinetic} and \eqref{eq:LD_kinetic_inc}, respectively, to
%---------------------
\begin{subequations} \label{Kinetic-Equation}
\begin{align}
	\frac{\partial \phi}{\partial \dot\Fv} + \frac{\partial \check\Psi}{\partial \Fv} &= \mathbf 0 
	&& \text{for compressible viscoelastic solids,}\label{eq:comp_kinetic}\\
	\frac{\partial \phi}{\partial \dot\Fv} + \frac{\partial \check\Psi}{\partial \Fv} &= q \Fv^{-\star}
	&& \text{for incompressible viscoelastic solids.}\label{eq:incomp_kinetic}
\end{align}
\end{subequations}
%---------------------
Eq. \eqref{eq:comp_kinetic} is identical to Eq.~(7-b) as it appears in \citep{Kumar2016} and Eq. \eqref{eq:incomp_kinetic} is equivalent to Eq.~(3.10) as it appears in \citep{LeTallec1993}.

\end{remark}
%---------------------

%-----------------------------
%-----------------------------
\subsection{Thermodynamics of viscoelasticity}

%-----------------------------
%-----------------------------
\subsubsection{The first law of thermodynamics}

The first law of thermodynamics postulates the existence of a state functional, namely the internal energy, which satisfies the following balance of energy \citep{truesdell1952mechanical, gurtin1974modern, MarsdenHughes1983, yavari2006spatial}
%-----------------------------
\begin{equation}\label{eq:Thermo_First}
\frac{d}{dt}\int_{\mathcal{U}} \left(\mathcal{E}+\frac{1}{2}\rho_o \Vert\mathbf V\Vert^2_{\mathbf g}\right)dV=\int_{\mathcal{U}}\rho_o \left(\llangle\boldsymbol{\mathsf{B}},\mathbf{V}\rrangle_{\mathbf{g}}+R\right)dV+\int_{\partial \mathcal{U}}\left(\llangle\boldsymbol{\mathsf{T}},\mathbf{V}\rrangle_{\mathbf{g}}+H\right)dA\,,
\end{equation}
%-----------------------------
where $\mathcal{E}=\hat{\mathcal E}(X, \mathcal N, \mathbf C^\flat, \Ce^\flat, \mathbf G)$ is the material internal energy density (per unit undeformed volume), $R=R(X,t)$ is the heat supply per unit mass, and $H$ is the heat flux across a material surface; which may be written as $H=-\llangle \mathbf{Q},\mathbf{N}\rrangle_{\mathbf{G}}\,$, where $\mathbf Q=\mathbf Q(X,\Theta,d\Theta,\mathbf C^\flat,\mathbf G)$ is the external heat flux per unit material area, $\mathbf{N}$ is the $\mathbf{G}$-unit normal to the boundary $\partial\mathcal B\,$, and $\Theta=\Theta(X,t)$ is the temperature field.
In local form, the energy balance \eqref{eq:Thermo_First} reads\footnote{Note that the localization of the energy balance \eqref{eq:Thermo_First} is typically presented without the last term appearing in \eqref{eq:loc_Thermo_First} that vanishes after imposing the balance of linear momentum. At this point in this work, even though we have already proven the balance of linear momentum \eqref{eq:inc_LD_LinM} in terms of the tensorial derivatives of the free energy, we have not yet proven the Doyle-Ericksen formula (the stress constitutive equation) relating them to the stress tensors, and may hence not yet impose that the last term in \eqref{eq:loc_Thermo_First} vanishes.}

%-----------------------------
\begin{equation}\label{eq:loc_Thermo_First}
	\dot{\mathcal E} = \mathbf S\!:\! \mathbf D - \operatorname{Div} \mathbf Q +\rho_o R
	+ \llangle \operatorname{Div}\mathbf{P}+ \rho_o(\boldsymbol{\mathsf{B}} 
	- \mathbf{A}),\mathbf{V} \rrangle_{\mathbf{g}} \,,
\end{equation}
%-----------------------------
where a dotted quantity denotes its total time derivative, $\mathbf P$ is the first Piola-Kirchhoff stress tensor\textemdash$\boldsymbol{\mathsf{T}} = \mathbf P \mathbf N\,$, $\mathbf S=\mathbf F^{-1}\mathbf P$ is the second Piola-Kirchhoff stress tensor, and $\mathbf D = \frac{1}{2}\dot{\mathbf C}^\flat$ is the material rate of deformation tensor.

%-----------------------------
%-----------------------------
\subsubsection{The second law of thermodynamics}

The second law of thermodynamics postulates the existence of a state functional, namely the entropy, which satisfies the material Clausius-Duhem inequality \citep{truesdell1952mechanical, gurtin1974modern, MarsdenHughes1983}
%-----------------------------
\begin{equation}\label{Thermo_Second}
\frac{d}{dt}\int_{\mathcal{U}} \mathcal{N}dV\geq\int_{\mathcal{U}}\rho_o \frac{R}{\Theta}dV+\int_{\partial\mathcal{U}}\frac{H}{\Theta}dA\,,
\end{equation}
%-----------------------------
where $\mathcal{N}= \hat{\mathcal N}(X, \Theta, \mathbf C^\flat, \Ce^\flat, \mathbf G)$ is the material entropy density (per unit undeformed volume).
In localized form, the material Clausius-Duhem inequality \eqref{Thermo_Second} reads
%-----------------------------
\begin{equation}
\label{loc_Thermo_Second}
\dot\eta = \dot{\mathcal{N}}\Theta + \Theta\operatorname{Div}\left(\frac{\mathbf Q}{\Theta}\right) - \rho_o R \geq 0\,,
\end{equation}
%-----------------------------
where $\dot\eta$ is the rate of energy dissipation.

%-----------------------------
%-----------------------------
\subsubsection{Constitutive relations and balance laws}
\label{sec:const_rel}

The free energy density $\Psi$ is the Legendre transform of the internal energy density $\mathcal E$ with respect to the conjugate variables of temperature $\Theta$ and $\mathcal N\,$, i.e.,
%-----------------------------
\begin{equation}\label{eq:Legendre_trans}
\Psi = \mathcal E - \Theta \mathcal N\,,
\end{equation}
%-----------------------------
 where ${\mathcal E = \hat{\mathcal E}(X, \mathcal N, \mathbf C^\flat, \Ce^\flat, \mathbf G)}\,$, ${\mathcal N = \hat{\mathcal N}(X, \Theta, \mathbf C^\flat, \Ce^\flat, \mathbf G)}\,$, and ${\Psi = \hat\Psi(X, \Theta, \mathbf C^\flat, \Ce^\flat, \mathbf G)}\,$.
It hence follows from \eqref{loc_Thermo_Second} that
%-----------------------------
\begin{equation}\label{eq:loc_Thermo_Second_1}
\dot\eta = \dot{\mathcal E} -\dot\Psi - \dot\Theta \mathcal N+ \operatorname{Div}\mathbf Q - \frac{1}{\Theta} \langle d\Theta, \mathbf Q \rangle -\rho_o R \geq 0\,.
\end{equation}
%-----------------------------
Using \eqref{eq:loc_Thermo_First} in \eqref{eq:loc_Thermo_Second_1} and expanding $\dot\Psi\,$, one finds
%-----------------------------
\begin{equation}\label{eq:loc_Thermo_Second_2}
\dot\eta = \frac{1}{2} \mathbf S\!:\!\dot{\mathbf C}^\flat -\frac{\partial \hat\Psi}{\partial \Theta}\dot\Theta -\frac{\partial \hat\Psi}{\partial \mathbf C^\flat}\!:\!\dot{\mathbf C}^\flat -\frac{\partial \hat\Psi}{\partial\Ce^\flat} \!:\! \dot\Ce^\flat -\dot\Theta \mathcal N - \frac{1}{\Theta} \langle d\Theta, \mathbf Q \rangle 
+\llangle \operatorname{Div}\mathbf{P}+ \rho_o(\boldsymbol{\mathsf{B}} - \mathbf{A}),\mathbf{V} \rrangle_{\mathbf{g}}
\geq 0\,.
\end{equation}
%-----------------------------
It can be seen that\footnote{Note that $\dot\Ce^\flat=\Fv^{-\star}\dot{\mathbf C}^\flat\Fv^{-1} - \Ce^\flat\dot\Fv\Fv^{-1} - \Fv^{-\star}\dot\Fv\Ce^\flat\,$, which follows from $\Ce^\flat=\Fv_*\mathbf C^\flat=\Fv^{-\star}\mathbf C^\flat\Fv^{-1}\,$.}
%-----------------------------
\begin{equation}\nonumber
	\frac{\partial \hat\Psi}{\partial\Ce^\flat} \!:\! \dot\Ce^\flat 
	= \Fv^{-1}\frac{\partial \hat\Psi}{\partial\Ce^\flat}\Fv^{-\star} \!:\! \dot{\mathbf C}^\flat 
	- 2\Ce^\flat\frac{\partial \hat\Psi}{\partial\Ce^\flat}\Fv^{-\star} \!:\! \dot\Fv
	\,,
\end{equation}
%-----------------------------
and it follows that \eqref{eq:loc_Thermo_Second_2} is simplified to read
%-----------------------------
\begin{equation}\label{eq:loc_Thermo_Second_3}
\begin{aligned}
	\dot\eta & = \left(\mathcal N +\frac{\partial \hat\Psi}{\partial \Theta}\right)\dot\Theta
	+ \frac{1}{2}\left(\mathbf S - 2\frac{\partial \hat\Psi}{\partial \mathbf C^\flat} 
	- 2\Fv^{-1}\frac{\partial \hat\Psi}{\partial\Ce^\flat}\Fv^{-\star} \right)\!:\!\dot{\mathbf C}^\flat \\
	&\quad +\llangle \operatorname{Div}\mathbf{P}+ \rho_o(\boldsymbol{\mathsf{B}} - \mathbf{A}),\mathbf{V} \rrangle_{\mathbf{g}}
	+ 2\Ce^\flat\frac{\partial \hat\Psi}{\partial\Ce^\flat}\Fv^{-\star} \!:\! \dot\Fv
	- \frac{1}{\Theta} \langle d\Theta, \mathbf Q \rangle \geq 0\,.
\end{aligned}
\end{equation}
%-----------------------------
The inequality \eqref{eq:loc_Thermo_Second_3} above must hold for all deformations $\varphi$ and temperature fields $\Theta\,$. Hence, it follows that\footnote{Following \eqref{eq:LD_kinetic}, one has $2\Ce^\flat({\partial \hat{\Psi}_{\text{NEQ}}}/{\partial \Ce^\flat})\Fv^{-\star} = {\partial \phi(X,\mathbf{F},\Fv,\dot\Fv,\mathbf{G},\mathbf{g})}/{\partial \dot\Fv}\,$; and recall that $\mathbf Q = \mathbf Q(X,\Theta,d\Theta,\mathbf C,\mathbf G)\,$. Therefore, the coefficients of $\dot\Fv$ and $d\Theta$ in the inequality \eqref{eq:loc_Thermo_Second_3} depend on $\dot\Fv$ and $d\Theta\,$, respectively; hence, they may not be identically equal to zero in spite of the arbitrariness of $\dot\Fv$ and $d\Theta\,$. Also, recall from \eqref{eq:Psi_C_decomp} that $\hat\Psi(X,\Theta,\mathbf C^\flat, \Ce^\flat, \mathbf G)=\hat\Psi_{\text{EQ}}(X,\Theta,\mathbf C^\flat, \mathbf G)+\hat\Psi_{\text{NEQ}}(X,\Theta,\Ce^\flat, \mathbf G)\,$.}
%-----------------------------
\begin{subequations}\label{eq:Cons_Eq}
\begin{align}
	\mathcal N &= -\frac{\partial \hat\Psi}{\partial \Theta}\,,\\
	\label{eq:Cons_Eq_stress}
	\mathbf S &= 2\frac{d \hat\Psi}{d \mathbf C^\flat} = 2\frac{\partial \hat\Psi_{\text{EQ}}}{\partial \mathbf C^\flat}
	+ 2\Fv^{-1}\frac{\partial \hat\Psi_{\text{NEQ}}}{\partial\Ce^\flat}\Fv^{-\star}\,.
\end{align}
\end{subequations}
%-----------------------------
The first Piola-Kirchhoff $\mathbf P\,$, the Cauchy $\boldsymbol \sigma\,$, and the convected stress $\boldsymbol \Sigma$ tensors\footnote{Recall that $\mathbf S = \mathbf F^{-1}  \mathbf P = J \boldsymbol \Sigma = J \mathbf F^{-1} \boldsymbol \sigma \mathbf F^{-\star}\,$.} may hence be written as 
%-----------------------------
\begin{subequations}\label{eq:other_stress}
\begin{align}
	\mathbf P
	&= \mathbf g^\sharp \frac{d \Psi}{d \mathbf F}
	= \mathbf g^\sharp \frac{\partial \Psi_{\text{EQ}}}{\partial \mathbf F} 
	+\mathbf{g}^{\sharp}\frac{\partial \Psi_{\text{NEQ}}}{\partial \Fe}\,\Fv^{-\star}
	\,,\label{eq:other_stress_P}\\
	\boldsymbol \sigma
	&= \frac{2}{J}\frac{d \Psi}{d \mathbf g}
	= \frac{2}{J}\frac{\partial \Psi_{\text{EQ}}}{\partial \mathbf g} 
	+ \frac{2}{J}\frac{\partial \Psi_{\text{NEQ}}}{\partial \mathbf g}
	\,,\\
	\boldsymbol\Sigma
	&= \frac{2}{J}\frac{d \hat\Psi}{d \mathbf C^\flat} 
	= \frac{2}{J}\frac{\partial \hat\Psi_{\text{EQ}}}{\partial \mathbf C^\flat} 
	+ \frac{2}{J}\Fv^{-1}\frac{\partial \hat\Psi_{\text{NEQ}}}{\partial\Ce^\flat}\Fv^{-\star}\,.
\end{align}
\end{subequations}
%-----------------------------
Substituting the stress constitutive equation \eqref{eq:other_stress_P} for $\mathbf P$ into \eqref{eq:LD_LinM} yields the balance of linear momentum in the two-point tensorial form and the traction boundary condition:
%---------------------
\begin{subequations}\label{eq:Bal_LinM}
\begin{align}
	\operatorname{Div}\mathbf P + \rho_o \boldsymbol{\mathsf{B}} &= \rho_o \mathbf A\,,\label{eq:Bal_LinM1}\\
	\left.\mathbf P\right|_{\partial\mathcal B} \mathbf N &= \boldsymbol{\mathsf{T}}\,.
\end{align}
\end{subequations}
%---------------------
From \eqref{eq:Cons_Eq_stress} and since $\mathbf C^{\star}=\mathbf C\,$, one finds the balance of angular momentum in two-point tensorial form
%-----------------------------
\begin{equation}
	\mathbf P^{\star} \mathbf F^{-\star} = \mathbf F^{-1} \mathbf P\,.
\end{equation}
%-----------------------------
In spatial form, the balance of linear and angular momenta and the traction boundary condition read
%---------------------
\begin{subequations}\label{eq:s_Bal_LinM}
\begin{align}
	\operatorname{div}\boldsymbol \sigma + \rho \boldsymbol{\mathsf{b}} &= \rho \mathbf a\,,\\
	\boldsymbol \sigma^\star &=\boldsymbol \sigma\,,\\
	\left.\boldsymbol \sigma\right|_{\partial\mathcal B} \mathbf n &= \boldsymbol{\mathsf{t}}\,,
\end{align}
\end{subequations}
%---------------------
where $\boldsymbol{\mathsf{b}}=\boldsymbol{\mathsf{B}}\circ\varphi_t^{-1}\,$, $\boldsymbol{\mathsf{t}} = J \boldsymbol{\mathsf{T}}\circ\varphi_t^{-1}\,$, and $\mathbf n=\mathbf N\circ\varphi_t^{-1}\,$.
%---------------------
In convected form, the balance of linear and angular momenta and the traction boundary condition read
%---------------------
\begin{subequations}\label{eq:C_Bal_LinM}
\begin{align}
	\operatorname{Div}_{\mathbf C^\flat}\boldsymbol \Sigma + \varphi_t^*(\rho \boldsymbol{\mathsf{b}}) &= \varphi_t^*(\rho \mathbf a)\,,\label{eq:C_Bal_LinM1}\\
	\boldsymbol \Sigma^\star &=\boldsymbol \Sigma\,,\\
	\left.\boldsymbol \Sigma\right|_{\partial\mathcal B} \mathbf N &= \varphi_t^*\boldsymbol{\mathsf{t}}\,,
\end{align}
\end{subequations}
%---------------------
where $\operatorname{Div}_{\mathbf{C}^{\flat}}$ denotes the convected Levi-Civita divergence operator, i.e., the divergence operator associated with the Levi-Civita connection of the convected manifold $(\mathcal B, \mathbf C^\flat)\,$.

Within the scope of this work, we assume that the viscoelastic body undergoes an isothermal process, i.e., ${d\Theta=0}\,$; hence, using \eqref{eq:Cons_Eq} and \eqref{eq:Bal_LinM}, the Clausius-Duhem inequality \eqref{eq:loc_Thermo_Second_3} reduces to read
%-----------------------------
\begin{equation}\label{eq:loc_Thermo_Second_CD}
\dot\eta = 2\Ce^\flat\frac{\partial \hat\Psi_{\text{NEQ}}}{\partial\Ce^\flat}\Fv^{-\star} \!:\! \dot\Fv \geq 0\,.
\end{equation}
%-----------------------------

\paragraph{Incompressible viscoelastic solids} For an incompressible viscoelastic solid, the Legendre transform \eqref{eq:Legendre_trans} is modified to take into account the volume preserving constraints \eqref{eq:incompressibility} as follows
%-----------------------------
\begin{equation}\label{eq:Legendre_trans_inc}
\Psi -p(J-1) - q(\Jv-1)= \mathcal E - \Theta \mathcal N\,.
\end{equation}
%-----------------------------
Similarily to \eqref{eq:del_J} and \eqref{eq:del_Jv}, one finds
%-----------------------------
\begin{equation}\label{eq:Legendre_trans_inc}
	\dot J =\frac{1}{2}J \mathbf C^{-\sharp}\!:\!\dot{\mathbf C}^\flat\quad\text{and}\qquad
	 \dot\Jv= \Jv \Fv^{-\star}\!:\!\dot\Fv\,.
\end{equation}
%-----------------------------
One hence simplifies the constitutive relations \eqref{eq:Cons_Eq} and \eqref{eq:other_stress} to read
%-----------------------------
\begin{subequations}\label{eq:Cons_Eq_inc}
\begin{align}
	\mathcal N &= -\frac{\partial \hat\Psi}{\partial \Theta}\,,\\
	\label{eq:Cons_Eq_stress_inc}
	\mathbf S &= 2\frac{d \hat\Psi}{d \mathbf C^\flat}-p\mathbf C^{-\sharp}
	= 2\frac{\partial \hat\Psi_{\text{EQ}}}{\partial \mathbf C^\flat} + 2\Fv^{-1}\frac{\partial \hat\Psi_{\text{NEQ}}}{\partial\Ce^\flat}\Fv^{-\star} -p\mathbf C^{-\sharp}\,,\\
	\mathbf P
	&= \mathbf g^\sharp \frac{d \Psi}{d \mathbf F}-p \mathbf g^\sharp \mathbf F^{-\star}
	= \mathbf g^\sharp \frac{\partial \Psi_{\text{EQ}}}{\partial \mathbf F} +\mathbf{g}^{\sharp}\frac{\partial \Psi_{\text{NEQ}}}{\partial \Fe}\,\Fv^{-\star}-p \mathbf g^\sharp \mathbf F^{-\star}
	\,,\\
	\boldsymbol \sigma
	&= 2\frac{d \Psi}{d \mathbf g}-p \mathbf g^\sharp
	= 2\frac{\partial \Psi_{\text{EQ}}}{\partial \mathbf g} + 2\frac{\partial \Psi_{\text{NEQ}}}{\partial \mathbf g}-p \mathbf g^\sharp
	\,,\\
	\boldsymbol\Sigma
	&= 2\frac{d \hat\Psi}{d \mathbf C^\flat} -p\mathbf C^{-\sharp}
	= 2\frac{\partial \hat\Psi_{\text{EQ}}}{\partial \mathbf C^\flat} + 2\Fv^{-1}\frac{\partial \hat\Psi_{\text{NEQ}}}{\partial\Ce^\flat}\Fv^{-\star}-p\mathbf C^{-\sharp}\,.
\end{align}
\end{subequations}
%-----------------------------
The Clausius-Duhem inequality \eqref{eq:loc_Thermo_Second_CD} is rewritten as
%-----------------------------
\begin{equation}\label{eq:loc_Thermo_Second_CD_inc}
\dot\eta = 2\Ce^\flat\frac{\partial \hat\Psi_{\text{NEQ}}}{\partial\Ce^\flat}\Fv^{-\star} \!:\! \dot\Fv +q \Fv^{-\star}  \geq 0\,.
\end{equation}
%-----------------------------

%-----------------------------
\begin{remark}\label{rm:non-neg_entr}
From the convexity of the dissipation potential $\phi$ as defined in \eqref{eq:convex}, if one assumes that for fixed $\Fv\,$, $\phi$ is minimized for $\dot\Fv=\mathbf{0}\,$, it follows that
%---------------------
\begin{equation} \label{eq:convex_c}
	 \frac{\partial \phi}{\partial \dot\Fv}:\dot\Fv \geq 0 \,,
\end{equation}
%---------------------
which is consistent with the Clausius-Duhem inequality stating that the rate of energy dissipation is non-negative, i.e., $\dot\eta \geq 0\,$.\footnote{As a matter of fact, \eqref{eq:convex_c} may be alternatively found following \eqref{eq:LD_kinetic} and \eqref{eq:loc_Thermo_Second_CD}\textemdash or \eqref{eq:LD_kinetic_inc} and \eqref{eq:loc_Thermo_Second_CD_inc} for the incompressible case.}
\end{remark}
%-----------------------------

From here on, consistent with the isothermal process assumption, we may drop the temperature dependance in both the free energy and the dissipation potential, i.e., $\Psi=\Psi(X,\mathbf F,\Fe,\mathbf G,\mathbf g)=\hat\Psi(X,\mathbf C^\flat,\Ce^\flat,\mathbf G)$ and ${\phi=\phi(X,\mathbf{F},\Fv,\dot\Fv,\mathbf{G},\mathbf{g})=\hat\phi(X,\mathbf{C}^\flat,\Fv,\dot\Fv,\mathbf{G})}\,$.

%-----------------------------
%-----------------------------
\subsection{Stress in the intermediate configurations in anelasticity versus viscoelasticity}
\label{InterConfig}

\paragraph{Anelasticity.}
In anelasticity, the energy density has the form $W=\mathring{W}(X,\Fe,\mathbf{G},\mathbf{g})\,$. The first Piola-Kirchhoff stress is calculated as
%---------------------
\begin{equation}
	\mathbf{P}=\mathbf{g}^{\sharp}\frac{\partial W}{\partial \mathbf{F}}
	=\mathbf{g}^{\sharp}\frac{\partial \mathring{W}}{\partial \Fe}\frac{\partial \Fe}{\partial \mathbf{F}}
	=\mathbf{g}^{\sharp}\frac{\partial \mathring{W}}{\partial \Fe}\Fa^{-\star}
	\,.
\end{equation}
%---------------------
In components $P^{aA}=g^{ab} \,\partial \mathring{W}/ \cFe^b{}_M\,(\cFa^{-1})^A{}_M\,$. Stress in the intermediate configuration is calculated as
%---------------------
\begin{equation}
	\mathbf{P}_{\text{int}}
	=\mathbf{g}^{\sharp}\frac{\partial \mathring{W}}{\partial \Fe}\Bigg|_{\Fe=\operatorname{id}}\Fa^{-\star}=\mathbf{0}
	\,,
\end{equation}
%---------------------
as $\partial \mathring{W}/\partial \Fe$ vanishes in the absence of a local elastic deformation.

\paragraph{Viscoelasticity.}
In viscoelasticity, recall that the free energy density is written as $\Psi(X,\mathbf{F},\Fe,\mathbf{G},\mathbf{g})=\Psi_{\text{EQ}}(X,\mathbf{F},\mathbf{G},\mathbf{g})+\Psi_{\text{NEQ}}(X,\Fe,\mathbf{G},\mathbf{g})\,$, and hence stress is calculated as 
%---------------------
\begin{equation}\label{stress}
	\mathbf{P}=\mathbf{g}^{\sharp}\frac{d \Psi}{d \mathbf{F}}
	=\mathbf{g}^{\sharp}\frac{\partial \Psi_{\text{EQ}}}{\partial \mathbf{F}}
	+\mathbf{g}^{\sharp}\frac{\partial \Psi_{\text{NEQ}}}{\partial \Fe}\,\Fv^{-\star}	\,.
\end{equation}
%---------------------
Stress in the intermediate configuration is then calculated as
%---------------------
\begin{equation}
	\mathbf{P}_{\text{int}}
	=\mathbf{g}^{\sharp}\frac{d \Psi}{d \mathbf{F}}\Bigg|_{\Fe=\operatorname{id}}
	=\mathbf{g}^{\sharp}\frac{\partial \Psi_{\text{EQ}}}{\partial \mathbf{F}}\Bigg|_{\mathbf{F}=\Fv}
	+\mathbf{g}^{\sharp}\frac{\partial \Psi_{\text{NEQ}}}{\partial \Fe}\Bigg|_{\Fe=\operatorname{id}}\Fv^{-\star}
	=\mathbf{g}^{\sharp}\frac{\partial \Psi_{\text{EQ}}}{\partial \mathbf{F}}\Bigg|_{\mathbf{F}=\Fv}
	\,,
\end{equation}
%---------------------
as $\partial \Psi_{\text{NEQ}}/\partial \Fe$ vanishes when there is no local elastic deformation. This clearly shows that in viscoelasticity the local intermediate configuration is, in general, not stress free (see Fig.~\ref{Local-Configurations}).

\begin{remark}
If a small neighborhood of a material point in the current configuration is unloaded, the instantaneous unloaded length is calculated using the metric $\Fe_*\mathbf G\,$. However, this is not the natural length; the natural length is found in the reference configuration and is measured by $\mathbf G\,$. In the intermediate configuration, the natural length is calculated using the metric $\Fv_*\mathbf{G}\,$. As the intermediate configuration is not stress-free, the metric $\Fv_*\mathbf{G}$ is not of much physical significance and does not appear anywhere in the present theory. However, there is another metric that is of physical significance for the non-equilibrium free energy, see \S\ref{Sec:ViscousMetric}.
\end{remark}

%-----------------------------
\begin{remark}\label{rmrk:F_LargeTimes}
Let us consider a viscoelastic body that undergoes a motion in the time interval $[0,\infty)\,$. At time $t=T\,$, the applied loads (and/or boundary displacements) are held fixed. Thus, for $t>T\,$, $\mathbf{F}(X,t)=\mathbf{F}(X,T)=:\bar{\mathbf{F}}(X)\,$. However, note that the elastic and viscous deformation gradients are still time dependent, i.e.,
%-----------------------------
\begin{equation}
	\bar{\mathbf{F}}(X)=\Fe(X,t)\,\Fv(X,t)\,,\qquad t>T
	\,.
\end{equation}
%-----------------------------
In terms of the physical components $\widehat{\bar{\mathbf{F}}}(X)=\widehat{\Fe}(X,t)\,\widehat{\Fv}(X,t)\,$, where ${\widehat{\bar{F}}}^a{}_A=\sqrt{g_{aa}G^{AA}}\,F^a{}_A\,$, ${\widehat{\cFe}}^a{}_A=\sqrt{g_{aa}G^{AA}}\,\cFe^a{}_A\,$, and ${\widehat{\cFv}}^A{}_B=\cFv^A{}_B$ (no summation on repeated indices) \citep{Truesdell1953physical}.
We are interested in the evolution of $\Fe$ and $\Fv$ when $t\to\infty\,$.
Recall that $\Psi(X,\mathbf{F},\Fe,\mathbf{G},\mathbf{g})=\Psi_{\text{EQ}}(X,\mathbf{F},\mathbf{G},\mathbf{g})+\Psi_{\text{NEQ}}(X,\Fe,\mathbf{G},\mathbf{g})\,$, and that stress has ƒconsequently the following additive decomposition
%---------------------
\begin{equation}
	\mathbf{P}=\mathbf{P}_{\text{EQ}}+\mathbf{P}_{\text{NEQ}}\,,\qquad
	\mathbf{P}_{\text{EQ}}=\mathbf{g}^{\sharp}\frac{\partial \Psi_{\text{EQ}}}{\partial \mathbf{F}}\,,\qquad
	\mathbf{P}_{\text{NEQ}}=\mathbf{g}^{\sharp}\frac{\partial \Psi_{\text{NEQ}}}{\partial \Fe}\,\Fv^{-\star}	\,.
\end{equation}
%---------------------
For $t\to\infty\,$, $\mathbf{P}=\mathbf{P}_{\text{EQ}}\,$, i.e., $\mathbf{P}_{\text{NEQ}}=\mathbf{0}\,$. This implies that
%---------------------
\begin{equation}
	\lim_{t\to\infty}\frac{\partial \Psi_{\text{NEQ}}}{\partial \Fe}=\mathbf{0}	\,.
\end{equation}
%---------------------
Therefore, $\lim_{t\to\infty}\widehat{\Fe}(X,t)=\mathbf{I}\,$, and hence $\lim_{t\to\infty}\Fv(X,t)=\widehat{\bar{\mathbf{F}}}(X)\,$. This is the nonlinear analogue of what is observed in a stress-relaxation experiment for the standard solid model~\citep{Zener1948} that is briefly revised next (see~\citep{SimoHughes2006}), see Fig.~\ref{Standard-Solid}. The non-equilibrium stress has the following relation with the viscous strain: $P_{\text{NEQ}}(t)=\eta\,\dot{\epsilon}_v(t)\,$.
From equilibrium, $P_{\text{NEQ}}(t)=E_{\text{NEQ}}\,\epsilon_e(t)=E_{\text{NEQ}}\left(\epsilon(t)-\epsilon_v(t)\right)\,$. Thus, from these two equations, one obtains
 %---------------------
\begin{equation}
	\dot{\epsilon}_v(t)+\frac{1}{\eta}\epsilon_v(t)=\frac{1}{\eta}\epsilon(t)\,,\qquad \tau=\frac{\eta}{E_{\text{NEQ}}}
	\,.
\end{equation}
%---------------------
 Also, note that $P(t)=P_{\text{EQ}}(t)+P_{\text{NEQ}}(t)=E_{\text{EQ}}\,\epsilon(t)+E_{\text{NEQ}}\left(\epsilon(t)-\epsilon_v(t)\right)=E\epsilon(t)-E_{\text{NEQ}}\,\epsilon_v(t)\,$, where $E=E_{\text{EQ}}+E_{\text{NEQ}}\,$.
Now, suppose that the total strain is fixed, i.e., $\epsilon(t)=\epsilon_0$ and $\epsilon_v(0)=0\,$. Hence, $\epsilon_v(t)=\epsilon_0\left[1-e^{-\frac{t}{\tau}}\right]$ and $\epsilon_e(t)=\epsilon_0\,e^{-\frac{t}{\tau}}\,$. It is observed that $\lim_{t\to\infty}\epsilon_e(t)=0$ and $\lim_{t\to\infty}\epsilon_v(t)=\epsilon_0\,$.
%-----------------------------
%-----------------------------
\begin{figure}[t!]
\centering
\includegraphics[width=0.4\textwidth]{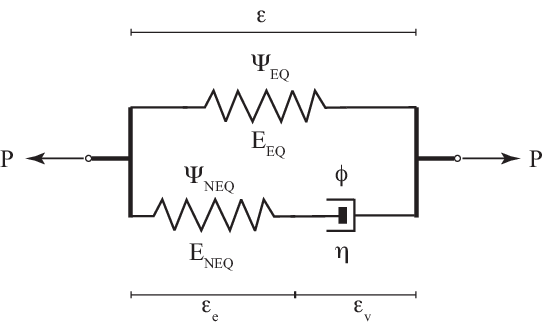}
\vspace*{0.0in}
\caption{Standard solid: $\Psi_{\text{EQ}}\,$, $\Psi_{\text{NEQ}}\,$, and $\phi$ are the nonlinear analogues of $E_{\text{EQ}}\,$, $E_{\text{NEQ}}\,$, and $\eta\,$, respectively.} 
\label{Standard-Solid}
\end{figure}
%-----------------------------
%-----------------------------
\end{remark}
%-----------------------------

%-----------------------------
%-----------------------------
\section{Material Symmetry} \label{Sec:MaterialSymmetry}

In this section, we discuss material symmetry in both anelasticity and viscoelasticity.

%-----------------------------
%-----------------------------
\subsection{Material symmetry in nonlinear anelasticity}

For an elastic solid, let us assume an energy functional of the form $\mathring{W}=\mathring{W}(X,\mathbf{F},\mathring{\mathbf{G}},\mathbf{g})\,$, where $\mathbf{g}$ is the metric of the Euclidean ambient space and $\mathring{\mathbf{G}}$ is the induced metric on the body, which is the material metric in the absence of eigenstrains. 
The material symmetry group $\mathring{\mathcal{G}}_X$ at a point $X$ with respect to the Euclidean reference configuration $(\mathcal{B},\mathring{\mathbf{G}})$ is defined as\footnote{$\mathring{\mathcal{G}}_X\leqslant \mathrm{Orth}(\mathring{\mathbf{G}})$ indicates that $\mathring{\mathcal{G}}_X$ is a subgroup of $\mathrm{Orth}(\mathring{\mathbf{G}})\,$.}
%-----------------------------
\begin{equation} 
	\mathring{\mathbf{K}}_*\mathring{W}(X,\mathbf{F},\mathring{\mathbf{G}},
	\mathring{\mathbf{g}})=
	\mathring{W}(X,\mathring{\mathbf{K}}^*\mathbf{F},
	\mathring{\mathbf{K}}^*\mathring{\mathbf{G}},\mathring{\mathbf{g}})
	=\mathring{W}(X,\mathbf{F},\mathring{\mathbf{G}},\mathring{\mathbf{g}})\,, 
	\qquad \forall\,\,
	\mathring{\mathbf{K}}\in \mathring{\mathcal{G}}_X\leqslant 
	\mathrm{Orth}(\mathring{\mathbf{G}})\,,
\end{equation}
%-----------------------------
for all deformation gradients $\mathbf{F}\,$, where $\mathrm{Orth}(\mathring{\mathbf{G}})=\{\mathbf{Q}: T_X\mathcal{B}\to T_X\mathcal{B}~|~ {\mathbf{Q}}^{\star}\mathring{\mathbf{G}}{\mathbf{Q}}=\mathring{\mathbf{G}} \}\,$.
Note that $\mathring{\mathbf{K}}^*\mathbf{F}=\mathbf{F}\mathring{\mathbf{K}}$ and $\mathring{\mathbf{K}}^*\mathring{\mathbf{G}}=\mathring{\mathbf{K}}^{\star}\mathring{\mathbf{G}}\mathring{\mathbf{K}}=\mathring{\mathbf{G}}\,$. Thus
%-----------------------------
\begin{equation} \label{An-Material-Sym}
	\mathring{W}(X,\mathbf{F}\mathring{\mathbf{K}},\mathring{\mathbf{G}},
	\mathring{\mathbf{g}})
	=\mathring{W}(X,\mathbf{F},\mathring{\mathbf{G}},\mathring{\mathbf{g}})\,, \qquad \forall\,\,
	\mathring{\mathbf{K}}\in \mathring{\mathcal{G}}_X\leqslant \mathrm{Orth}(\mathring{\mathbf{G}})\,.
\end{equation}
%-----------------------------

\begin{figure}[t!]
\centering
\includegraphics[width=0.90\textwidth]{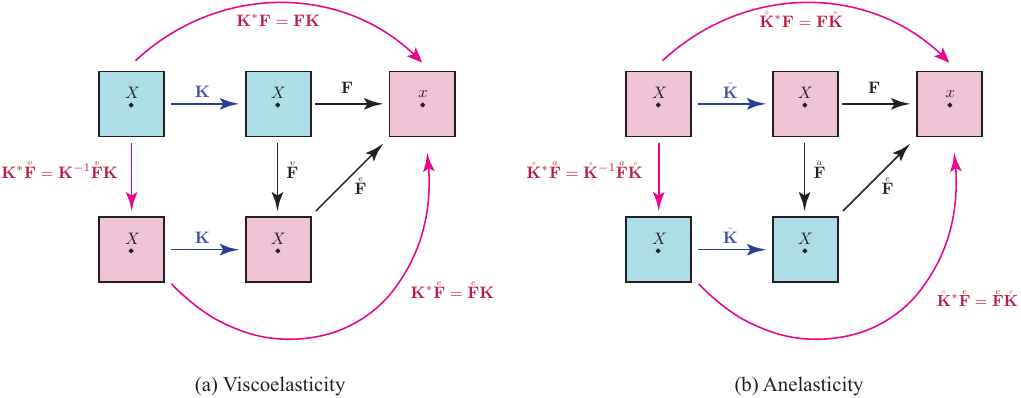}
\vspace*{-0.0in}
\caption{The actions of the symmetry group on the total, elastic, and anelastic deformation gradients. The blue and pink squares indicate locally stress-free and locally stressed configurations, respectively.} 
\label{Symmetry-Transformation}
\end{figure}

For an anelastic body, $\mathbf{F}=\Fe\Fa\,$, where $\Fe$ and $\Fa$ are the local elastic and anelastic deformations, respectively. Energy explicitly depends on the local elastic deformation, i.e., $W=\mathring{W}(X,\Fe,\mathring{\mathbf{G}},\mathbf{g})\,$.
\citet{YavariSozio2023} defined the following energy functional\footnote{In \citep{YavariSozio2023}, $\mathring{\mathbf{g}}$ was used for the metric of the Euclidean ambient space and $\mathbf{g}$ was reserved for the metric of the spatial intermediate configuration.}
%-----------------------------
\begin{equation} 
	W(X,\mathbf{F},\Fa,\mathring{\mathbf{G}},\mathbf{g})
	=W(X,\Fe\Fa,\Fa,\mathring{\mathbf{G}},\mathbf{g})
	=\mathring{W}(X,\mathbf{F}\Fa^{-1},\mathring{\mathbf{G}},\mathbf{g})
	=\mathring{W}(X,\Fe,\mathring{\mathbf{G}},\mathbf{g})\,.
\end{equation}
%-----------------------------
The following is Eq. (3.20) in \citep{YavariSozio2023}:
%-----------------------------
\begin{equation}  \label{YASO_Anelasticity-Sym-Group}
\begin{aligned}
	W(X,\Fe\Fa,\Fa,\mathring{\mathbf{G}},\mathring{\mathbf{g}}) 
	& =\mathring{W}(X,\Fe,\mathring{\mathbf{G}},\mathring{\mathbf{g}}) \\
	& =\mathring{W}(X,\Fe\mathring{\mathbf{K}},\mathring{\mathbf{G}},\mathring{\mathbf{g}})  \\
	& =W(X,\Fe\mathring{\mathbf{K}}\Fa,\Fa,\mathring{\mathbf{G}},\mathring{\mathbf{g}}) \\
	& =W(X,\mathbf{F}\Fa^{-1}\mathring{\mathbf{K}}\Fa,\Fa,\mathring{\mathbf{G}},\mathring{\mathbf{g}}) \\
	&=W(X,\mathbf{F}\mathbf{K},\Fa,\mathring{\mathbf{G}},\mathring{\mathbf{g}})
	\,, 
	\qquad\quad \forall\,\,\mathring{\mathbf{K}}\in \mathring{\mathcal{G}}_X\,.
\end{aligned}
\end{equation}
%-----------------------------
The first equality is the definition of $W$ and the second equality is a consequence of material symmetry \eqref{An-Material-Sym}. The third equality is incorrect; instead of $\Fa\,$, $\mathring{\mathbf{K}}^*\Fa=\mathring{\mathbf{K}}^{-1}\Fa \mathring{\mathbf{K}}$ should be used in the second and the third dependent variable entries (see Fig.~\ref{Symmetry-Transformation}b). Thus, their Eq. (3.20) should be corrected to read
%-----------------------------
\begin{equation}  \label{Anelasticity-Sym-Group}
\begin{aligned}
	W(X,\Fe\Fa,\Fa,\mathring{\mathbf{G}},\mathring{\mathbf{g}}) 
	& =\mathring{W}(X,\Fe,\mathring{\mathbf{G}},\mathring{\mathbf{g}}) \\
	& =\mathring{W}(X,\Fe\mathring{\mathbf{K}},\mathring{\mathbf{G}},\mathring{\mathbf{g}})  \\
	& =W(X,\Fe\mathring{\mathbf{K}}\mathring{\mathbf{K}}^{-1}\Fa \mathring{\mathbf{K}},
	\mathring{\mathbf{K}}^{-1}\Fa \mathring{\mathbf{K}},\mathring{\mathbf{G}},\mathring{\mathbf{g}}) \\
	& =W(X,\mathbf{F} \mathring{\mathbf{K}},
	\mathring{\mathbf{K}}^{-1}\Fa \mathring{\mathbf{K}},\mathring{\mathbf{G}},\mathring{\mathbf{g}}) 
	\,, 
	\qquad\qquad \forall\,\,\mathring{\mathbf{K}}\in \mathring{\mathcal{G}}_X\,.
\end{aligned}
\end{equation}
%-----------------------------
\citet{YavariSozio2023} suggested a connection between \eqref{YASO_Anelasticity-Sym-Group} and Noll's rule; but it turns out that the corrected Eq.~\eqref{Anelasticity-Sym-Group} bears no such connection. Noll's rule states that under a material diffeomorphism, the symmetry group of an elastic body transforms naturally, i.e.,~through push forward. More specifically, consider a transformation $\boldsymbol{\mathsf{F}}(X):T_X\mathcal{B} \to T_X\mathcal{B}\,$. Let us denote the material symmetry group of the elastic solid with respect to $(T_X\mathcal{B},\mathring{\mathbf{G}})$ by $\mathring{\mathcal{G}}_X\,$, and that with respect to $(T_X\mathcal{B},\boldsymbol{\mathsf{F}}_*\mathring{\mathbf{G}})$ by $\mathcal{G}_X\,$. Noll's rule says that $\mathcal{G}_X=\boldsymbol{\mathsf{F}}_*\mathring{\mathcal{G}}_X=\boldsymbol{\mathsf{F}}\,\mathring{\mathcal{G}}_X \boldsymbol{\mathsf{F}}^{-1}\,$. However, notice that \eqref{Anelasticity-Sym-Group} has nothing to do with Noll's rule. It simply tells us how symmetry group acts on deformation gradient and the anelastic local deformation. It should be noted that Eqs. (3.21)-(3.23) in~\citep{YavariSozio2023} are incorrect as well. However, what follows after their Eq. (3.23) is correct. This mistake did not affect any of the conclusions in section \textsection 3.5 of~\citep{YavariSozio2023}.

%-----------------------------
%-----------------------------
%-----------------------------
%-----------------------------

%-----------------------------
%-----------------------------
\subsection{Material symmetry in nonlinear viscoelasticity}

The material symmetry group $\mathcal{G}_X$ of a viscoelastic solid with the equilibrium free energy functional $\Psi_{\text{EQ}}=\Psi_{\text{EQ}}(X,\mathbf{F},\mathbf{G},\mathbf{g})\,$, the non-equilibrium free energy $\Psi_{\text{NEQ}}=\Psi_{\text{NEQ}}(X,\Fe,\mathbf{G},\mathbf{g})\,$, and the dissipation potential $\phi=\phi(X,\mathbf{F},\Fv,\dot\Fv,\mathbf{G},\mathbf{g})$ at a point $X$ with respect to the Euclidean reference configuration $(\mathcal{B},\mathbf{G})$ is defined as\footnote{$\mathcal{G}_X\leqslant \mathrm{Orth}(\mathbf{G})$ indicates that $\mathcal{G}_X$ is a subgroup of $\mathrm{Orth}(\mathbf{G})\,$.}
%-----------------------------
\begin{equation} \label{Material-Sym}
\begin{dcases}
	\mathbf{K}_*\Psi_{\text{EQ}}(X,\mathbf{F},\mathbf{G},\mathbf{g})=
	\Psi_{\text{EQ}}(X,\mathbf{F}\mathbf{K},\mathbf{G},\mathbf{g})
	=\Psi_{\text{EQ}}(X,\mathbf{F},\mathbf{G},\mathbf{g})\,,\\
	\mathbf{K}_*\Psi_{\text{NEQ}}(X,\Fe,\mathbf{G},\mathbf{g})=
	\Psi_{\text{NEQ}}(X,\Fe\mathbf{K},\mathbf{G},\mathbf{g}) =\Psi_{\text{NEQ}}(X,\Fe,\mathbf{G},\mathbf{g})\,,\\
	\mathbf{K}_*\phi(X,\mathbf{F},\Fv,\dot\Fv,\mathbf{G},\mathbf{g})=
\phi(X,\mathbf{F}\mathbf{K},\mathbf{K}^*\Fv,\mathbf{K}^*\dot\Fv,\mathbf{G},\mathbf{g})
	=\phi(X,\mathbf{F},\Fv,\dot\Fv,\mathbf{G},\mathbf{g})\,,
\end{dcases}
\quad \forall\,\,\mathbf{K}\in \mathcal{G}_X\leqslant \mathrm{Orth}(\mathbf{G})\,,
\end{equation}
%-----------------------------
for all deformation gradients $\mathbf{F}$ and viscous deformation gradients $\Fv\,$, where $\mathbf{K}^*\Fv=\mathbf{K}^{-1}\Fv \mathbf{K}$ (see Fig.~\ref{Symmetry-Transformation}a) and $\mathrm{Orth}(\mathbf{G})=\{\mathbf{Q}: T_X\mathcal{B}\to T_X\mathcal{B}~|~ \mathbf{Q}^{\star}\mathbf{G}\mathbf{Q}=\mathbf{G} \}\,$.

%-----------------------------
%-----------------------------
\subsubsection{Structural tensors, viscous metric, and viscous structural tensors}\label{Sec:ViscousMetric}

The $\mu^{th}$ power Kronecker product $\langle\mathbf{Q}\rangle_{\mu}$ of a $\mathbf{G}$-orthogonal transformation $\mathbf{Q}$ for a $\mu^{th}$ order tensor $\boldsymbol{\Lambda}$ is defined as $(\langle\mathbf{Q}\rangle_{\mu}\boldsymbol\Lambda)^{\bar{A}_1\hdots\bar{A}_{\mu}}=\mathrm Q^{\bar{A}_1}\mbox{}_{A_1}\hdots \mathrm Q^{\bar{A}_{\mu}}\mbox{}_{A_{\mu}}\, {\Lambda}^{A_1\hdots A_{\mu}}\,$. In particular, $\langle\mathbf{Q}\rangle_{m}\left(\mathbf{W}_1\otimes\hdots\otimes\mathbf{W}_m\right)=\mathbf{Q}\mathbf{W}_1\otimes\hdots\otimes\mathbf{Q}\mathbf{W}_m\,$, where $\mathbf{W}_i\in T_X\mathcal{B}\,$, $i=1,\hdots,m\,$. A symmetry group $\mathcal{G} \leqslant\mathrm{Orth}(\mathbf{G})$ may be characterized via a finite collection of structural tensors $\boldsymbol{\Lambda}_i$ of order $\mu_i\,$, $i=1,\dots,N$ as follows~\citep{liu1982,boehler1987,zheng1993,zheng1994theory,lu2000covariant,MazzucatoRachele2006}
%------------------------
\begin{equation} \label{invar}
	\mathbf{Q} \in \mathcal{G} \leqslant\mathrm{Orth}(\mathbf{G}) \iff 
	\langle \mathbf{Q} \rangle_{\mu_i} \boldsymbol{\Lambda}_i=\ \boldsymbol{\Lambda}_i\,,\forall i=1,\dots,N\,.
\end{equation}
%-----------------------
In other words, the set of structural tensors is a basis for the space of $\mathcal{G}$-invariant tensors. We denote the collection of structural tensors by $\boldsymbol{\Lambda}\,$. When $\boldsymbol{\Lambda}$ is added to the arguments of the (free or dissipation) energy functional, the energy functional becomes an isotropic functional of its arguments---the so-called \emph{principle of isotropy of space}~\citep{Boehler1979}. Now, the energy functional being isotropic, the corresponding set of isotropic invariants can be used to simplify its dependence on its arguments. A theorem proved by Hilbert in $1890$~\citep{Hilbert1993} (see also~\citep{Olive2017}) tells us that any finite collection of tensors has a finite set of isotropic invariants---the  \emph{integrity basis} for the set of isotropic invariants of the collection~\citep{Spencer1971}.
Therefore, since the free energy functionals $\Psi_{\text{EQ}}$ and $\Psi_{\text{NEQ}}$ are isotropic functionals of symmetric tensors, i.e.,~$\Psi_{\text{EQ}}=\hat\Psi_{\text{EQ}}(X,\Theta,\mathbf C^\flat,\boldsymbol\Lambda_1,\hdots,\boldsymbol\Lambda_N,\mathbf G)\,$, and $\Psi_{\text{NEQ}}=\hat\Psi_{\text{NEQ}}(X,\Theta,\Ce^\flat,\boldsymbol\Lambda_1,\hdots,\boldsymbol\Lambda_N,\mathbf G)\,$, one writes $\Psi_{\text{EQ}} = \overline{\Psi}(X,I_1,\hdots,I_m)\,$, where $\left\{I_1,\hdots,I_m\right\}$ is the integrity basis for the set of isotropic invariants of $\left\{\mathbf C^\flat,\boldsymbol\Lambda_1,\hdots,\boldsymbol\Lambda_N\right\}$; and $\Psi_{\text{NEQ}} = \widetilde{\Psi}(X,\Ie_1,\hdots,\Ie_m)\,$, where $\left\{\Ie_1,\hdots,\Ie_m\right\}$ is the integrity basis for the set of isotropic invariants of $\left\{\Ce^\flat,\boldsymbol\Lambda_1,\hdots,\boldsymbol\Lambda_N\right\}\,$. However, the dissipation potential $\phi=\hat{\phi}(X,\Theta,\mathbf{C}^{\flat},\boldsymbol\Lambda_1,\hdots,\boldsymbol\Lambda_N,\Fv,\dot\Fv,\mathbf{G})$ is an isotropic functional of the symmetric tensors $\left\{\mathbf C^\flat,\boldsymbol\Lambda_1,\hdots,\boldsymbol\Lambda_N\right\}$ and two generally non-symmetric material tensors {$\Fv$} and $\dot\Fv$\,; hence, the classical representation theorems cannot be used.

Next, we show that the dependence of the non-equilibrium free energy on {$\Fe$} can be reduced to a dependence on the total deformation gradient $\mathbf{F}\,$. From \eqref{eq:Psi_F_decomp}, recall that $\Psi_{\text{NEQ}}=\Psi_{\text{NEQ}}(X,\Theta,\Fe,\mathbf{G},\mathbf{g})\,$. For an anisotropic solid we have a collection of structural tensors denoted by $\boldsymbol{\Lambda}\,$. Let us add this collection to the list of arguments of the non-equilibrium free energy and write
%---------------------
\begin{equation}
	\Psi_{\text{NEQ}}=\Psi_{\text{NEQ}}(X,\Theta,\Fe,\mathbf{G},\boldsymbol{\Lambda},\mathbf{g})\,.
\end{equation}
%---------------------
Now, $\Psi_{\text{NEQ}}$ is a materially-covariant functional \citep{Lu2012,YavariSozio2023}, i.e., for any invertible linear transformation $\mathbf{T}:T_X\mathcal{B}\to T_X\mathcal{B}\,$, one has
%-----------------------------
\begin{equation} 
	\Psi_{\text{NEQ}}(X,\Theta,\mathbf{T}^*\Fe, \mathbf{T}^*\mathbf{G},
	\mathbf{T}^*\boldsymbol{\Lambda},\mathbf{g})
	=\Psi_{\text{NEQ}}(X,\Theta,\Fe,\mathbf{G},\boldsymbol{\Lambda},\mathbf{g})
	\,.
\end{equation}
%-----------------------------
Noting that $\Fe=\Fv_*\mathbf{F}\,$, and choosing $\mathbf{T}=\Fv\,$, material covariance implies that 
%-----------------------------
\begin{equation} 
	\Psi_{\text{NEQ}}(X,\Theta,\Fe,\mathbf{G},\boldsymbol{\Lambda},\mathbf{g})=
	\Psi_{\text{NEQ}}(X,\Theta,\Fv^*\Fv_*\mathbf{F}, \Fv^*\mathbf{G},
	\Fv^*\boldsymbol{\Lambda},\mathbf{g})
	=\Psi_{\text{NEQ}}(X,\Theta,\mathbf{F}, \Gv,\Lambdav,\mathbf{g})
	\,,
\end{equation}
%-----------------------------
where $\Gv=\Fv^*\mathbf{G}\,$, and $\Lambdav=\Fv^*\boldsymbol{\Lambda}\,$. Thus, in summary, we have
%---------------------
\begin{equation}
	\Psi=\Psi(X,\Theta,\mathbf F, \Fe, \mathbf{G},\boldsymbol{\Lambda},\mathbf{g})
	=\Psi_{\text{EQ}}(X,\Theta,\mathbf F, \mathbf{G},\boldsymbol{\Lambda},\mathbf{g})
	+\Psi_{\text{NEQ}}(X,\Theta,\mathbf{F}, \Gv,\Lambdav,\mathbf{g})\,.
\end{equation}
%---------------------
This means that the non-equilibrium free energy is a function of the total deformation gradient as long as the viscous metric $\Gv$ and viscous structural tensors $\Lambdav$ are used. Objectivity implies that\footnote{This is consistent with \eqref{eq:Psi_C_decomp} when structural tensors are included. First, note that $\Ce=\Fv_*\mathbf{C}^{\flat}\,$. Thus
%---------------------
\begin{equation} 
	\hat\Psi_{\text{NEQ}}=\hat\Psi_{\text{NEQ}}(X,\Theta,\Ce^\flat, \mathbf G, \boldsymbol{\Lambda})
	=\hat\Psi_{\text{NEQ}}(X,\Theta,\Fv_*\mathbf{C}^{\flat}, \mathbf G, \boldsymbol{\Lambda})
	=\hat\Psi_{\text{NEQ}}(X,\Theta,\mathbf{C}^{\flat}, \Fv^*\mathbf G,\Fv^*\boldsymbol{\Lambda})
	=\hat\Psi_{\text{NEQ}}(X,\Theta,\mathbf{C}^{\flat}, \Gv,\Lambdav)	\,.
\end{equation}
%---------------------
}
%---------------------
\begin{equation}
	\Psi=\hat{\Psi}_{\text{EQ}}(X,\Theta,\mathbf{C}^{\flat}, \mathbf{G},\boldsymbol{\Lambda})
	+\hat{\Psi}_{\text{NEQ}}(X,\Theta,\mathbf{C}^{\flat},\Gv,\Lambdav)\,.
\end{equation}
%---------------------

Next, we use the integrity basis for isotropic, transversely isotropic, orthotropic, and monoclinic viscoelastic solids, and explicitly write their respective stress constitutive relations and kinetic equations.

%-----------------------------
%-----------------------------
\subsubsection{Isotropic solids}

\paragraph{Stress constitutive equations.}
For isotropic solids, $\Psi_{\text{EQ}}$ and $\Psi_{\text{NEQ}}$ depend only on the principal invariants of $\mathbf{C}^{\flat}\,$, and $\Ce^{\flat}\,$, respectively, i.e.,
%---------------------
\begin{equation}
\label{Psi_const_iso}
	\Psi_{\text{EQ}} = \overline{\Psi}(X,I_1,I_2,I_3)   \,,\qquad
	\Psi_{\text{NEQ}}  = \widetilde{\Psi}(X,\Ie_1,\Ie_2,\Ie_3)  \,,
\end{equation}
%---------------------
where\footnote{\label{Cayley-Hamilton}The characteristic polynomial of $\mathbf{C}$ reads: 
%---------------------
\begin{equation}
	\lambda^3-I_1\,\lambda^2+I_2\,\lambda -I_3=0\,,\qquad
	I_1=\operatorname{tr}\mathbf{C}\,, \qquad
	I_2=(\det\mathbf{C})\operatorname{tr}\mathbf{C}^{-1}\,, \qquad I_3 =\det\mathbf{C}\,.
\end{equation}
%---------------------
The Cayley-Hamilton theorem tells us that $\mathbf{C}^3-I_1\,\mathbf{C}^2+I_2\,\mathbf{C} -I_3\,\mathbf{I}=\mathbf{0}\,$. Multiplying both sides by $\mathbf{C}^{-1}\,$, one concludes that $I_3\,\mathbf{C}^{-1}=\mathbf{C}^2-I_1\,\mathbf{C}+I_2\,\mathbf{I} \,$. This, in particular, implies that $I_2=\frac{1}{2}\left(I_1^2-\operatorname{tr}\mathbf{C}^2\right)\,$.
}
%-----------------------------
\begin{equation} 
\begin{aligned}
	& I_1=\operatorname{tr}\mathbf{C}=\mathrm{C}^A{}_A\,, &&\quad
	I_2=\frac{1}{2}\left(I_1^2-\operatorname{tr}\mathbf{C}^2\right)
	=\frac{1}{2}\left(I_1^2-\mathrm{C}^A{}_B\,\mathrm{C}^B{}_A\right)\,,
	&&\quad  I_3 =\det\mathbf{C}\,,\\
	& \Ie_1=\operatorname{tr}\Ce=\cCe^A{}_A\,,&&\quad
	\Ie_2=\frac{1}{2}\left(\Ie_1^2-\operatorname{tr}\Ce^2\right)
	=\frac{1}{2}\left(\Ie_1^2-\cCe^A{}_B\,\cCe^B{}_A\right)\,,
	&&\quad  \Ie_3 =\det\Ce\,.
\end{aligned}
\end{equation}
%-----------------------------
Note that $\mathbf{C}^{\flat}=\Fv^*\Ce^{\flat}=\Fv^{\star}\Ce^{\flat}\Fv\,$, or equivalently $\Ce^{\flat}=\Fv^{-\star}\mathbf{C}^{\flat}\Fv^{-1}\,$.
The second Piola-Kirchhoff stress is written as
%-----------------------------
\begin{equation} 
	\mathbf{S}=2\frac{d \hat{\Psi}}{d \mathbf{C}^\flat}
	=2\frac{d \hat{\Psi}_{\text{EQ}}}{d\mathbf{C}^\flat}+2\frac{d \hat{\Psi}_{\text{NEQ}}}{d\mathbf{C}^\flat}
	=2\frac{\partial \hat{\Psi}_{\text{EQ}}}{\partial\mathbf{C}^\flat}
	+2\frac{\partial \hat{\Psi}_{\text{NEQ}}}{\partial \Ce^{\flat}}\cdot\frac{\partial \Ce^{\flat}}{\partial\mathbf{C}^\flat}
	=2\frac{\partial \hat{\Psi}_{\text{EQ}}}{\partial\mathbf{C}^\flat}
	+2\Fv^{-1}\,\frac{\partial \hat{\Psi}_{\text{NEQ}}}{\partial \Ce^{\flat}}\,\Fv^{-\star}\,.
\end{equation}
%-----------------------------
In terms of the principal invariants, one writes
%---------------------------------
\begin{equation}
	\mathbf{S}=2\frac{d \hat{\Psi}}{d\mathbf{C}^\flat}
	=\sum_{j=1}^{3}2\overline{\Psi}_j\frac{\partial I_j}{\partial\mathbf{C}^\flat}
	+\Fv^{-1} \left[\sum_{j=1}^{3}2\widetilde{\Psi}_j\frac{\partial \Ie_j}{\partial\Ce^\flat}\right]\Fv^{-\star}
	\,,
\end{equation}
%---------------------------------
where
%---------------------------------
\begin{equation}
	\overline{\Psi}_j=\overline{\Psi}_j(X,I_1,I_2,I_3):=\frac{\partial \overline{\Psi}}{\partial I_j}\,,\qquad
	\widetilde{\Psi}_j=\widetilde{\Psi}_j(X,\Ie_1,\Ie_2,\Ie_3):=\frac{\partial \widetilde{\Psi}}{\partial \Ie_j}\,, 		\qquad j=1,2,3\,.
\end{equation}
%---------------------------------
Therefore\footnote{See Appendix \ref{appendix:derivatives} for the derivatives of the principal invariants of $\mathbf{C}$ and $\Ce\,$.}
%---------------------------------
\begin{equation}
\begin{aligned}
	\mathbf{S} &=2\overline{\Psi}_1\,\mathbf{G}^\sharp+2\overline{\Psi}_2\,(I_2\,\mathbf{C}^{-\sharp}-I_3\,\mathbf{C}^{-2\sharp})
	+2\overline{\Psi}_3\,I_3\,\mathbf{C}^{-\sharp} \\
	& \quad+\Fv^{-1} \left[ 2\widetilde{\Psi}_1\,\mathbf{G}^\sharp
	+2\widetilde{\Psi}_2\,\left(\Ie_2\,\Ce^{-\sharp}-\Ie_3\,\Ce^{-2\sharp}\right)
	+2\widetilde{\Psi}_3\,\Ie_3\,\Ce^{-\sharp} \right]\Fv^{-\star}\,.
\end{aligned}
\end{equation}
%---------------------------------
Note that $\Fv^{-1}\mathbf{G}^\sharp \Fv^{-\star}=\Fv^*\mathbf{G}^{\sharp}\,$, $\Fv^{-1}\Ce^{-\sharp} \Fv^{-\star}=\Fv^*\Ce^{-\sharp}\,$, and $\Fv^{-1}\Ce^{-2\sharp} \Fv^{-\star}=\Fv^*\Ce^{-2\sharp}\,$. Thus
%---------------------------------
\begin{equation}
\begin{aligned}
	\mathbf{S} &=2\overline{\Psi}_1\,\mathbf{G}^{\sharp}+2\overline{\Psi}_2\,(I_2\,\mathbf{C}^{-\sharp}-I_3\,\mathbf{C}^{-2\sharp})
	+2\overline{\Psi}_3\,I_3\,\mathbf{C}^{-\sharp} \\
	& \quad+ 2\widetilde{\Psi}_1\,\Fv^*\mathbf{G}^\sharp
	+2\widetilde{\Psi}_2\Fv^*\left(\Ie_2\Ce^{-\sharp}-\Ie_3\Ce^{-2\sharp}\right)
	+2\widetilde{\Psi}_3\,\Ie_3\,\Fv^*\Ce^{-\sharp} \,.
\end{aligned}
\end{equation}
%---------------------------------
The Cauchy stress is related to the second Piola-Kirchhoff stress as $\boldsymbol{\sigma}=\frac{1}{\sqrt{I_3}}\mathbf{F}\mathbf{S}\mathbf{F}^{\star}\,$. Recall that
$\mathbf{F}\mathbf{G}^{\sharp}\mathbf{F}^{\star}=\varphi_*\mathbf{G}^{\sharp}=\mathbf{b}^{\sharp}\,$,
$\mathbf{F}\mathbf{C}^{-\sharp}\mathbf{F}^{\star}=\mathbf{F}\mathbf{C}^{-1}\mathbf G^\sharp\mathbf{F}^{\star}=\mathbf{F}(\mathbf{F}^{-1}\mathbf{F}^{-\mathsf T})\mathbf{F}^{\mathsf T}\mathbf g^\sharp=\mathbf{g}^{\sharp}\,$, and
%---------------------
\begin{equation}
	\mathbf{F}\mathbf{C}^{-2\sharp}\mathbf{F}^{\star}
	=\mathbf{F}\mathbf{C}^{-1}\mathbf{C}^{-1}\mathbf{G}^{\sharp}\mathbf{F}^{\star}
	=\mathbf{F}(\mathbf{F}^{-1}\mathbf{F}^{-\mathsf T})(\mathbf{F}^{-1}
	\mathbf{F}^{-\mathsf T})
	(\mathbf{F}^{\mathsf T}\mathbf g^\sharp)
	=\mathbf{F}^{-\mathsf T}\mathbf{F}^{-1}\mathbf g^\sharp
	=\mathbf c^\sharp\,.
\end{equation}
%---------------------
Thus
%---------------------------------
\begin{equation}
\begin{aligned}
	\boldsymbol{\sigma} &= \frac{2}{\sqrt{I_3}} \left[ \left(I_2\,\overline{\Psi}_2+I_3\,\overline{\Psi}_3\right)\mathbf{g}^{\sharp} 
	+\overline{\Psi}_1\,\mathbf{b}^{\sharp}-I_3\,\overline{\Psi}_2\,\mathbf{c}^{\sharp} \right]\\
	& \quad+ \frac{2}{\sqrt{I_3}}\,\Fe \left[ \widetilde{\Psi}_1\,\mathbf{G}^\sharp
	+\left(\Ie_2\,\widetilde{\Psi}_2+\Ie_3\,\widetilde{\Psi}_3\right)\Ce^{-\sharp}
	-\Ie_3\,\widetilde{\Psi}_2\,\Ce^{-2\sharp} \right]\Fe^{\star} \\
	&= \frac{2}{\sqrt{I_3}} \left[ \left(I_2\,\overline{\Psi}_2+I_3\,\overline{\Psi}_3
	+\Ie_2\,\widetilde{\Psi}_2+\Ie_3\,\widetilde{\Psi}_3\right)\mathbf{g}^{\sharp} 
	+\overline{\Psi}_1\,\mathbf{b}^{\sharp}+\widetilde{\Psi}_1\,\be^{\sharp}-I_3\,\overline{\Psi}_2\,\mathbf{c}^{\sharp}
	-\Ie_3\,\widetilde{\Psi}_2\,\ce^{\sharp} \right]\,.
\end{aligned}
\end{equation}
%---------------------------------

For an incompressible isotropic solid, $I_3=\Ie_3=1\,$, and hence
%---------------------------------
\begin{equation}
\label{eq:sigma_iso}
	\boldsymbol{\sigma} = -p\,\mathbf{g}^{\sharp} 
	+2\overline{\Psi}_1\,\mathbf{b}^{\sharp}+2\widetilde{\Psi}_1\,\be^{\sharp}
	-2\,\overline{\Psi}_2\,\mathbf{c}^{\sharp}
	-2\,\widetilde{\Psi}_2\,\ce^{\sharp} \,,
\end{equation}
%---------------------------------
where $p$ is the Lagrange multiplier associated with the incompressibility constraint $J=\sqrt{I_3}=1\,$. 

\paragraph{Dissipation potential.}
For an isotropic viscoelastic solid, the dissipation potential must be invariant under the orthogonal group, i.e.,
%-----------------------------
\begin{equation}\label{eq:phi_sym}
	\phi(X,\mathbf{F}\mathbf{K},\mathbf{K}^{-1}\Fv \mathbf{K},
	\mathbf{K}^{-1}\dot\Fv\mathbf{K},
	\mathbf{G},\mathbf{g})=
	\phi(X,\mathbf{F},\Fv,\dot\Fv,\mathbf{G},\mathbf{g})\,,
	\qquad \forall\,\,\mathbf{K}\in \mathrm{Orth}(\mathbf{G})\,,
\end{equation}
%-----------------------------
for all deformation gradients $\mathbf{F}$ and viscous deformation gradients $\Fv\,$.
Notice that even for an isotropic viscoelastic solid, the dependence of $\phi$ on $\Fv$ and $\dot{\Fv}$ cannot always be reduced to a dependence on the symmetric tensors $\Cv^\flat$ and $\dot{\Cv}^\flat\,$. 
Let $\lambda_i^2$ ($i=1, 2, 3$) be the eigenvalues of the symmetric tensor $\mathbf{C}^{\sharp}\,$. Let us denote the corresponding unit eigenvectors by $\mathbf{W}_i$ ($i=1, 2, 3$). Thus, $\mathbf{C}^{\sharp}=\lambda_1^2\,\mathbf{W}_1\otimes\mathbf{W}_1+\lambda_2^2\,\mathbf{W}_2\otimes\mathbf{W}_2+\lambda_3^2\,\mathbf{W}_3\otimes\mathbf{W}_3\,$. The dissipation potential will be a functional of $I_1, I_2, I_3\,$, and the following $18$ spectral invariants \citep{Shariff2022}:
%-----------------------------
\begin{equation}
	F_{ij}=\llangle\mathbf{W}_i,\Fv \mathbf{W}_j\rrangle_{\mathbf G}\,,\qquad \widetilde{F}_{ij}=\llangle\mathbf{W}_i,\dot\Fv \mathbf{W}_j\rrangle_{\mathbf G}\,,
	\qquad i,j=1,2,3\,,
\end{equation}
%-----------------------------
i.e., $\hat{\phi}=\bar{\phi}(I_1,I_2,I_3,F_{11},F_{12},\cdots,F_{33},\widetilde{F}_{11},\widetilde{F}_{12},\cdots,\widetilde{F}_{33})\,$.\footnote{This functional form essentially means that, even in the isotropic case, while the dissipation potential depends only on the three principal invariants of the right Cauchy-Green deformation tensor instead of its $6$ components, there is no reduction in its dependence on the non-symmetric tensors $\Fv$ and $\dot\Fv$; it still depends on all their $18$ components. Note, however, that when these components are written with respect to the eigenbasis $\{\mathbf W_1,\mathbf W_2,\mathbf W_3\}\,$, they are invariant under the orthogonal group.}

%-----------------------------
\begin{remark}
If one assumes that the dissipation potential has the functional form $\phi(X,\Fv,\dot\Fv,\mathbf{G})\,$, then for an isotropic solid,  $\phi$ is an isotropic functional of two non-symmetric material tensors $\Fv$ and $\dot\Fv\,$. Let $\mu_i$ ($i=1, 2, 3$) be the eigenvalues of the symmetric tensor $\Cv^{\sharp}\,$, where $\Cv=\Fv^{\mathsf T}\Fv\,$.
Let us denote the corresponding eigenvectors by $\mathbf{U}_i$ ($i=1, 2, 3$). Thus, $\Cv^{\sharp}=\mu_1\mathbf{U}_1\otimes\mathbf{U}_1+\mu_2\mathbf{U}_2\otimes\mathbf{U}_2+\mu_3\mathbf{U}_3\otimes\mathbf{U}_3\,$. The dissipation potential will be a functional of the following $18$ spectral invariants \citep{Shariff2022}:
%-----------------------------
\begin{equation}
	F_{ij}=\llangle\mathbf{U}_i,\Fv \mathbf{U}_j\rrangle_{\mathbf G}\,,\qquad \widetilde{F}_{ij}=\llangle\mathbf{U}_i,\dot\Fv \mathbf{U}_j\rrangle_{\mathbf G}\,,
	\qquad i,j=1,2,3\,,
\end{equation}
%-----------------------------
i.e., $\phi=\tilde{\phi}(F_{11},F_{12},\cdots,F_{33},\widetilde{F}_{11},\widetilde{F}_{12},\cdots,\widetilde{F}_{33})\,$.
\end{remark}
%-----------------------------

\paragraph{Kinetic equation.}
Following \eqref{eq:part_Ie}, one may write
%---------------------
\begin{equation}
\frac{\partial \Psi}{\partial \Ce^\flat} =\frac{\partial \Psi_{\text{NEQ}}}{\partial \Ce^\flat}
= \widetilde\Psi_1\frac{\partial \Ie_1}{\partial \Ce^\flat} + \widetilde\Psi_2\frac{\partial \Ie_2}{\partial \Ce^\flat} + \widetilde\Psi_3\frac{\partial \Ie_3}{\partial \Ce^\flat}
= \widetilde\Psi_1\mathbf{G}^{\sharp} + \widetilde\Psi_2 \left(\Ie_2\Ce^{-\sharp}-\Ie_3\Ce^{-2\sharp}\right) + \widetilde\Psi_3 \Ie_3\Ce^{-\sharp}
\,.
\end{equation}
%---------------------
Hence, it follows from \eqref{eq:LD_kinetic} that the kinetic equation for compressible isotropic viscoelastic solids reads\footnote{\label{Kin-Eq_Equi}Similarly to what was observed earlier in Footnote \ref{Cayley-Hamilton}, the Cayley-Hamilton theorem for $\Ce$ tells us that $\Ie_2\mathbf{I}-\Ie_3\Ce^{-1}=\Ie_1\Ce-\Ce^2\,$, which then changes the kinetic equation to the following equivalent form
%---------------------
\begin{equation}
	\frac{\partial \phi}{\partial \dot\Fv} 
	-2\widetilde{\Psi}_1\Ce^{\flat}\mathbf{G}^{\sharp}\Fv^{-\star}
	+2\widetilde{\Psi}_2\left[\Ce^{2\flat}-\Ie_1\Ce^{\flat} \right]\mathbf{G}^\sharp\Fv^{-\star}
	-2\Ie_3\widetilde{\Psi}_3\Fv^{-\star}
	 = \mathbf{0}\,.
\end{equation}
%---------------------
}
%---------------------
\begin{equation}
	\frac{\partial \phi}{\partial \dot\Fv} 
	-2\widetilde{\Psi}_1\Ce^{\flat}\mathbf{G}^{\sharp}\Fv^{-\star}
	-2\widetilde{\Psi}_2\left[\Ie_2\mathbf{I}-\Ie_3\mathbf{G}\Ce^{-\sharp} \right]\Fv^{-\star}
	-2\Ie_3\widetilde{\Psi}_3\Fv^{-\star}
	 = \mathbf{0}\,;
\end{equation}
%---------------------
and in the case of incompressible isotropic viscoelastic solids, the kinetic equation \eqref{eq:LD_kinetic_inc} is written as
%---------------------
\begin{equation} \label{eq:Kinetic-Equation-Isotropic_inc}
	\frac{\partial \phi}{\partial \dot\Fv}
	-2\widetilde{\Psi}_1\Ce^{\flat}\mathbf{G}^{\sharp}\Fv^{-\star}
	-2\widetilde{\Psi}_2\left[\Ie_2\mathbf{I}-\mathbf{G}\Ce^{-\sharp} \right]\Fv^{-\star}
	 = q \Fv^{-\star}\,.
\end{equation}
%---------------------

%-----------------------------
%-----------------------------
\subsubsection{Transversely isotropic solids}

A transversely isotropic solid at a material point $X \in\mathcal{B}$ has a material preferred direction that is specified by a unit vector $\mathbf{N}(X)\,$, which is normal to the plane of isotropy at that point.

\paragraph{Stress constitutive relation.}
The equilibrium and non-equilibrium free energies become isotropic functionals of their arguments when the structural tensor $\mathbf{A}=\mathbf{N}\otimes\mathbf{N}$ is added to the list of their arguments \citep{Doyle1956,spencer1982formulation,lu2000covariant}.\footnote{The functionals $\Psi_{\text{EQ}}(X,\Theta,\mathbf F, \mathbf{A}, \mathbf G)\,$, $\Psi_{\text{NEQ}}(X,\Theta,\Fe, \mathbf{A}, \mathbf G)\,$, and $\Psi(X,\Theta,\mathbf F, \Fe, \mathbf{A}, \mathbf G, \mathbf g)$ are isotropic.}
Equivalently,
%---------------------
\begin{equation}
\label{Psi_const_trans-iso}
	\Psi_{\text{EQ}} = \overline{\Psi}(X,I_1,I_2,I_3,I_4,I_5)   \,,\qquad
	\Psi_{\text{NEQ}}  = \widetilde{\Psi}(X,\Ie_1,\Ie_2,\Ie_3,\Ie_4,\Ie_5)  \,,
\end{equation}
%---------------------
where
%---------------------
\begin{equation} 
\begin{aligned}
	& I_1=\mathrm{tr}\,\mathbf{C}=\mathrm{C}^A{}_A\,,&&
	I_2=\mathrm{det}\,\mathbf{C}~\mathrm{tr}~\mathbf{C}^{-1}
	=\mathrm{det}(\mathrm{C}^A{}_B)(\mathrm{C}^{-1})^D{}_D\,,&&
	I_3=\mathrm{det}\mathbf{C}=\mathrm{det}(\mathrm{C}^A{}_B)\,, \\
	& I_4=\mathbf{N}\cdot\mathbf{C}\cdot\mathbf{N}=N^AN^B\,\mathrm{C}_{AB}\,,&&
	I_5=\mathbf{N}\cdot\mathbf{C}^2\cdot\mathbf{N}
	=N^AN^B\,\mathrm{C}_{BM}\,\mathrm{C}^M{}_A\,,
\end{aligned}
\end{equation}
%---------------------
and
%---------------------
\begin{equation} 
\begin{aligned}
	& \Ie_1=\mathrm{tr}\,\Ce=\cCe^A{}_A\,,&&
	\Ie_2=\mathrm{det}\,\Ce~\mathrm{tr}~\Ce^{-1}
	=\mathrm{det}(\cCe^A{}_B)(\cCe^{-1})^D{}_D\,,&&
	\Ie_3=\mathrm{det}\Ce=\mathrm{det}(\cCe^A{}_B)\,, \\
	& \Ie_4=\mathbf{N}\cdot\Ce\cdot\mathbf{N}=N^AN^B\,\cCe_{AB}\,,&&
	\Ie_5=\mathbf{N}\cdot\Ce^2\cdot\mathbf{N}=N^AN^B\,\cCe_{BM}\,\cCe^M{}_A\,.
\end{aligned}
\end{equation}
%---------------------
Note that for the extra invariants
%---------------------
\begin{equation}
	\frac{\partial I_4}{\partial\mathbf{C}^\flat}=\mathbf{N}\otimes\mathbf{N}\,,\qquad
	\frac{\partial I_5}{\partial\mathbf{C}^\flat}
	=\mathbf{N}\otimes(\mathbf{G}^{\sharp}\mathbf{C}^{\flat}\mathbf{N})
	+(\mathbf{G}^{\sharp}\mathbf{C}^{\flat}\mathbf{N})\otimes\mathbf{N}\,,
\end{equation}
%---------------------
and
%---------------------
\begin{equation}\label{eq:diff_Ie4-5}
	\frac{\partial \Ie_4}{\partial\Ce^\flat}=\mathbf{N}\otimes\mathbf{N}\,,\qquad
	\frac{\partial \Ie_5}{\partial\Ce^\flat}
	=\mathbf{N}\otimes(\mathbf{G}^{\sharp}\Ce^{\flat}\mathbf{N})
	+(\mathbf{G}^{\sharp}\Ce^{\flat}\mathbf{N})\otimes\mathbf{N}\,.
\end{equation}
%---------------------
The second Piola-Kirchhoff stress has the following representation
%---------------------------------
\begin{equation}
	\mathbf{S}=2\frac{\partial \hat{\Psi}}{\partial\mathbf{C}^\flat}
	=\sum_{j=1}^{5}2\overline{\Psi}_j\frac{\partial I_j}{\partial\mathbf{C}^\flat}
	+\Fv^{-1} \left[\sum_{j=1}^{5}2\widetilde{\Psi}_j\frac{\partial \Ie_j}{\partial\Ce^\flat}\right]\Fv^{-\star}
	\,,
\end{equation}
%---------------------------------
where
%---------------------------------
\begin{equation}
	\overline{\Psi}_j=\overline{\Psi}_j(X,I_1,I_2,I_3,I_4,I_5):=\frac{\partial \overline{\Psi}}{\partial I_j}\,,
	\qquad
	\widetilde{\Psi}_j=\widetilde{\Psi}_j(X,\Ie_1,\Ie_2,\Ie_3,\Ie_4,\Ie_5)
	:=\frac{\partial \widetilde{\Psi}}{\partial \Ie_j}\,, \qquad j=1,\cdots,5\,.
\end{equation}
%---------------------------------
Thus
%---------------------------------
\begin{equation}
\begin{aligned}
	\mathbf{S} &=2\overline{\Psi}_1\,\mathbf{G}^{\sharp}+2\overline{\Psi}_2\,(I_2\,\mathbf{C}^{-\sharp}-I_3\,\mathbf{C}^{-2\sharp})
	+2\overline{\Psi}_3\,I_3\,\mathbf{C}^{-\sharp}+2\overline{\Psi}_4\,\mathbf{N}\otimes\mathbf{N}
	+2\overline{\Psi}_5\left[\mathbf{N}\otimes(\mathbf{G}^{\sharp}\mathbf{C}^{\flat}\mathbf{N})
	+(\mathbf{G}^{\sharp}\mathbf{C}^{\flat}\mathbf{N})\otimes\mathbf{N} \right] \\
	& \quad+ 2\widetilde{\Psi}_1\,\Fv^*\mathbf{G}^\sharp
	+2\widetilde{\Psi}_2\,\left(\Ie_2\,\Fv^*\Ce^{-\sharp}-\Ie_3\,\Fv^*\Ce^{-2\sharp}\right)
	+2\widetilde{\Psi}_3\,\Ie_3\,\Fv^*\Ce^{-\sharp} 
	+2\widetilde{\Psi}_4\, \Fv^{-1} \left[\mathbf{N}\otimes\mathbf{N}\ \right]\Fv^{-\star} \\
	& \quad+2\widetilde{\Psi}_5\,\Fv^{-1} 
	\left[ \mathbf{N}\otimes(\mathbf{G}^{\sharp}\Ce^{\flat}\mathbf{N})
	+(\mathbf{G}^{\sharp}\Ce^{\flat}\mathbf{N})\otimes\mathbf{N} \right]\Fv^{-\star} \\
	&=2\overline{\Psi}_1\,\mathbf{G}^{\sharp}+2\overline{\Psi}_2\,(I_2\,\mathbf{C}^{-\sharp}-I_3\,\mathbf{C}^{-2\sharp})
	+2\overline{\Psi}_3\,I_3\,\mathbf{C}^{-\sharp}+2\overline{\Psi}_4\,\mathbf{N}\otimes\mathbf{N}
	+2\overline{\Psi}_5\left[\mathbf{N}\otimes(\mathbf{G}^{\sharp}\mathbf{C}^{\flat}\mathbf{N})
	+(\mathbf{G}^{\sharp}\mathbf{C}^{\flat}\mathbf{N})\otimes\mathbf{N} \right]\\
	& \quad+ 2\widetilde{\Psi}_1\,\Fv^*\mathbf{G}^\sharp
	+2\widetilde{\Psi}_2\,\left(\Ie_2\,\Fv^*\Ce^{-\sharp}-\Ie_3\,\Fv^*\Ce^{-2\sharp}\right)
	+2\widetilde{\Psi}_3\,\Ie_3\,\Fv^*\Ce^{-\sharp} 
	+2\widetilde{\Psi}_4\,  (\Fv^{-1}\mathbf{N}) \otimes (\Fv^{-1}\mathbf{N}) \\
	& \quad+2\widetilde{\Psi}_5
	\left[(\Fv^{-1}\mathbf{N})\otimes (\Fv^{-1}\mathbf{G}^{\sharp}\Ce^{\flat}\mathbf{N})
	+(\Fv^{-1}\mathbf{G}^{\sharp}\Ce^{\flat}\mathbf{N}) \otimes(\Fv^{-1}\mathbf{N}) \right]
	 \,.
\end{aligned}
\end{equation}
%---------------------------------
Note that $\mathbf{F}\left[\mathbf{N}\otimes(\mathbf{G}^{\sharp}\mathbf{C}^{\flat}\mathbf{N})+(\mathbf{G}^{\sharp}\mathbf{C}^{\flat}\mathbf{N})\otimes\mathbf{N} \right]\mathbf{F}^{\star}=\mathbf{n}\otimes \mathbf{F}(\mathbf{G}^{\sharp}\mathbf{C}^{\flat}\mathbf{N})+\mathbf{F}(\mathbf{G}^{\sharp}\mathbf{C}^{\flat}\mathbf{N})\otimes\mathbf{n}\,$. Also notice that 
%---------------------------------
\begin{equation}
	\mathbf{F}(\mathbf{G}^{\sharp}\mathbf{C}^{\flat}\mathbf{N})
	=\mathbf{F}\mathbf{G}^{\sharp}\mathbf{F}^{\star}\mathbf{g}\mathbf{F}\mathbf{N}
	=(\mathbf{F}\mathbf{G}^{\sharp}\mathbf{F}^{\star})\mathbf{g}\mathbf{n}=\mathbf{b}^{\sharp}\mathbf{g}\mathbf{n}
	\,.
\end{equation}
%---------------------------------
Similarly
%---------------------------------
\begin{equation}
	\Fe(\mathbf{G}^{\sharp}\Ce^{\flat}\mathbf{N})
	=\be^{\sharp}\mathbf{g}\Ne
	\,,
\end{equation}
%---------------------------------
where $\Ne=\Fe\mathbf{N}\,$.
The Cauchy stress has the following representation 
%---------------------------------
\begin{equation}
\begin{aligned}
	\boldsymbol{\sigma} 
	&= \frac{2}{\sqrt{I_3}} \Bigg\{ \left(I_2\,\overline{\Psi}_2+I_3\,\overline{\Psi}_3
	+\Ie_2\,\widetilde{\Psi}_2+\Ie_3\,\widetilde{\Psi}_3\right)\mathbf{g}^{\sharp} 
	+\overline{\Psi}_1\,\mathbf{b}^{\sharp}+\widetilde{\Psi}_1\,\be^{\sharp}-I_3\,\overline{\Psi}_2\,\mathbf{c}^{\sharp}
	-\Ie_3\,\widetilde{\Psi}_2\,\ce^{\sharp} \\
	& \quad +\overline{\Psi}_4\,\mathbf{n}\otimes\mathbf{n}
	+\overline{\Psi}_5\left[\mathbf{n}\otimes(\mathbf{b}^{\sharp}\mathbf{g}\mathbf{n})
	+(\mathbf{b}^{\sharp}\mathbf{g}\mathbf{n})\otimes\mathbf{n} \right] 
	+\widetilde{\Psi}_4\,\Ne\otimes\Ne
	+\widetilde{\Psi}_5 \Big[\Ne\otimes (\be^{\sharp}\mathbf{g}\Ne)+(\be^{\sharp}\mathbf{g}\Ne)\otimes\Ne\Big]
	\Bigg\}\,.
\end{aligned}
\end{equation}
%---------------------------------
For an incompressible isotropic solid, $I_3=\Ie_3=1\,$, and hence
%---------------------------------
\begin{equation}
\begin{aligned}
	\boldsymbol{\sigma} 
	&=-p\,\mathbf{g}^{\sharp} 
	+2\overline{\Psi}_1\,\mathbf{b}^{\sharp}+2\widetilde{\Psi}_1\,\be^{\sharp}
	-2\overline{\Psi}_2\,\mathbf{c}^{\sharp}
	-2\,\widetilde{\Psi}_2\,\ce^{\sharp} 
	+2\overline{\Psi}_4\,\mathbf{n}\otimes\mathbf{n}
	+2\overline{\Psi}_5\left[\mathbf{n}\otimes(\mathbf{b}^{\sharp}\mathbf{g}\mathbf{n})
	+(\mathbf{b}^{\sharp}\mathbf{g}\mathbf{n})\otimes\mathbf{n} \right] \\
	&\quad +2\widetilde{\Psi}_4\,\Ne\otimes\Ne
	+2\widetilde{\Psi}_5 \Big[\Ne\otimes (\be^{\sharp}\mathbf{g}\Ne)+(\be^{\sharp}\mathbf{g}\Ne)\otimes\Ne\Big]
	\,,
\end{aligned}
\end{equation}
%---------------------------------
where $p$ is the Lagrange multiplier associated with the incompressibility constraint $J=\sqrt{I_3}=1\,$. 

\paragraph{Dissipation potential.}
For a transversely isotropic viscoelastic solid, when the structural tensor $\mathbf{A}=\mathbf{N}\otimes\mathbf{N}$ is added to the list of the arguments of the dissipation potential $\phi\,$, it becomes an isotropic functional of its arguments $\phi=\hat{\phi}(X,\mathbf{C}^{\flat},\Fv,\dot\Fv,\mathbf{A},\mathbf{G},\mathbf{g})\,$. Although one may not use the standard representation theorem as for the free energy functionals, the dissipation potential will be a functional of some standard invariants and a set of spectral invariants, similarly to the dissipation potential of isotropic viscoelastic solids.

\paragraph{Kinetic equation.}
Following \eqref{eq:part_Ie} and \eqref{eq:diff_Ie4-5}, one may write
%---------------------
\begin{equation}
\begin{split}
\frac{\partial \Psi}{\partial \Ce^\flat}
&=\frac{\partial \Psi_{\text{NEQ}}}{\partial \Ce^\flat} = \widetilde\Psi_1\frac{\partial \Ie_1}{\partial \Ce^\flat} + \widetilde\Psi_2\frac{\partial \Ie_2}{\partial \Ce^\flat} + \widetilde\Psi_3\frac{\partial \Ie_3}{\partial \Ce^\flat} + \widetilde\Psi_4\frac{\partial \Ie_4}{\partial \Ce^\flat} + \widetilde\Psi_5\frac{\partial \Ie_5}{\partial \Ce^\flat}\\
&= \widetilde\Psi_1\mathbf{G}^{\sharp} + \widetilde\Psi_2 \left(\Ie_2\Ce^{-\sharp}-\Ie_3\Ce^{-2\sharp}\right) + \widetilde\Psi_3 \Ie_3\Ce^{-\sharp} + \widetilde\Psi_4\mathbf{N}\otimes\mathbf{N} + \widetilde\Psi_5(\mathbf{N}\otimes(\Ce\mathbf{N}) + (\Ce\mathbf{N})\otimes\mathbf{N})
\,.
\end{split}
\end{equation}
%---------------------
Hence, it follows from \eqref{eq:LD_kinetic} that the kinetic equation for compressible transversely isotropic viscoelastic solids reads
%---------------------
\begin{equation}
\begin{aligned}
	\frac{\partial \phi}{\partial \dot\Fv} 
	&-2\widetilde{\Psi}_1\Ce^{\flat}\mathbf{G}^{\sharp}\Fv^{-\star}
	-2\widetilde{\Psi}_2\left[\Ie_2\mathbf{I}-\Ie_3\mathbf{G}\Ce^{-\sharp} \right]\Fv^{-\star}
	-2\Ie_3\widetilde{\Psi}_3\Fv^{-\star} \\
	& -2\widetilde\Psi_4\Ce^\flat\mathbf{N}\otimes\Fv^{-1}\mathbf{N}
	-2\widetilde\Psi_5\left[\Ce^\flat\mathbf{N}\otimes(\Fv^{-1}\Ce\mathbf{N}) 
	+ (\mathbf G \Ce^2\mathbf{N})\otimes\Fv^{-1}\mathbf{N}\right]
	= \mathbf{0}\,;
\end{aligned}
\end{equation}
%---------------------
and in the case of incompressible transversely isotropic viscoelastic solids, the kinetic equation \eqref{eq:LD_kinetic_inc} is written as
%---------------------
\begin{equation}
\begin{aligned}
	 \frac{\partial \phi}{\partial \dot\Fv}
	&-2\widetilde{\Psi}_1\Ce^{\flat}\mathbf{G}^{\sharp}\Fv^{-\star}
	-2\widetilde{\Psi}_2\left[\Ie_2\mathbf{I}-\mathbf{G}\Ce^{-\sharp} \right]\Fv^{-\star} 
	-2\widetilde\Psi_4\Ce^\flat\mathbf{N}\otimes\Fv^{-1}\mathbf{N}\\
	& -2\widetilde\Psi_5 \left[\Ce^\flat\mathbf{N}\otimes(\Fv^{-1}\Ce\mathbf{N}) 
	+ (\mathbf G \Ce^2\mathbf{N})\otimes\Fv^{-1}\mathbf{N}\right]= q \Fv^{-\star}\,.
\end{aligned}
\end{equation}
%---------------------

%-----------------------------
%-----------------------------
\subsubsection{Orthotropic solids}

An orthotropic solid at a material point $X\in\mathcal{B}$ has reflection symmetry with respect to three mutually perpendicular planes with $\mathbf G$-orthonormal vectors $\mathbf{N}_1(X)\,$, $\mathbf{N}_2(X)\,$, and $\mathbf{N}_3(X)\,$, i.e., $\llangle\mathbf{N}_i(X),\mathbf{N}_j(X)\rrangle_{\mathbf G}=\delta_{ij}\,$.
A choice for structural tensors is the set $\boldsymbol\Lambda=\{\mathbf{A}_1=\mathbf{N}_1\otimes\mathbf{N}_1, \mathbf{A}_2=\mathbf{N}_2\otimes\mathbf{N}_2, \mathbf{A}_3=\mathbf{N}_3\otimes\mathbf{N}_3\}\,$. However, $\mathbf{A}_1+\mathbf{A}_2+\mathbf{A}_3=\mathbf{I}\,$, and hence only two of them are independent.

\paragraph{Stress constitutive equations.}
One can take $\mathbf{A}_1$ and $\mathbf{A}_2$ to be the independent structural tensors of the set $\boldsymbol\Lambda\,$. When these two tensors are added to the list of the arguments of the equilibrium and non-equilibrium free energies, they become isotropic functionals of their arguments.\footnote{The functionals $\Psi_{\text{EQ}}(X,\Theta,\mathbf F, \mathbf{A}_1, \mathbf{A}_2 , \mathbf G)\,$, $\Psi_{\text{NEQ}}(X,\Theta,\Fe, \mathbf{A}_1, \mathbf{A}_2, \mathbf G)\,$, and $\Psi(X,\Theta,\mathbf F, \Fe, \mathbf{A}_1, \mathbf{A}_2, \mathbf G, \mathbf g)$ are isotropic.} This is equivalent to the free energy functionals each depending on seven invariants:
%---------------------
\begin{equation}
	\Psi_{\text{EQ}} = \overline{\Psi}(X,I_1,I_2,I_3,I_4,I_5,I_6,I_7)   \,,\qquad
	\Psi_{\text{NEQ}}  = \widetilde{\Psi}(X,\Ie_1,\Ie_2,\Ie_3,\Ie_4,\Ie_5,\Ie_6,\Ie_7)  \,,
\end{equation}
%---------------------
where 
%---------------------
\begin{equation} \label{Orthotropic-Invariants}
\begin{aligned}
	& I_1=\mathrm{tr}\,\mathbf{C}\,,&&\quad I_2=\mathrm{det}\,\mathbf{C}\,\mathrm{tr}\,
	\mathbf{C}^{-1}\,,&&\quad
	I_3=\mathrm{det}\,\mathbf{C}\,,\\
	& I_4=\mathbf{N}_1\cdot\mathbf{C}\cdot\mathbf{N}_1\,,&&\quad
	I_5=\mathbf{N}_1\cdot\mathbf{C}^2\cdot\mathbf{N}_1\,,\\
	& I_6=\mathbf{N}_2\cdot\mathbf{C}\cdot\mathbf{N}_2\,,
	&&\quad I_7=\mathbf{N}_2\cdot\mathbf{C}^2\cdot\mathbf{N}_2\,,
\end{aligned}
\end{equation}
%---------------------
and
%---------------------
\begin{equation}  \label{Orthotropic-Invariants2}
\begin{aligned}
	& \Ie_1=\mathrm{tr}\,\Ce\,,&&\quad 
	\Ie_2=\mathrm{det}\,\Ce\,\mathrm{tr}\,\Ce^{-1}\,,&&\quad
	\Ie_3=\mathrm{det}\,\Ce\,,\\
	& \Ie_4=\mathbf{N}_1\cdot\Ce\cdot\mathbf{N}_1\,,&&\quad
	I_5=\mathbf{N}_1\cdot\Ce^2\cdot\mathbf{N}_1\,,\\
	& \Ie_6=\mathbf{N}_2\cdot\Ce\cdot\mathbf{N}_2\,,
	&&\quad \Ie_7=\mathbf{N}_2\cdot\Ce^2\cdot\mathbf{N}_2\,,
\end{aligned}
\end{equation}
%---------------------
The Cauchy stress has the following representation 
%---------------------------------
\begin{equation}
\begin{aligned}
	\boldsymbol{\sigma} 
	&= \frac{2}{\sqrt{I_3}} \Bigg\{ \left(I_2\,\overline{\Psi}_2+I_3\,\overline{\Psi}_3
	+\Ie_2\,\widetilde{\Psi}_2+\Ie_3\,\widetilde{\Psi}_3\right)\mathbf{g}^{\sharp} 
	+\overline{\Psi}_1\,\mathbf{b}^{\sharp}+\widetilde{\Psi}_1\,\be^{\sharp}-I_3\,\overline{\Psi}_2\,\mathbf{c}^{\sharp}
	-\Ie_3\,\widetilde{\Psi}_2\,\ce^{\sharp} \\
	& \qquad\qquad +2\overline{\Psi}_4\,\mathbf{n}_1\otimes\mathbf{n}_1
	+2\overline{\Psi}_5\left[\mathbf{n}_1\otimes(\mathbf{b}^{\sharp}\mathbf{g}\mathbf{n}_1)
	+(\mathbf{b}^{\sharp}\mathbf{g}\mathbf{n}_1)\otimes\mathbf{n}_1 \right] \\
	& \qquad\qquad +2\overline{\Psi}_6\,\mathbf{n}_2\otimes\mathbf{n}_2
	+2\overline{\Psi}_7\left[\mathbf{n}_2\otimes(\mathbf{b}^{\sharp}\mathbf{g}\mathbf{n}_2)
	+(\mathbf{b}^{\sharp}\mathbf{g}\mathbf{n}_2)\otimes\mathbf{n}_2 \right] \\
	& \qquad\qquad +2\widetilde{\Psi}_4\,\Ne_1\otimes\Ne_1
	+2\widetilde{\Psi}_5 \Big[\Ne_1\otimes (\be^{\sharp}\mathbf{g}\Ne_1)+(\be^{\sharp}\mathbf{g}\Ne_1)\otimes\Ne_1\Big]\\
	& \qquad\qquad +2\widetilde{\Psi}_6\,\Ne_2\otimes\Ne_2
	+2\widetilde{\Psi}_7 \Big[\Ne_2\otimes (\be^{\sharp}\mathbf{g}\Ne_2)+(\be^{\sharp}\mathbf{g}\Ne_2)\otimes\Ne_2\Big]
	\Bigg\}\,,
\end{aligned}
\end{equation}
%---------------------------------
where $\Ne_1=\Fe\mathbf{N}_1\,$, $\Ne_2=\Fe\mathbf{N}_2\,$, and
%---------------------------------
\begin{equation}
	\overline{\Psi}_j=\overline{\Psi}_j(X,I_1,I_2,I_3,I_4,I_5,I_6,I_7):=\frac{\partial \overline{\Psi}}{\partial I_j}\,,\quad
	\widetilde{\Psi}_j=\widetilde{\Psi}_j(X,\Ie_1,\Ie_2,\Ie_3,\Ie_4,\Ie_5,\Ie_6,\Ie_7)
	:=\frac{\partial \widetilde{\Psi}}{\partial \Ie_j}\,, \quad j=1,\cdots,7\,.
\end{equation}
%---------------------------------
For an incompressible isotropic solid, $I_3=\Ie_3=1\,$, and hence
%---------------------------------
\begin{equation}
\begin{aligned}
	\boldsymbol{\sigma} 
	&=-p\,\mathbf{g}^{\sharp} 
	+2\overline{\Psi}_1\,\mathbf{b}^{\sharp}+2\widetilde{\Psi}_1\,\be^{\sharp}-\overline{\Psi}_2\,\mathbf{c}^{\sharp}
	-2\,\widetilde{\Psi}_2\,\ce^{\sharp} \\
	& \quad +2\overline{\Psi}_4\,\mathbf{n}_1\otimes\mathbf{n}_1
	+2\overline{\Psi}_5\left[\mathbf{n}_1\otimes(\mathbf{b}^{\sharp}\mathbf{g}\mathbf{n}_1)
	+(\mathbf{b}^{\sharp}\mathbf{g}\mathbf{n}_1)\otimes\mathbf{n}_1 \right] \\
	& \quad +2\overline{\Psi}_6\,\mathbf{n}_2\otimes\mathbf{n}_2
	+2\overline{\Psi}_7\left[\mathbf{n}_2\otimes(\mathbf{b}^{\sharp}\mathbf{g}\mathbf{n}_2)
	+(\mathbf{b}^{\sharp}\mathbf{g}\mathbf{n}_2)\otimes\mathbf{n}_2 \right] \\
	&\quad +2\widetilde{\Psi}_4\,\Ne_1\otimes\Ne_1
	+2\widetilde{\Psi}_5 \Big[\Ne_1\otimes (\be^{\sharp}\mathbf{g}\Ne_1)+(\be^{\sharp}\mathbf{g}\Ne_1)\otimes\Ne_1\Big]\\
	&\quad +2\widetilde{\Psi}_6\,\Ne_2\otimes\Ne_2
	+2\widetilde{\Psi}_7 \Big[\Ne_2\otimes (\be^{\sharp}\mathbf{g}\Ne_2)+(\be^{\sharp}\mathbf{g}\Ne_2)\otimes\Ne_2\Big]
	\,,
\end{aligned}
\end{equation}
%---------------------------------
where $p$ is the Lagrange multiplier associated with the incompressibility constraint $J=\sqrt{I_3}=1\,$. 

\paragraph{Dissipation potential.}
For an orthotropic viscoelastic solid, when two elements of the set of structural tensors $\boldsymbol\Lambda$ are added to the list of the arguments of the dissipation potential $\phi\,$, it becomes an isotropic functional of its arguments, e.g., $\phi=\hat{\phi}(X,\mathbf{C}^{\flat},\Fv,\dot\Fv,\mathbf{A}_1, \mathbf{A}_2, \mathbf{G},\mathbf{g})\,$. Although one may not use the standard representation theorem as for the free energy functionals, the dissipation potential will be a functional of some standard invariants and a set of spectral invariants, similarly to the dissipation potential of the isotropic viscoelastic solids.

\paragraph{Kinetic equation.}
Following \eqref{eq:part_Ie} and \eqref{eq:diff_Ie4-5}, one may write
%---------------------
\begin{equation}
\begin{split}
\frac{\partial \Psi}{\partial \Ce^\flat}
&=\frac{\partial \Psi_{\text{NEQ}}}{\partial \Ce^\flat} = \widetilde\Psi_1\frac{\partial \Ie_1}{\partial \Ce^\flat} + \widetilde\Psi_2\frac{\partial \Ie_2}{\partial \Ce^\flat} + \widetilde\Psi_3\frac{\partial \Ie_3}{\partial \Ce^\flat} + \widetilde\Psi_4\frac{\partial \Ie_4}{\partial \Ce^\flat} + \widetilde\Psi_5\frac{\partial \Ie_5}{\partial \Ce^\flat} + \widetilde\Psi_6\frac{\partial \Ie_6}{\partial \Ce^\flat} + \widetilde\Psi_7\frac{\partial \Ie_7}{\partial \Ce^\flat}\\
&= \widetilde\Psi_1\mathbf{G}^{\sharp} + \widetilde\Psi_2 \left(\Ie_2\Ce^{-\sharp}-\Ie_3\Ce^{-2\sharp}\right) + \widetilde\Psi_3 \Ie_3\Ce^{-\sharp} + \widetilde\Psi_4\mathbf{N}_1\otimes\mathbf{N}_1 + \widetilde\Psi_5(\mathbf{N}_1\otimes(\Ce\mathbf{N}_1) + (\Ce\mathbf{N}_1)\otimes\mathbf{N}_1)\\
&\quad+ \widetilde\Psi_6\mathbf{N}_2\otimes\mathbf{N}_2 + \widetilde\Psi_7(\mathbf{N}_2\otimes(\Ce\mathbf{N}_2) + (\Ce\mathbf{N}_2)\otimes\mathbf{N}_2)
\,.
\end{split}
\end{equation}
%---------------------
Hence, it follows from \eqref{eq:LD_kinetic} that the kinetic equation for compressible orthotropic viscoelastic solids reads
%---------------------
\begin{equation}
\begin{aligned}
	 \frac{\partial \phi}{\partial \dot\Fv} 
	&-2\widetilde{\Psi}_1\Ce^{\flat}\mathbf{G}^{\sharp}\Fv^{-\star}
	-2\widetilde{\Psi}_2\left[\Ie_2\mathbf{I}-\Ie_3\mathbf{G}\Ce^{-\sharp} \right]\Fv^{-\star}
	-2\Ie_3\widetilde{\Psi}_3\Fv^{-\star} \\
	&  -2\widetilde\Psi_4\Ce^\flat\mathbf{N}_1\otimes\Fv^{-1}\mathbf{N}_1
	-2\widetilde\Psi_5(\Ce^\flat\mathbf{N}_1\otimes(\Fv^{-1}\Ce\mathbf{N}_1) 
	+ (\mathbf G \Ce^2\mathbf{N}_1)\otimes\Fv^{-1}\mathbf{N}_1) \\
	&  -2\widetilde\Psi_6\Ce^\flat\mathbf{N}_2\otimes\Fv^{-1}\mathbf{N}_2
	-2\widetilde\Psi_7(\Ce^\flat\mathbf{N}_2\otimes(\Fv^{-1}\Ce\mathbf{N}_2) 
	+ (\mathbf G \Ce^2\mathbf{N}_2)\otimes\Fv^{-1}\mathbf{N}_2)= \mathbf{0}\,;
\end{aligned}
\end{equation}
%---------------------
and in the case of incompressible orthotropic viscoelastic solids, the kinetic equation \eqref{eq:LD_kinetic_inc} is written as
%---------------------
\begin{equation}
\begin{aligned}
	 \frac{\partial \phi}{\partial \dot\Fv}
	&-2\widetilde{\Psi}_1\Ce^{\flat}\mathbf{G}^{\sharp}\Fv^{-\star}
	-2\widetilde{\Psi}_2\left[\Ie_2\mathbf{I}-\mathbf{G}\Ce^{-\sharp} \right]\Fv^{-\star} \\
	&  -2\widetilde\Psi_4\Ce^\flat\mathbf{N}_1\otimes\Fv^{-1}\mathbf{N}_1
	-2\widetilde\Psi_5(\Ce^\flat\mathbf{N}_1\otimes(\Fv^{-1}\Ce\mathbf{N}_1) 
	+ (\mathbf G \Ce^2\mathbf{N}_1)\otimes\Fv^{-1}\mathbf{N}_1) \\
	&  -2\widetilde\Psi_6\Ce^\flat\mathbf{N}_2\otimes\Fv^{-1}\mathbf{N}_2
	-2\widetilde\Psi_7(\Ce^\flat\mathbf{N}_2\otimes(\Fv^{-1}\Ce\mathbf{N}_2) 
	+ (\mathbf G \Ce^2\mathbf{N}_2)\otimes\Fv^{-1}\mathbf{N}_2)= q \Fv^{-\star}\,.
\end{aligned}
\end{equation}
%---------------------

%-----------------------------
%-----------------------------
\subsubsection{Monoclinic solids}

A monoclinic solid at a material point $X\in\mathcal{B}$ has three material preferred directions $\{\mathbf{N}_1(X),\mathbf{N}_2(X),\mathbf{N}_3(X)\}$ such that $\mathbf{N}_1\cdot\mathbf{N}_2\neq 0$ and $\mathbf{N}_3$ is normal to the plane of $\mathbf{N}_1$ and $\mathbf{N}_2$ \citep{merodio2020finite}.

\paragraph{Stress constitutive equations.}
The equilibrium and non-equilibrium free energies of a monoclinic solid depend on nine invariants \citep{Spencer1986}:
%---------------------
\begin{equation}
	\Psi_{\text{EQ}} = \overline{\Psi}_{\text{EQ}}(X,I_1,I_2,I_3,I_4,I_5,I_6,I_7,I_8,I_9)   \,,\qquad
	\Psi_{\text{NEQ}}  
	= \overline{\Psi}_{\text{NEQ}}(X,\Ie_1,\Ie_2,\Ie_3,\Ie_4,\Ie_5,\Ie_6,\Ie_7,\Ie_8,\Ie_9)  \,.
\end{equation}
%---------------------
For each free energy, the first seven invariants are identical to those of orthotropic solids \eqref{Orthotropic-Invariants} and \eqref{Orthotropic-Invariants2}. The three extra invariants are
%---------------------
\begin{equation}
	I_8=\mathcal{I}\,\mathbf{N}_1\cdot\mathbf{C}\cdot\mathbf{N}_2\,,\qquad
	\Ie_8=\mathcal{I}\,\mathbf{N}_1\cdot\Ce\cdot\mathbf{N}_2\,,\qquad
	I_9=\Ie_9=\mathcal{I}^2\,,\qquad \mathcal{I}=\mathbf{N}_1\cdot\mathbf{N}_2\,.
\end{equation}  
%---------------------
Note that
%---------------------
\begin{equation}\label{eq:mono_inv}
	\frac{\partial I_8}{\partial\mathbf{C}^\flat}
	=\frac{1}{2}\mathcal{I}\,(\mathbf{N}_1\otimes\mathbf{N}_2+\mathbf{N}_2\otimes\mathbf{N}_1)\,,\quad
	\frac{\partial I_9}{\partial\mathbf{C}^\flat}=\mathbf{0}\,,\quad
	\frac{\partial \Ie_8}{\partial\Ce^\flat}
	=\frac{1}{2}\mathcal{I}\,(\mathbf{N}_1\otimes\mathbf{N}_2+\mathbf{N}_2\otimes\mathbf{N}_1)\,,\quad
	\frac{\partial \Ie_9}{\partial\Ce^\flat}=\mathbf{0}\,.
\end{equation}
%---------------------
The Cauchy stress has the following representation 
%---------------------------------
\begin{equation}
\begin{aligned}
	\boldsymbol{\sigma} 
	&= \frac{2}{\sqrt{I_3}} \Bigg\{ \left(I_2\,\overline{\Psi}_2+I_3\,\overline{\Psi}_3
	+\Ie_2\,\widetilde{\Psi}_2+\Ie_3\,\widetilde{\Psi}_3\right)\mathbf{g}^{\sharp} 
	+\overline{\Psi}_1\,\mathbf{b}^{\sharp}+\widetilde{\Psi}_1\,\be^{\sharp}-I_3\,\overline{\Psi}_2\,\mathbf{c}^{\sharp}
	-\Ie_3\,\widetilde{\Psi}_2\,\ce^{\sharp} \\
	& \qquad\qquad +2\overline{\Psi}_4\,\mathbf{n}_1\otimes\mathbf{n}_1
	+2\overline{\Psi}_5\left[\mathbf{n}_1\otimes(\mathbf{b}^{\sharp}\mathbf{g}\mathbf{n}_1)
	+(\mathbf{b}^{\sharp}\mathbf{g}\mathbf{n}_1)\otimes\mathbf{n}_1 \right] \\
	& \qquad\qquad +2\overline{\Psi}_6\,\mathbf{n}_2\otimes\mathbf{n}_2
	+2\overline{\Psi}_7\left[\mathbf{n}_2\otimes(\mathbf{b}^{\sharp}\mathbf{g}\mathbf{n}_2)
	+(\mathbf{b}^{\sharp}\mathbf{g}\mathbf{n}_2)\otimes\mathbf{n}_2 \right]
	+\mathcal{I}\,\overline{\Psi}_8\left(\mathbf{n}_1\otimes\mathbf{n}_2+\mathbf{n}_2\otimes\mathbf{n}_1\right)
	 \\
	& \qquad\qquad +2\widetilde{\Psi}_4\,\Ne_1\otimes\Ne_1
	+2\widetilde{\Psi}_5 \Big[\Ne_1\otimes (\be^{\sharp}\mathbf{g}\Ne_1)+(\be^{\sharp}\mathbf{g}\Ne_1)\otimes\Ne_1\Big]\\
	& \qquad\qquad +2\widetilde{\Psi}_6\,\Ne_2\otimes\Ne_2
	+2\widetilde{\Psi}_7 \Big[\Ne_2\otimes (\be^{\sharp}\mathbf{g}\Ne_2)+(\be^{\sharp}\mathbf{g}\Ne_2)\otimes\Ne_2\Big]
	+\mathcal{I}\,\widetilde{\Psi}_8\left(\Ne_1\otimes\Ne2+\Ne_2\otimes\Ne_1\right)
	\Bigg\}\,.
\end{aligned}
\end{equation}
%---------------------------------
For an incompressible monoclinic solid
%---------------------------------
\begin{equation}
\begin{aligned}
	\boldsymbol{\sigma} 
	&=-p\,\mathbf{g}^{\sharp} 
	+2\overline{\Psi}_1\,\mathbf{b}^{\sharp}+2\widetilde{\Psi}_1\,\be^{\sharp}-\overline{\Psi}_2\,\mathbf{c}^{\sharp}
	-2\,\widetilde{\Psi}_2\,\ce^{\sharp} \\
	& \quad +2\overline{\Psi}_4\,\mathbf{n}_1\otimes\mathbf{n}_1
	+2\overline{\Psi}_5\left[\mathbf{n}_1\otimes(\mathbf{b}^{\sharp}\mathbf{g}\mathbf{n}_1)
	+(\mathbf{b}^{\sharp}\mathbf{g}\mathbf{n}_1)\otimes\mathbf{n}_1 \right] \\
	& \quad +2\overline{\Psi}_6\,\mathbf{n}_2\otimes\mathbf{n}_2
	+2\overline{\Psi}_7\left[\mathbf{n}_2\otimes(\mathbf{b}^{\sharp}\mathbf{g}\mathbf{n}_2)
	+(\mathbf{b}^{\sharp}\mathbf{g}\mathbf{n}_2)\otimes\mathbf{n}_2 \right] 
	+\mathcal{I}\,\overline{\Psi}_8\left(\mathbf{n}_1\otimes\mathbf{n}_2+\mathbf{n}_2\otimes\mathbf{n}_1\right)
	\\
	&\quad +2\widetilde{\Psi}_4\,\Ne_1\otimes\Ne_1
	+2\widetilde{\Psi}_5 \Big[\Ne_1\otimes (\be^{\sharp}\mathbf{g}\Ne_1)+(\be^{\sharp}\mathbf{g}\Ne_1)\otimes\Ne_1\Big]\\
	&\quad +2\widetilde{\Psi}_6\,\Ne_2\otimes\Ne_2
	+2\widetilde{\Psi}_7 \Big[\Ne_2\otimes (\be^{\sharp}\mathbf{g}\Ne_2)+(\be^{\sharp}\mathbf{g}\Ne_2)\otimes\Ne_2\Big]
	+\mathcal{I}\,\widetilde{\Psi}_8\left(\Ne_1\otimes\Ne2+\Ne_2\otimes\Ne_1\right)
	\,.
\end{aligned}
\end{equation}
%---------------------------------

\paragraph{Dissipation potential.}
For a monoclinic viscoelastic solid, when the full set of structural tensors
%---------------------
\begin{equation}
\boldsymbol\Lambda=\{\mathbf{A}_1=\mathbf{N}_1\otimes\mathbf{N}_1, \mathbf{A}_2=\mathbf{N}_2\otimes\mathbf{N}_2, \mathbf{A}_3=\mathbf{N}_3\otimes\mathbf{N}_3\}
\end{equation}
%---------------------------------
 is added to the list of the arguments of the dissipation potential $\phi\,$, it becomes an isotropic functional of its arguments, i.e., $\phi=\hat{\phi}(X,\mathbf{C}^{\flat},\Fv,\dot\Fv,\mathbf{A}_1, \mathbf{A}_2, \mathbf{A}_3, \mathbf{G},\mathbf{g})\,$. Although one may not use the standard representation theorem as for the free energy functionals, the dissipation potential will be a functional of some standard invariants and a set of spectral invariants, similarly to the dissipation potential of isotropic viscoelastic solids.

\paragraph{Kinetic equation.}
Following \eqref{eq:part_Ie}, \eqref{eq:diff_Ie4-5}, and \eqref{eq:mono_inv}, one may write
%---------------------
\begin{equation}
\begin{split}
\frac{\partial \Psi}{\partial \Ce^\flat}
&=\frac{\partial \Psi_{\text{NEQ}}}{\partial \Ce^\flat} = \widetilde\Psi_1\frac{\partial \Ie_1}{\partial \Ce^\flat} + \widetilde\Psi_2\frac{\partial \Ie_2}{\partial \Ce^\flat} + \widetilde\Psi_3\frac{\partial \Ie_3}{\partial \Ce^\flat} + \widetilde\Psi_4\frac{\partial \Ie_4}{\partial \Ce^\flat} + \widetilde\Psi_5\frac{\partial \Ie_5}{\partial \Ce^\flat} + \widetilde\Psi_6\frac{\partial \Ie_6}{\partial \Ce^\flat} + \widetilde\Psi_7\frac{\partial \Ie_7}{\partial \Ce^\flat}+ \widetilde\Psi_8\frac{\partial \Ie_8}{\partial \Ce^\flat} + \widetilde\Psi_9\frac{\partial \Ie_9}{\partial \Ce^\flat}\\
&= \widetilde\Psi_1\mathbf{G}^{\sharp} + \widetilde\Psi_2 \left(\Ie_2\Ce^{-\sharp}-\Ie_3\Ce^{-2\sharp}\right) + \widetilde\Psi_3 \Ie_3\Ce^{-\sharp} + \widetilde\Psi_4\mathbf{N}_1\otimes\mathbf{N}_1 + \widetilde\Psi_5(\mathbf{N}_1\otimes(\Ce\mathbf{N}_1) + (\Ce\mathbf{N}_1)\otimes\mathbf{N}_1)\\
&\quad+ \widetilde\Psi_6\mathbf{N}_2\otimes\mathbf{N}_2 + \widetilde\Psi_7(\mathbf{N}_2\otimes(\Ce\mathbf{N}_2) + (\Ce\mathbf{N}_2)\otimes\mathbf{N}_2)+ \frac{1}{2}(\mathbf{N}_1\cdot\mathbf{N}_2)\widetilde\Psi_8(\mathbf{N}_1\otimes\mathbf{N}_2+\mathbf{N}_2\otimes\mathbf{N}_1)
\,.
\end{split}
\end{equation}
%---------------------
Hence, it follows from \eqref{eq:LD_kinetic} that the kinetic equation for compressible monoclinic viscoelastic solids is written as
%---------------------
\begin{equation}
\begin{aligned}
	\frac{\partial \phi}{\partial \dot\Fv} 
	&-2\widetilde{\Psi}_1\Ce^{\flat}\mathbf{G}^{\sharp}\Fv^{-\star}
	-2\widetilde{\Psi}_2\left[\Ie_2\mathbf{I}-\Ie_3\mathbf{G}\Ce^{-\sharp} \right]\Fv^{-\star}
	-2\Ie_3\widetilde{\Psi}_3\Fv^{-\star} \\
	&  -2\widetilde\Psi_4\Ce^\flat\mathbf{N}_1\otimes\Fv^{-1}\mathbf{N}_1
	-2\widetilde\Psi_5(\Ce^\flat\mathbf{N}_1\otimes(\Fv^{-1}\Ce\mathbf{N}_1) 
	+ (\mathbf G \Ce^2\mathbf{N}_1)\otimes\Fv^{-1}\mathbf{N}_1) \\
	& -2\widetilde\Psi_6\Ce^\flat\mathbf{N}_2\otimes\Fv^{-1}\mathbf{N}_2
	-2\widetilde\Psi_7(\Ce^\flat\mathbf{N}_2\otimes(\Fv^{-1}\Ce\mathbf{N}_2) 
	+ (\mathbf G \Ce^2\mathbf{N}_2)\otimes\Fv^{-1}\mathbf{N}_2) \\
	&  -(\mathbf{N}_1\cdot\mathbf{N}_2)\widetilde\Psi_8(\Ce^\flat\mathbf{N}_1\otimes
	\Fv^{-1}\mathbf{N}_2+\Ce^\flat\mathbf{N}_2\otimes\Fv^{-1}\mathbf{N}_1)= \mathbf{0}\,;
\end{aligned}
\end{equation}
%---------------------
and in the case of incompressible monoclinic viscoelastic solids, the kinetic equation \eqref{eq:LD_kinetic_inc} reads
%---------------------
\begin{equation}
\begin{aligned}
	\frac{\partial \phi}{\partial \dot\Fv}
	&-2\widetilde{\Psi}_1\Ce^{\flat}\mathbf{G}^{\sharp}\Fv^{-\star}
	-2\widetilde{\Psi}_2\left[\Ie_2\mathbf{I}-\mathbf{G}\Ce^{-\sharp} \right]\Fv^{-\star} \\
	& -2\widetilde\Psi_4\Ce^\flat\mathbf{N}_1\otimes\Fv^{-1}\mathbf{N}_1
	-2\widetilde\Psi_5(\Ce^\flat\mathbf{N}_1\otimes(\Fv^{-1}\Ce\mathbf{N}_1) 
	+ (\mathbf G \Ce^2\mathbf{N}_1)\otimes\Fv^{-1}\mathbf{N}_1) \\
	&  -2\widetilde\Psi_6\Ce^\flat\mathbf{N}_2\otimes\Fv^{-1}\mathbf{N}_2
	-2\widetilde\Psi_7(\Ce^\flat\mathbf{N}_2\otimes(\Fv^{-1}\Ce\mathbf{N}_2) 
	+ (\mathbf G \Ce^2\mathbf{N}_2)\otimes\Fv^{-1}\mathbf{N}_2) \\
	&  -(\mathbf{N}_1\cdot\mathbf{N}_2)\widetilde\Psi_8(\Ce^\flat\mathbf{N}_1\otimes
	\Fv^{-1}\mathbf{N}_2+\Ce^\flat\mathbf{N}_2\otimes\Fv^{-1}\mathbf{N}_1)= q \Fv^{-\star}\,.
\end{aligned}
\end{equation}
%---------------------

%-----------------------------
%-----------------------------
\subsection{Invariance in anelasticity and viscoelasticity: A critical discussion of some of the existing works} \label{Sec:CriticalDiscussion}

The notion of invariance and its interpretation in the presence of inelastic deformations has eluded mechanicians over the past few decades. Invariance is a central notion in physics, and particularly in mechanics; there is a deep connection between balance laws and symmetries. Noether's theorems tell us that any symmetry of the Lagrangian density (or action) corresponds to a conserved quantity or a balance law~\citep{Kosmann2011,MarsRat2013MechSym}.
For example, invariance under time shifts corresponds to the balance of energy. As another example, for continuum mechanics formulated in a Euclidean ambient space, the balance of angular momentum corresponds to invariance under rigid body rotations in the ambient space. On the other hand, local invariance in the reference configuration is related to material symmetry.

\paragraph{The work of Green and Naghdi.}
\citet{GreenNaghdi1971} observed that for any proper orthogonal tensor $\mathbf{Q}\,$, the multiplicative decomposition of the deformation gradient can be written as $\mathbf{F}=\Fe\Fp=\Fe\mathbf{Q}\mathbf{Q}^{\mathsf{T}}\Fp\,$, and hence {$\Fe$} and $\Fp$ can be replaced by $\Fe\mathbf{Q}$ and $\mathbf{Q}^{\mathsf{T}}\Fp\,$, respectively. However, it should be noted that replacing {$\Fe$} by $\Fe\mathbf{Q}$ implies that $\mathbf{Q}$ is an element of the material symmetry group $\mathcal{G}\,$. Assuming that $\mathbf{Q}$ is any proper orthogonal tensor (or rotation) is equivalent to assuming that the material is isotropic. In other words,~\citet{GreenNaghdi1971}'s argument is incorrect for anisotropic solids; there is a $\mathcal{G}$-ambiguity in the multiplicative decomposition and not an $SO(3)$-ambiguity, see also~\citet{YavariSozio2023}.

\paragraph{The work of Simo.}
In formulating finite plasticity,~\citet{Simo1988} considered the multiplicative decomposition of the deformation gradient into elastic and plastic parts: $\mathbf{F}=\Fe\Fp\,$. He considered coordinate charts $\{x^i\}$ and $\{X^I\}$ for the current and reference configurations, respectively. The spatial metric has components $g_{ij}$ and the metric of the reference configuration has components $G_{IJ}\,$. Looking at the coordinate representation of $\Fp$ in Eq.(1.2b) in~\citep{Simo1988}, clearly it is assumed that $\Fp$ is a linear map from the tangent space of the reference configuration to itself ($\Fp:T_X\mathcal{B}\to T_X\mathcal{B}$ in our notation). This means that the ``intermediate configuration'' is identified with $T_X\mathcal{B}\,$. After Eq.(1.2b), it is explicitly mentioned that ``where we have endowed the intermediate configuration with the metric tensor $\mathbf{G}$\,''. In other words, the same metric is used in both the reference and intermediate configurations.~\citet{Simo1988} assumes a free energy function of the form $\psi=\hat{\psi}(\mathbf{g},\Fe,\mathbf{F})$ (an explicit dependence on $\mathbf{G}$ is suppressed perhaps because the flat metric $\mathbf{G}$ is induced from the spatial metric $\mathbf{g}$). Then ``invariance under rigid-body motions superposed onto the intermediate configuration'' is assumed that~\citet{Simo1988} writes as
%-----------------------------
\begin{equation} \label{Simo-Invariance}
	\hat{\psi}(\mathbf{g},\Fe\mathbf{Q},\mathbf{F})=\hat{\psi}(\mathbf{g},\Fe,\mathbf{F})\,,\quad
	\forall \mathbf{Q}\in SO(3)\,,
\end{equation}
%-----------------------------
i.e.,~for any rotation $\mathbf{Q}$ in the ``intermediate configuration''. Recall that $\Fe:T_X \mathcal{B}\to\ T_x\mathcal{C}$ and $\mathbf{F}:T_X\mathcal{B}\to\ T_x\mathcal{C}\,$, i.e.,~{$\Fe$} and $\mathbf{F}$ have the same tensor character, and hence \eqref{Simo-Invariance} does not make sense; $\mathbf{F}$ must be transformed as well. Simo was aware that not including $\mathbf{F}$ as an argument of the free energy in \eqref{Simo-Invariance} implies material isotropy. He introduced $\mathbf{F}$ as an argument in the free energy that is unchanged under ``rotations in the intermediate configuration'' in order to avoid material isotropy~\citep[Remark 1.6]{Simo1988}. However, assuming invariance with respect to the ``intermediate configuration'' is equivalent to material symmetry under all rotations and indeed precludes anisotropic response. In his last piece of work before passing that was posthumously published,\footnote{We thank Sanjay Govindjee for bringing this reference to our attention.}  Simo remarked that~\citep[Remark 34.2]{Simo1998Handbook}: ``The entire issue depends on an a priori specification of the class of admissible rotations $\mathbf{Q}$ for such transformations. This question is related to a constitutive assumption on the symmetry group of the material and appears to have little to do with any fundamental principle in continuum physics''.

\paragraph{The work of Gurtin and Anand.}
\citet{Gurtin2005} studied material symmetry in the presence of local plastic deformations. They call the target space of $\Fp$ ``relaxed space'', which is usually called ``intermediate configuration''. They treated it as an entity completely independent from the reference and current configurations; independent in the sense that the ``relaxed space'' is not affected by either referential or spatial transformations. This assumption leads to the definition of two symmetry groups, namely, ``referential symmetry group'' and ``relaxational symmetry group''. More specifically, they assumed a free energy $\psi=\hat{\psi}(\Fp,\Fe)\,$. The two symmetry groups are defined as
%-----------------------------
\begin{equation} \label{Ref-Rel}
\begin{aligned}
	\hat{\psi}(\Fp,\Fe) &=\hat{\psi}(\Fp\mathbf{H},\Fe)\,,\quad && 
	\forall \mathbf{H}\in\mathcal{G}^{\text{ref}}\,,\\
	\hat{\psi}(\Fp,\Fe) &=\hat{\psi}(\mathbf{H}^{-1}\Fp,\Fe\mathbf{H})\,,\quad 
	&& \forall \mathbf{H}\in\mathcal{G}^{\text{rel}}
	\,.
\end{aligned}
\end{equation}
%-----------------------------
An ``intermediate configuration'' or a ``relaxed space'' is defined pointwise, and for a given point, it is a linear space. The total deformation gradient is a linear map between tangent spaces of the material and the ambient space manifolds: $\mathbf{F}:T_X\mathcal{B}\to T_x\mathcal{S}\,$. There are only two spaces (manifolds) in anelasticity and viscoelasticity: the ambient space manifold $\mathcal{S}$ (which is usually assumed to be the Euclidean $3$-space)\footnote{In some applications the ambient space could be curved, in general. See~\citet{Yavari2016} for a detailed discussion and examples.} and the material manifold $\mathcal{B}$ (which is an embedded $3$-submanifold of the Euclidean ambient space).
In the decomposition $\mathbf{F}=\Fe\Fp\,$, {$\Fe$} and $\Fp$ are linear maps. Their domain and target spaces can be either $T_X\mathcal{B}$ or $T_x\mathcal{C}$ (see  Fig.~\ref{rmrk:V_intermediate}). In other words, assuming a third linear space distinct from $T_X\mathcal{B}$ and $T_x\mathcal{C}$ does not have physical relevance.
Therefore, the correct symmetry group should be defined as
%-----------------------------
\begin{equation} 
\begin{aligned}
	\hat{\psi}(\Fp,\Fe) =\hat{\psi}(\mathbf{H}^{-1}\Fp\mathbf{H},\Fe\mathbf{H})\,,\quad 
	\forall \mathbf{H}\in\mathcal{G}
	\,.
\end{aligned}
\end{equation}
%-----------------------------

\paragraph{The work of Kumar and Lopez-Pamies.}
\citet{Kumar2016} assumed that {$\Fe$} and {$\Fv$} are compatible\textemdash see their Eq. (4). In addition to the reference $\Omega_{0}$ and current $\Omega$ configurations, a global intermediate configuration $\Omega_v$ was also assumed\textemdash see their Fig. 2. Under a material symmetry $\mathbf{K}\,$, they assumed that {$\Fv$} is transformed to $\Fv \mathbf{K}\,$, and hence, $\Fe=\mathbf F \Fv^{-1}$ remains unchanged, since $(\mathbf F \mathbf K) (\Fv \mathbf K)^{-1}=\Fe\,$. It was finally concluded that $\Psi_{\text{NEQ}}$ is unaltered by material symmetry. First, it should be noted that there is no reason to expect that {$\Fe$} (and consequently {$\Fv$}) is compatible. In other words, a global Euclidean intermediate configuration does not exist, in general. Further, knowing that $\Fv(X):T_X\mathcal{B}\to T_X\mathcal{B}\,$, a change of material configuration by $\mathbf K$ transforms {$\Fv$} to $\mathbf{K}^{-1}\Fv \mathbf{K}$ and {$\Fe$} to $\Fe \mathbf{K}$\, (see Fig.~\ref{Symmetry-Transformation}a).
In~\citep{Kumar2016}, it was assumed that $\Psi_{\text{EQ}}=\Psi_{\text{EQ}}(X,\mathbf F,\mathbf G,\mathbf g)$ and $\phi=\phi(X,\Fv,\dot\Fv,\mathbf{G})$ have the same symmetry group, but $\Psi_{\text{NEQ}}=\Psi_{\text{NEQ}}(X,\Fe,\mathbf G,\mathbf g)$ was excluded seemingly because it was assumed that {$\Fe$} was not affected by material symmetries. In light of the discussion in \S\ref{Additive_decomp}, $\Fe(X):T_X\mathcal{B}\to T_x\mathcal{C}\,$, and it hence seems natural to assume that $\Psi_{\text{NEQ}}$ has the same symmetry group as well. 

\paragraph{The work of Ciambella and Nardinocchi.}
In a recent paper~\citep{Ciambella2021} aiming to formulate a theory of anisotropic viscoelasticity, the multiplicative decomposition {$\mathbf{F}=\Fe\Fv$} was used. The authors recognized that {$\Fe$} and {$\Fv$} are incompatible, in general. However, they confused viscoelasticity with
anelasticity and assumed that {$\Fv$} defines a local relaxed configuration (after their Eq. (2.1), they say that the viscous deformation gradient acts on a small piece of the body and maps it ``into its relaxed (or natural) state at time $t$\,''.). This is an incorrect assumption. Their choice of the free energy in their Eq. (3.13) is identical to what one would see in anelasticity. \citet{Ciambella2021} also claimed that a theory of nonlinear viscoelasticity has to be ``structurally frame indifferent''. They based this claim on the work of~\citet{GreenNaghdi1971}. In summary, invariance in the ``intermediate configuration'' or ``structural invariance'' is not physically meaningful. The above-mentioned fundamentally questionable assumptions, unfortunately, make the formulation presented in~\citep{Ciambella2021} flawed.

\paragraph{What have we learned?}
The source of confusion in the literature has been a lack of understanding of the mathematical nature of ``intermediate configuration''. A body $\mathcal{B}$ is an embedded topological submanifold of the Euclidean ambient space $\mathcal{S}\,$. In nonlinear elasticity, $\mathcal{B}$ is equipped with a metric that is induced from the ambient space. This defines a Euclidean material manifold. In anelasticity and viscoelasticity, ``intermediate configuration'' has traditionally been defined locally; a local intermediate configuration is a linear space with a Euclidean metric. One should note that in the case of the direct Bilby-Kr\"oner-Lee decomposition, an intermediate configuration (manifold) has the same topology as $\mathcal{B}\,$. However, an intermediate configuration cannot be isometrically embedded in the Euclidean space because the material metric is non-Euclidean, in general. In anelasticity and viscoelasticity, there are only two manifolds that are of physical significance: (i) the ambient space manifold $\mathcal{S}\,$, which is the Euclidean $3$-space, and (ii) the material manifold $\mathcal{B}\,$, which is an embedded topological submanifold of $\mathcal{S}\,$. Any local invariance is either defined for $x\in\mathcal{S}$ on $T_x\mathcal{S}\,$, or for $X\in\mathcal{B}$ on $T_X\mathcal{B}\,$. The former invariance is the material-frame-indifference (objectivity), and the latter is related to material symmetry; any ``intermediate configuration invariance'' is nothing but a material symmetry, in the case of the direct Bilby-Kr\"oner-Lee decomposition.

\vskip 0.1in
\noindent 
Table \ref{Table-Summary} summarizes some of the important fields, constitutive equations, and governing equations of nonlinear anisotropic viscoelasticity.

%-----------------------------
%-----------------------------
%-----------------------------
\begin{table}
\begin{center}
\hrule
\begin{tabular}{ll}
\multicolumn{2}{c}{\bf Nonlinear Anisotropic Viscoelasticity}  \\
\\
{\bf Kinematics} & \\
$ {\displaystyle 
 \mathbf{F}=\Fe\Fv } $ &  $\Fv(X):T_X\mathcal{B}\to T_X\mathcal{B}$\\
& $\Fe(X):T_X\mathcal{B}\to T_x\mathcal{C}$ \\ 
& \\
%& \\
{\bf Free energy} & {\bf Isotropic solids} \\
$\Psi=\Psi_{\text{EQ}}+\Psi_{\text{NEQ}}$  & $\Psi_{\text{EQ}}=\Psi_{\text{EQ}}(X,\mathbf{F},\mathbf{G},\mathbf{g})=\hat{\Psi}_{\text{EQ}}(X,\mathbf{C}^{\flat},\mathbf{G})$ \\
 & $\Psi_{\text{NEQ}}=\Psi_{\text{NEQ}}(X,\Fe,\mathbf{G},\mathbf{g})=\hat{\Psi}_{\text{NEQ}}(X,\Ce^{\flat},\mathbf{G})$ \\
 \\
 $\boldsymbol{\Lambda}$: structural tensors  & {\bf Anisotropic solids} \\
$\Lambdav=\Fv^*\boldsymbol{\Lambda}$: viscous structural tensors & $\Psi_{\text{EQ}}=\Psi_{\text{EQ}}(X,\Theta,\mathbf F, \mathbf{G},\boldsymbol{\Lambda},\mathbf{g})
=\hat{\Psi}_{\text{EQ}}(X,\Theta,\mathbf{C}^{\flat}, \mathbf{G},\boldsymbol{\Lambda})$ \\
 $\Gv=\Fv^*\mathbf{G}$: viscous material metric & $\Psi_{\text{NEQ}}
 =\Psi_{\text{NEQ}}(X,\Theta,\mathbf{F}, \Gv,\Lambdav,\mathbf{g})
 =\hat{\Psi}_{\text{NEQ}}(X,\Theta,\mathbf{C}^{\flat},\Gv,\Lambdav)$  \\
 \\
{\bf Dissipation potential} & \\
 $\Bv= -\partial \phi/\partial \dot\Fv$~(dissipative viscous-force)  & $\phi=\phi(X,\mathbf{F},\Fv,\dot\Fv,\mathbf{G},\mathbf{g})=\hat{\phi}(X,\mathbf{C}^{\flat},\Fv,\dot\Fv,\mathbf{G})$ \\
 \\

 {\bf Material symmetry group} & 
 \\
$
\begin{cases}
	\Psi_{\text{EQ}}(X,\mathbf{F}\mathbf{K},\mathbf{G},\mathbf{g})
	=\Psi_{\text{EQ}}(X,\mathbf{F},\mathbf{G},\mathbf{g}) \\
	\Psi_{\text{NEQ}}(X,\Fe\mathbf{K},\mathbf{G},\mathbf{g}) =\Psi_{\text{NEQ}}(X,\Fe,\mathbf{G},\mathbf{g}) \\
	\phi(X,\mathbf{F}\mathbf{K},\mathbf{K}^*\Fv,\mathbf{K}^*\dot\Fv,\mathbf{G},\mathbf{g})
	=\phi(X,\mathbf{F},\Fv,\dot\Fv,\mathbf{G},\mathbf{g})
\end{cases}$ 
& $\forall\,\,\mathbf{K}\in \mathcal{G}_X\leqslant \mathrm{Orth}(\mathbf{G})$  \\

\\

 {\bf The Clausius-Duhem inequality} & \\
 
$\begin{aligned}
& \dot\eta = 2\Ce^\flat\frac{\partial \hat\Psi_{\text{NEQ}}}{\partial\Ce^\flat}\Fv^{-\star} \!:\! \dot\Fv \geq 0 && \text{(comp.)}\\
&\dot\eta = 2\Ce^\flat\frac{\partial \hat\Psi_{\text{NEQ}}}{\partial\Ce^\flat}\Fv^{-\star} : \dot\Fv +q \Fv^{-\star} \geq 0 && \text{(incom.)}
\end{aligned} $  \\
\\

 {\bf The Balance of linear momentum} & \\
 
$\begin{aligned}
 &\operatorname{Div}\left[2\mathbf F \frac{\partial \hat{\Psi}}{\partial \mathbf C^\flat} 
	+2 \Fe\frac{\partial \hat{\Psi}}{\partial \Ce^\flat}\right] 
	+ \rho_o \boldsymbol{\mathsf{B}} = \rho_o \mathbf A && \text{(comp.)} \\
	& \operatorname{Div}\left[2\mathbf F \frac{\partial \hat{\Psi}}{\partial \mathbf C^\flat} +2 \Fe\frac{\partial \hat{\Psi}}{\partial \Ce^\flat}-p \mathbf g^\sharp \mathbf F^{-\star}\right] + \rho_o \boldsymbol{\mathsf{B}} = \rho_o \mathbf A && \text{(incomp.)}
	\end{aligned}$  \\
\\

 {\bf Kinetic equation} & \\

$\begin{aligned}
\partial \phi/\partial \dot\Fv + \partial \tilde\Psi/\partial \Fv &= \mathbf 0 && \text{(comp.)}  \\
\partial \phi/\partial \dot\Fv + \partial \tilde\Psi/\partial \Fv &= q \Fv^{-\star} && \text{(incomp.)}
\end{aligned}$ \\

\\

\end{tabular} 
\end{center}
\caption{Summary of the main fields, constitutive equations, and governing equations of nonlinear viscoelasticity.}
\label{Table-Summary}
\end{table}
%-----------------------------
%-----------------------------
%-----------------------------

%-----------------------------
%-----------------------------
\section{Examples} \label{Sec:Examples}

In this section, we study three examples of large deformations of isotropic and anisotropic viscoelastic solids. To simplify the kinematics we assume incompressible solids. The deformations considered in this section are subsets of the known universal deformations for incompressible isotropic~\citep{Ericksen1955,Yavari2021a} and anisotropic solids~\citep{YavariGoriely2021,yavari2023universal}. Universal deformations are those deformations that can be maintained in the absence of body forces for any material in a given class. For homogeneous compressible isotropic solids,~\citet{Ericksen1955} showed that the only universal deformations are homogeneous deformations. For homogeneous incompressible isotropic solids, in addition to isochoric homogeneous deformations,~\citet{Ericksen1954} found four families of universal deformations. A fifth family was later on discovered independently by~\citet{SinghPipkin1965} and~\citet{KlingbeilShield1966}. For some recent generalizations of Ericksen's problem to inhomogeneous and anisotropic solids, and anelasticity see~\citet{Yavari2021a,YavariGoriely2021,yavari2023universal,YavariGoriely2016}, and \citet{Goodbrake2020}.\footnote{The analogue of universal deformations in linear elasticity are universal displacements~\citep{Truesdell1966, Gurtin1972, Yavari2020, yavari2023universal, Yavari2022Anelastic-Universality, Yavari2023Fibers}.} The stress at any material point in a \emph{simple} material at time $t$ depends only on the history of the deformation gradient at that point up to time $t$~\citep{noll1958}.~\citet{Carroll1967} showed that the known universal deformations of homogeneous incompressible isotropic elastic solids are universal for simple materials as well. One should note that (simple) viscoelastic solids are a subclass of simple materials. It should, however, be noted that~\citet{Carroll1967} assumed that the total deformation is volume preserving. Here, we assume that both the local elastic and viscoelastic deformations are volume preserving.

%-----------------------------
%-----------------------------
\subsection{Quadratic dissipation potentials}
\label{SubSec:Quadr-Diss-Pot}

\citet{Kumar2016} assumed the following form for the dissipation potential
%-----------------------------
\begin{equation}\label{eq:quad_diss}
	\phi(X,\mathbf{F},\Fv,\dot\Fv,\mathbf{G})
	=\frac{1}{2}\,\dot\Fv\!:\!\boldsymbol{\mathbb{A}}(\mathbf{F},\Fe,\mathbf{G},\mathbf{g})\!:\!\dot\Fv
	=\frac{1}{2}\,\dot\cFv^{K}{}_{L}~ \mathbb{A}_{K}{}^L{}_M{}^N ~\dot\cFv^{M}{}_{N}\,,
\end{equation}
%-----------------------------
where ${\mathbb{A}}(\mathbf{F},\Fe,\mathbf{G},\mathbf{g})$ is a positive-definite fourth-order tensor.\footnote{Notice that \eqref{eq:quad_diss}, along with the positive definiteness of ${\mathbb{A}}\,$, trivially satisfies the Clausius--Duhem inequality \eqref{eq:loc_Thermo_Second_CD} (and \eqref{eq:loc_Thermo_Second_CD_inc}  for the incompressible case).} It is clear that only the major symmetric part of ${\mathbb{A}}$ contributes to dissipation, and indeed, by definition \eqref{eq:quad_diss} above, ${\mathbb{A}}$ has major symmetries. However, ${\mathbb{A}}$ does not necessarily have any minor symmetries.\footnote{Notice that for the dissipation potential $\phi(X,\mathbf{F},\Fv,\dot\Fv,\mathbf{G})=\frac{1}{2}\,\dot\Cv : {{\mathbb{B}}}(\mathbf{F},\Fe,\mathbf{G},\mathbf{g}) : \dot\Cv\,$, which is a particular case of \eqref{eq:quad_diss} for ${{{\mathbb{A}_{K}}^L}_M}^N = 4 \mathrm{G}_{KJ} {\cFv^J}_I \mathbb{B}^{ILAN} {\cFv^B}_A \mathrm{G}_{BM}\,$, the minor symmetries hold for ${\mathbb{B}}\,$.}

Adding the set of structural tensors $\boldsymbol \Lambda$ to its arguments, the dissipation potential $\phi=\phi(X,\mathbf{F},\Fv,\dot\Fv,\boldsymbol \Lambda,\mathbf{G})$ becomes an isotropic functional, i.e., 
%-----------------------------
\begin{equation}
	\mathbf{K}_*\phi(X,\mathbf{F},\Fv,\dot\Fv,\boldsymbol\Lambda,\mathbf{G})
	=\phi(X,\mathbf{K}^*\mathbf{F},\mathbf{K}^*\Fv,\mathbf{K}^*\dot\Fv,
	\mathbf{K}^*\boldsymbol\Lambda,\mathbf{G})=\phi(X,\mathbf{F},\Fv,
	\dot\Fv,\boldsymbol\Lambda,\mathbf{G})\,,\qquad\forall\,\mathbf{K}\in \mathrm{Orth}(\mathbf{G})\,.
\end{equation}
%-----------------------------
It hence follows that
%-----------------------------
\begin{equation}\label{eq:quad_diss_symm}
	\mathbf{K}^*\dot\Fv\!:\!\boldsymbol{\mathbb{A}}(\mathbf{K}^*\mathbf{F},\mathbf{K}^*\Fe,
	\mathbf{K}^*\boldsymbol \Lambda,\mathbf{G},\mathbf{g})\!:\!\mathbf{K}^*\dot\Fv 
	= \dot\Fv\!:\!\boldsymbol{\mathbb{A}}(\mathbf{F},\Fe,\boldsymbol\Lambda,\mathbf{G},
	\mathbf{g})\!:\!\dot\Fv\,,\qquad\forall\,\mathbf{F},\Fe,\dot\Fv\,,
	\forall\,\mathbf{K}\in \mathrm{Orth}(\mathbf{G})\,,
\end{equation}
%-----------------------------
where we recall that $\mathbf{K}^*\dot\Fv=\mathbf{K}^{-1}\dot\Fv\mathbf{K}\,$, $\mathbf{K}^*\mathbf{F}=\mathbf{F}\mathbf{K}\,$, and $\mathbf{K}^*=\Fe \mathbf{K}\,$.
Therefore\footnote{Note that in components, the left hand side of \eqref{eq:quad_diss_iso} reads
$$
[\mathbf{K}_* \boldsymbol{\mathbb{A}}]_{A}{}^B{}_C{}^D= \mathrm{K}^{-I}{}_A \mathrm{K}^{B}{}_J \mathrm{K}^{-K}{}_C \mathrm{K}^{D}{}_L {\mathbb{A}}_{I}{}^J{}_K{}^L\,.$$
}
%-----------------------------
\begin{equation}\label{eq:quad_diss_iso}
	\mathbf{K}_* \boldsymbol{\mathbb{A}}(\mathbf{F},\Fe,\boldsymbol\Lambda,\mathbf{G},
	\mathbf{g}) = \boldsymbol{\mathbb{A}}(\mathbf{F},\Fe,\boldsymbol\Lambda,,\mathbf{G},
	\mathbf{g})\,,
	\qquad \forall\,\mathbf{K}\in \mathrm{Orth}(\mathbf{G})\,.
\end{equation}
%-----------------------------
Hence, $\boldsymbol{\mathbb{A}}=\boldsymbol{\mathbb{A}}(\mathbf{F},\Fe,\boldsymbol\Lambda,\mathbf{G},\mathbf{g})$ is an isotropic tensor. The most general isotropic fourth-order tensor has the following representation \citep{Jog2006}
%-----------------------------
\begin{equation}
	\mathbb{A}_{KBMD}=\eta_1\,G_{KB}G_{MD}+\eta_2\,G_{KD}G_{BM}+\eta_3\,G_{KM}G_{BD}
	\,,
\end{equation}
%-----------------------------
where $\eta_i=\eta_i(\mathbf{F},\Fe,\boldsymbol\Lambda,\mathbf{G},\mathbf{g})=\hat{\eta_i}(\mathbf{C}^{\flat},\Ce^{\flat},\boldsymbol\Lambda,\mathbf{G})$ for $i=1,2,3\,$.
Thus
%-----------------------------
\begin{equation} \label{Quadratic-Dissipation}
	\mathbb{A}_{K}{}^L{}_M{}^N=\mathbb{A}_{KBMD}\,G^{BL}G^{DN}
	=\eta_1\,\delta_K^L \delta_M^N
	+\eta_2\,\delta _K^N \delta_M^L
	+\eta_3\,G_{KM}G^{LN}
	\,.
\end{equation}
%-----------------------------
The dissipation potential is written as
%-----------------------------
\begin{equation}\label{eq:iso-dissip}
\begin{aligned}
	\phi &=\frac{1}{2}\eta_1\left(\dot\cFv^A{}_A\right)^2+\frac{1}{2}\eta_2\,\dot\cFv^A{}_B\,\dot\cFv^B{}_A
	+\frac{1}{2}\eta_3\,G_{AC} \dot\cFv^C{}_D G^{DB} \dot\cFv^A{}_B \\
	&= \frac{1}{2}\eta_1 \left(\operatorname{tr}\dot\Fv\right)^2 + \frac{1}{2}\eta_2 \operatorname{tr}\left(\dot\Fv^2\right) + \frac{1}{2}\eta_3 \operatorname{tr}\left(\dot\Fv \dot\Fv^{\mathsf T}\right)\,.
\end{aligned}
\end{equation}
%-----------------------------
In order to find the necessary and sufficient conditions on $\eta_1, \eta_2\,$, and $\eta_3$ to ensure positive-definiteness of the tensor ${\mathbb{A}}\,$, we introduce new indices $\Gamma=\{AB\}$ such that $\{11,12,13,21,22,23,31,32,33\}\leftrightarrow \{1,2,3,4,5,6,7,8,9\}\,$. 
Now, in a Cartesian coordinate system, the dissipation potential is rewritten as $\phi=\frac{1}{2}\mathbb{C}^{\Gamma\Lambda}\,\dot{\mathsf{F}}_{\Gamma}\dot{\mathsf{F}}_{\Lambda}\,$. The tensor ${\mathbb{A}}$ is positive-definite if and only if the {$9\times9$} matrix ${\mathbb{C}}$ is positive-definite. It may be found that the eigenvalues of the matrix ${\mathbb{C}}$ are $ 3 \eta_1+\eta_2+\eta_3\,$, $\eta_2+\eta_3\,$, and $-\eta_2+\eta_3\,$.
Therefore, ${\mathbb{A}}$ is positive-definite if and only if\footnote{It is seen that $\eta_2=\eta_3$ is not acceptable, and hence, ${\mathbb{A}}$ cannot have minor symmetries.}
%---------------------
\begin{equation} 
	3 \eta_1+\eta_2+\eta_3>0\,,\quad \eta_2+\eta_3>0\,,\quad -\eta_2+\eta_3>0	\,.
\end{equation}
%---------------------

Choosing \eqref{Quadratic-Dissipation}, we have
%---------------------
\begin{equation}
	\frac{\partial \phi}{\partial \dot\cFv^A{}_B}= \eta_1\,\cFv^M{}_M\,\delta^B_A + \eta_2\,\dot\cFv^B{}_A
	+\eta_3\,G_{AM}\,\dot\cFv^M{}_N\,G^{NB}
	\,.
\end{equation}
%---------------------
Or
%---------------------
\begin{equation} \label{phi-isotropic}
	\frac{\partial \phi}{\partial \dot\Fv}
	=\eta_1(\operatorname{tr}\dot\Fv)\,\mathbf{I} 
	+ \eta_2\,\dot\Fv^{\star}
	+\eta_3\,\mathbf{G}\dot\Fv\mathbf{G}^{\sharp}
	\,,
\end{equation}
%---------------------
where $\operatorname{tr}\dot\Fv=\dot\cFv^C{}_C\,$. With this choice, the kinetic equation \eqref{eq:Kinetic-Equation-Isotropic_inc} is simplified to read 
%---------------------
\begin{equation}\label{eq:Quadr-Diss_Kin-Eq}
	\eta_1(\operatorname{tr}\dot\Fv)\,\mathbf{I} +\eta_2\,\dot\Fv^{\star}
	+\eta_3\,\mathbf{G}\dot\Fv\mathbf{G}^{\sharp}
	-2\widetilde{\Psi}_1\Ce^{\flat}\mathbf{G}^{\sharp}\Fv^{-\star}
	+2\widetilde{\Psi}_2\left[\Ce^{2\flat} -\Ie_1\Ce^{\flat} \right]\mathbf{G}^{\sharp}\Fv^{-\star}
	 = q \Fv^{-\star}\,.
\end{equation}
%---------------------

%-----------------------------
%-----------------------------
\subsection{Example 1: Finite extension of an incompressible isotropic circular cylindrical bar}

\paragraph{Kinematics.}
Let us consider a solid circular cylindrical bar subject to an axial loading. In its undeformed configuration, it has radius $R_{0}$ and length $L_{0}\,$. We consider longitudinal and radial deformations and assume the following deformation ansatz:

%-----------------------------
\begin{equation} \label{Deformation}
   r=r(R,t)\,,\qquad \theta=\Theta\,,\qquad z=\lambda(t) Z\,,
\end{equation}
%-----------------------------
where $\lambda(t)$ is the axial stretch.
In a displacement-control loading, the longitudinal stretch $\lambda(t)$ is given, while in a force-controlled loading, it is an unknown function to be determined. Acting on an initially stress-free unloaded bar, i.e.,~$\lambda(0)=1\,$, it is assumed that loading (either force or displacement-control) is slow enough such that the inertial effects can be ignored.
The deformation gradient reads
%-----------------------------
\begin{equation}
   \mathbf{F}=\mathbf{F}(R,t)=\begin{bmatrix}
  r_{,R}(R,t) & 0  & 0  \\
  0 & 1  & 0  \\
  0 & 0  & \lambda(t)
\end{bmatrix}\,.
\end{equation}
%-----------------------------
Incompressibility $J=1$ implies that $r(R,t)=\frac{R}{\sqrt{\lambda(t)}}\,$.
In terms of its physical components, the deformation gradient reads
%-----------------------------
\begin{equation}
\hat{\mathbf F}=\hat{\mathbf F}(R,t)=\begin{bmatrix}
  \frac{1}{\sqrt{\lambda(t)}} & 0  & 0  \\
  0 & \frac{1}{\sqrt{\lambda(t)}}  & 0  \\
  0 & 0  & \lambda(t)
\end{bmatrix}\,.
\end{equation}
%-----------------------------
We use a semi-inverse method and assume that the viscous deformation gradient has the following form
%-----------------------------
\begin{equation} \label{Fv-Finite-Extension}
   \Fv=\Fv(R,t)=\begin{bmatrix}
  a_v(R,t) & 0  & 0  \\
  0 & b_v(R,t)  & 0  \\
  0 & 0  & \lambda_v(R,t)
\end{bmatrix}\,.
\end{equation}
%-----------------------------
At the initial unloaded state, we have $\Fv(R,0)=\mathbf I\,$, i.e.,~$a_v(R,0)=b_v(R,0)=\lambda_v(R,0)=1\,$.
Incompressibility of the local viscous deformation implies that $\Jv(R,t)=a_v(R,t) \,b_v(R,t)\, \lambda_v(R,t)=1\,$,  and hence $b_v(R,t)=\frac{1}{a_v(R,t)\lambda_v(R,t)}\,$.
The physical components of the viscous deformation gradient read
%-----------------------------
\begin{equation}
	\hat a_v(R,t)=a_v(R,t)\,,\qquad
	\hat b_v(R,t)=b_v(R,t)=\frac{1}{a_v(R,t) \lambda_v(R,t)}\,, \qquad
	\hat\lambda_v(R,t)=\lambda_v(R,t)
	\,.
\end{equation}
%-----------------------------

Since $\mathbf{F}=\Fe\Fv\,$, it follows that the elastic deformation gradient has the following form
%-----------------------------
\begin{equation}
   \Fe=\Fe(R,t)=\begin{bmatrix}
  a_e(R,t) & 0  & 0  \\
  0 & b_e(R,t)   & 0  \\
  0 & 0  & \lambda_e(R,t)
\end{bmatrix}\,,
\end{equation}
%-----------------------------
where
%-----------------------------
\begin{equation}
	a_e(R,t)=\frac{1}{\lambda^{\frac{1}{2}}(t)a_v(R,t)}\,,\qquad
	b_e(R,t)=\frac{1}{b_v(R,t)}=a_v(R,t)\lambda_v(R,t)\,,\qquad
	\lambda_e(R,t)=\frac{\lambda(t)}{\lambda_v(R,t)}\,.
\end{equation}
%-----------------------------
The physical components read
%-----------------------------
\begin{equation}
	\hat a_e(R,t) = \frac{1}{\lambda^{\frac{1}{2}}(t)a_v(R,t)}\,,\qquad
	\hat b_e(R,t) = \frac{a_v(R,t)\lambda_v(R,t)}{\lambda^{\frac{1}{2}}(t)}\,, \qquad
	\hat \lambda_e(R,t) = \frac{\lambda(t)}{\lambda_v(R,t)}\,.
\end{equation}
%-----------------------------

%-----------------------------
\begin{remark} 
It should be noted that $\mathbf{F}$ is homogeneous. However, {$\Fv$}, and consequently {$\Fe$}, are not compatible, in general. Recall that incompatibility of {$\Fv$} is controlled by exterior derivative of {$\Fv$} (or its curl), i.e.,~$d\Fv\,$,  which has components ${\cFv^A}_{B,C}-{\cFv^A}_{C,B}$~\citep{Yavari2013}. Note that
%-----------------------------
\begin{equation}
	\cFv^2{}_{2,1}-\cFv^2{}_{1,2}=\frac{\partial b_v(R,t)}{\partial R}\,,\qquad
	\cFv^3{}_{3,1}-\cFv^3{}_{1,3}=\frac{\partial \lambda_v(R,t)}{\partial R}	\,.
\end{equation}
%-----------------------------
This means that $\Fv$ is compatible if and only if (a solid bar is simply-connected)
%-----------------------------
\begin{equation}
	\frac{\partial b_v(R,t)}{\partial R}=\frac{\partial \lambda_v(R,t)}{\partial R}=0	\,.
\end{equation}
%-----------------------------
\end{remark}
%-----------------------------

\paragraph{Kinetic equations.}
We assume an isotropic quadratic dissipation potential \eqref{eq:iso-dissip}. 
We obtain three independent kinetic equations for ${a}_v(R,t)\,$, ${\lambda}_v(R,t)\,$,  and $q(R,t)$\textemdash the Lagrange multiplier corresponding to viscous incompressibility. We then proceed to eliminate $q(R,t)$ from the system of kinetic equations and are left with the following two independent kinetic equations for ${a}_v(R,t)$ and ${\lambda}_v(R,t)$:
%-----------------------------

\begin{subequations}\label{eq:Ex1_GEq}
\begin{empheq}[left={\empheqlbrace\,}]{align}
	\begin{split}
	&\lambda \lambda_v \left[\eta_1\left(a_v^2 \lambda_v-1\right)^2+\left(\eta_2+\eta_3\right)
	\left(a_v^4 \lambda_v^2+1\right)\right]\dot{a}_v   \\
	&\quad  + \lambda a_v \left[\eta_1 \left(a_v^2 \lambda_v-1\right) 
	\left(a_v  \lambda_v^2-1\right)+\eta_2+\eta _3\right]\dot{\lambda}_v
	=2 a_v \lambda_v \left(1-a_v^4 \lambda_v^2\right) 
	\left[\lambda_v^2\widetilde{\Psi}_1+\lambda^2 \widetilde{\Psi}_2\right]\,,
	\end{split}\\
	\begin{split}
	&\lambda _v \left\{\left(\eta_1+\eta_2+\eta_3\right) a_v^3 \lambda_v-\eta_1\left(a_v^2 \lambda_v^2 + a_v - \lambda_v\right)\right\}\dot{a}_v  \\
	&\quad
	+a_v \left\{\eta_1 \left(a_v^2 \lambda_v^2 + \lambda_v -a_v \right)- \left(\eta_1+\eta_2+\eta_3\right) a_v \lambda_v^3 \right\}\dot{\lambda}_v    
	=-{2 \left(\lambda^3 a_v^2-\lambda_v^2\right) 
	\left[\lambda \widetilde{\Psi}_1+ a_v^2 \lambda_v^2\widetilde{\Psi}_2 \right]}/{\lambda^2}\,.
	\end{split}
\end{empheq}
\end{subequations}
%-----------------------------
Recall that following viscous incompressibility, i.e.,~$\Jv=1\,$,  we have $b_v=\frac{1}{a_v \lambda_v}\,$. We may then recast the kinetic equations~\eqref{eq:Ex1_GEq} above in terms of $b_v$ and $\lambda_v\,$. It is hence found that the kinetic equations in terms of $\left\{b_v,\lambda_v\right\}$ are identical to \eqref{eq:Ex1_GEq} in terms of $\left\{a_v,\lambda_v\right\}\,$,  such that the evolutions of $a_v$ and $b_v$ are both governed by the same equations; and since they are subject to the same initial condition $a_v(R,0)=b_v(R,0)=1\,$,  it follows that $b_v(R,t)=a_v(R,t)\,$,  and hence that $a_v=\lambda_v^{-\frac{1}{2}}\,$.
Therefore, \eqref{eq:Ex1_GEq} may be reduced to a single differential equation in terms of $\lambda_v$ as follows
%-----------------------------
\begin{equation}
\left[ 2\eta_1+\eta_2+\eta_3 - 4\eta_1 \lambda_v^{3/2} + 2\left(\eta_1+\eta_2+\eta_3\right) \lambda_v^3\right] \lambda^2 \dot{\lambda}_v
= 4 \left(\lambda^3-\lambda_v^3\right)\left[ \widetilde{\Psi}_1 \lambda + \widetilde{\Psi}_2 \lambda_v \right]\,.
\end{equation}
%-----------------------------

\paragraph{Stress and equilibrium equations.}
Following \eqref{eq:sigma_iso}, the non-zero components of the Cauchy stress are
%-----------------------------
\begin{subequations}
\begin{align}
	\sigma^{rr}(R,t) &=-p(R,t)
	+\frac{2\overline{\Psi}_1}{\lambda^(t)}
	-2\overline{\Psi}_2\,\lambda(t)+\frac{2\widetilde{\Psi}_1\,\lambda_v(R,t)}{\lambda(t)}
	-\frac{2\widetilde{\Psi}_2\,\lambda(t)}{\lambda_v(R,t)}\,,\\ 
	\sigma^{\theta\theta}(R,t) & = -\frac{p(R,t)\lambda(t)}{R^2}
	+\frac{2\overline{\Psi}_1}{R^2}
	-\frac{2\overline{\Psi}_2\,\lambda^2(t)}{R^2}
	+\frac{2\widetilde{\Psi}_1\,\lambda_v(R,t)}{R^2}
	-\frac{2\widetilde{\Psi}_2\,\lambda^2(t)}{R^2 \,\lambda_v(R,t)}
	 \,,\\
	\sigma^{zz}(R,t) & =-p(R,t)
	+2\overline{\Psi}_1\,\lambda^2(t)-\frac{2\overline{\Psi}_2}{\lambda^2(t)}
	+\frac{2\widetilde{\Psi}_1\,\lambda^2(t)}{\lambda_v^2(R,t)}
	-\frac{2\widetilde{\Psi}_2\,\lambda_v^2(R,t)}{\lambda^2(t)}
	\,,
\end{align}
\end{subequations}
%-----------------------------
where $p=p(R,t)$ is the Lagrange multiplier corresponding to incompressibility, i.e., $J=1\,$.
The only nontrivial equilibrium equation is $\sigma^{rr}{}_{,r}+\frac{1}{r}\sigma^{rr}-r\sigma^{\theta\theta}=0\,$. In terms of the referential coordinates, this reads 
%-----------------------------
\begin{equation}
	\frac{\partial}{\partial R}\sigma^{rr}(R,t) = 0\,.
\end{equation}
%-----------------------------
Since the lateral area of the cylinder is traction-free, it follows that $\sigma^{rr}(R_0,t)=0\,$. Therefore, $\sigma^{rr}(R,t) = 0$ and
%-----------------------------
\begin{equation}
	p(R,t) = \frac{2\overline{\Psi}_1}{\lambda^{2}(t)}-2\overline{\Psi}_2\,\lambda(t)
	+\frac{2\widetilde{\Psi}_1}{\lambda(t)\,a_v^2(R,t)}-2\widetilde{\Psi}_2\,\lambda(t)\,a_v^2(R,t)
	\,.
\end{equation}
%-----------------------------
Hence, it follows that $\sigma^{\theta\theta}(R,t)=0\,$, and the only non-zero stress component is $\sigma^{zz}\,$, which reads \footnote{Note that the longitudinal physical component of stress is given by $\hat{\sigma}^{zz}(R,t) = \sigma^{zz}(R,t)\,$.
} 
%-----------------------------
\begin{equation}
	\begin{split}
	\hat\sigma^{zz}(R,t) &= 
	2\left[\lambda^2(t)-\frac{1}{\lambda(t)}\right]\overline{\Psi}_1
	+2\left[\lambda(t)-\frac{1}{\lambda^2(t)}\right]\overline{\Psi}_2
	\\&\quad+2\left[\frac{\lambda^2(t)}{\lambda_v^2(R,t)}
	-\frac{\lambda_v(R,t)}{\lambda(t)}\right]\widetilde{\Psi}_1
	+2\left[\frac{\lambda(t)}{\lambda_v(R,t)}
	-\frac{\lambda_v^2(R,t)}{\lambda^2(t)}\right]\widetilde{\Psi}_2
	\,.
	 \end{split}
\end{equation}
%-----------------------------

The force at the two ends of the bar ($Z=0, L$) is written as
%-----------------------------
\begin{equation}\label{eq:AxialForce}
	F(t)=2\pi \int_{0}^{r(R_0)}\sigma^{zz}(R,t)r\,dr 
	= 2\pi \int_{0}^{R_0}\hat\sigma^{zz}(R,t)\lambda^{-1}(t)R\,dR\,,
\end{equation}
%-----------------------------
which is expanded to read
%-----------------------------
\begin{equation} \label{Ex1-AxialForce}
\begin{aligned}
	 & 2\left[\lambda(t)-\frac{1}{\lambda^2(t)}\right] \int_{0}^{R_0}R\,\overline{\Psi}_1\,dR 
	+2\left[1-\frac{1}{\lambda^3(t)}\right]\int_{0}^{R_0}R\,\overline{\Psi}_2\,dR \\
	& +2\int_{0}^{R_0}R
	\left[\frac{\lambda(t)}{\lambda_v^2(R,t)}-\frac{\lambda_v(R,t)}{\lambda^2(t)}\right]\widetilde{\Psi}_1\,dR
	+2\int_{0}^{R_0}R\left[\frac{1}{\lambda_v(R,t)}-\frac{\lambda_v^2(R,t)}{\lambda^3(t)}\right]\widetilde{\Psi}_2\,dR
	=\frac{F(t)}{2\pi}
	\,.
\end{aligned}
\end{equation}
%-----------------------------

%-----------------------------
\begin{example}
In this example, we assume a neo-Hookean viscoelastic solid 
%---------------------
\begin{equation} \label{Viscoealstic-neo-Hookean}
	\Psi_{\text{EQ}} = \overline{\Psi}(I_1,I_2)=\frac{\mu}{2}(I_1-3)   \,,\qquad
	\Psi_{\text{NEQ}}  = \widetilde{\Psi}(\Ie_1,\Ie_2)=\frac{\mu_e}{2}(\Ie_1-3)  \,.
\end{equation}
%---------------------
Thus, $\overline{\Psi}_1=\frac{1}{2}\mu\,$, $\widetilde{\Psi}_1=\frac{1}{2}\mu_e\,$, and $\overline{\Psi}_2=\widetilde{\Psi}_2=0\,$.
The kinetic equation is then simplified to read
%-----------------------------
\begin{equation}\label{Ex1_ODEs}
	\left[ 2\eta_1+\eta_2+\eta_3 - 4\eta_1 \lambda_v^{\frac{3}{2}} 
	+ 2\left(\eta_1+\eta_2+\eta_3\right) \lambda_v^3\right] \lambda\, \dot{\lambda}_v
	= 2 \mu_e \left(\lambda^3-\lambda_v^3\right) \,.
\end{equation}
%-----------------------------
It is seen that the ODE governing the time evolution of $\lambda_v(R,t)$ does not depend on $R$\,; and since the initial condition $\lambda_v(R,0) = 1$ does not depend on $R$ either, it follows that $\lambda_v$ does not depend on $R\,$,  i.e.,~$\lambda_v=\lambda_v(t)\,$.
By inspection of the dimensional quantities involved in the problem at hand, one may identify $\tau=\frac{\eta_1}{\mu}$ as a viscoelastic dissipation characteristic time of \eqref{Ex1_ODEs} above and the resulting time evolution of the viscous deformation gradient $\Fv(t)\,$.
In this case, the only non-zero physical stress component is independent of $R$ and reads
%-----------------------------
\begin{equation}
	\hat\sigma^{zz}(t)
	=  \mu\left(\lambda^2-\frac{1}{\lambda}\right) + \mu_e \left( \frac{\lambda^2}{\lambda_v^2} - \frac{\lambda_v}{\lambda} \right)
	\,.
\end{equation}
%-----------------------------

\paragraph{Displacement-control loading.}
We start with a displacement-control loading. Let us assume a loading such that the longitudinal stretch has the following form
%-----------------------------
\begin{equation}\label{disp_load}
\lambda(t)=
\begin{dcases}
1+(\lambda_0-1)\erf\left(\frac{t}{t_0}\right)\,,\quad & 0\leq t\leq t_1\,,\\
\lambda(t_1)+\frac{1-\lambda(t_1)}{2}\erf\left(\frac{t-t_1}{t_0}\right)\,,\quad & t_1 \leq t\leq t_2\,,\\
\lambda(t_2)+(1-\lambda(t_2))\erf\left(\frac{t-t_2}{t_0}\right)\,,\quad & t_2 \leq t\leq t_f\,,
\end{dcases}
\end{equation}
%-----------------------------
where $\erf$ is the error function, $t_{0}$ is the loading characteristic time, $t_1=25t_{0}\,$,  $t_2=50t_{0}\,$,  $t_f=75t_{0}\,$,  and $\lambda_{0}$ is the stretch at large times ${t_{0}\ll t<t_1}\,$. For a displacement-controlled loading, the following initial-value problem needs to be solved for $\lambda_v$\,:
%-----------------------------
\begin{equation}\label{Ex1_D_ODEs}
	\left[ 2\eta_1+\eta_2+\eta_3 - 4\eta_1 \lambda_v^{\frac{3}{2}} 
	+ 2\left(\eta_1+\eta_2+\eta_3\right) \lambda_v^3\right] \lambda\, \dot{\lambda}_v
	= 2 \mu_e \left(\lambda^3-\lambda_v^3\right)\,,\quad\lambda_v(0)=1\,.
\end{equation}
%-----------------------------

\paragraph{Force-control loading.}
For a force-control loading, the force required to maintain the deformation is exactly $F(t)$ as in \eqref{eq:AxialForce}. In the case of a neo-Hookean solid, \eqref{Ex1-AxialForce} is simplified to read
%-----------------------------
\begin{equation} 
	\frac{\mu  R_0^2}{2}\left[\lambda(t)-\frac{1}{\lambda^{2}(t)}\right]
	+ \frac{\mu_e R_0^2}{2}\left[\frac{\lambda(t)}{\lambda_v^2(t)}-\frac{\lambda_v(t)}{\lambda^2(t)}\right]
	= \frac{F(t)}{2 \pi }
	\,.
\end{equation}
%-----------------------------
Here, we assume the following loading
%-----------------------------
\begin{equation}\label{force_load}
F(t)=
\begin{dcases}
F_0\erf\left(\frac{t}{t_0}\right)\,,\quad & 0\leq t\leq t_1\,,\\
F_0+\frac{F(t_1)}{2}\erf\left(\frac{t-t_1}{t_0}\right)\,,\quad & t_1 \leq t\leq t_2\,,\\
\frac{F_0}{2}-F(t_2)\erf\left(\frac{t-t_2}{t_0}\right)\,,\quad & t_2 \leq t\leq t_f\,,
\end{dcases}
\end{equation}
%-----------------------------
where $t_{0}$ is the loading characteristic time, $t_1=25t_{0}\,$,  $t_2=50t_{0}\,$,  $t_f=75t_{0}\,$,  and $F_{0}$ is the force at large times ${t_{0}\ll t<t_1}\,$.
For a force-control loading, we solve the following initial-value problem for $\lambda_v$ and $\lambda$:
%-----------------------------
\begin{subequations}\label{eq:Force_Contr_t}
\begin{empheq}[left={\empheqlbrace\,}]{align}
	&\frac{\mu  R_0^2}{2}\left[\lambda(t)-\lambda^{-2}(t)\right]
	+ \frac{\mu_e R_0^2}{2}\left[\frac{\lambda(t)}{\lambda_v^2(R,t)}-\frac{\lambda_v(R,t)}{\lambda^2(t)}\right]
	= \frac{F(t)}{2 \pi }\,,
	\\
	&\left[ 2\eta_1+\eta_2+\eta_3 - 4\eta_1 \lambda_v^{3/2} + 2\left(\eta_1+\eta_2+\eta_3\right) \lambda_v^3\right] \lambda\, \dot{\lambda}_v
= 2 \mu_e \left(\lambda^3-\lambda_v^3\right)\,,
	\\
	&\lambda_v(0)=1\,.
\end{empheq}
\end{subequations}
%-----------------------------

%-----------------------------
\begin{figure}
\centering
\includegraphics[width=1\textwidth]{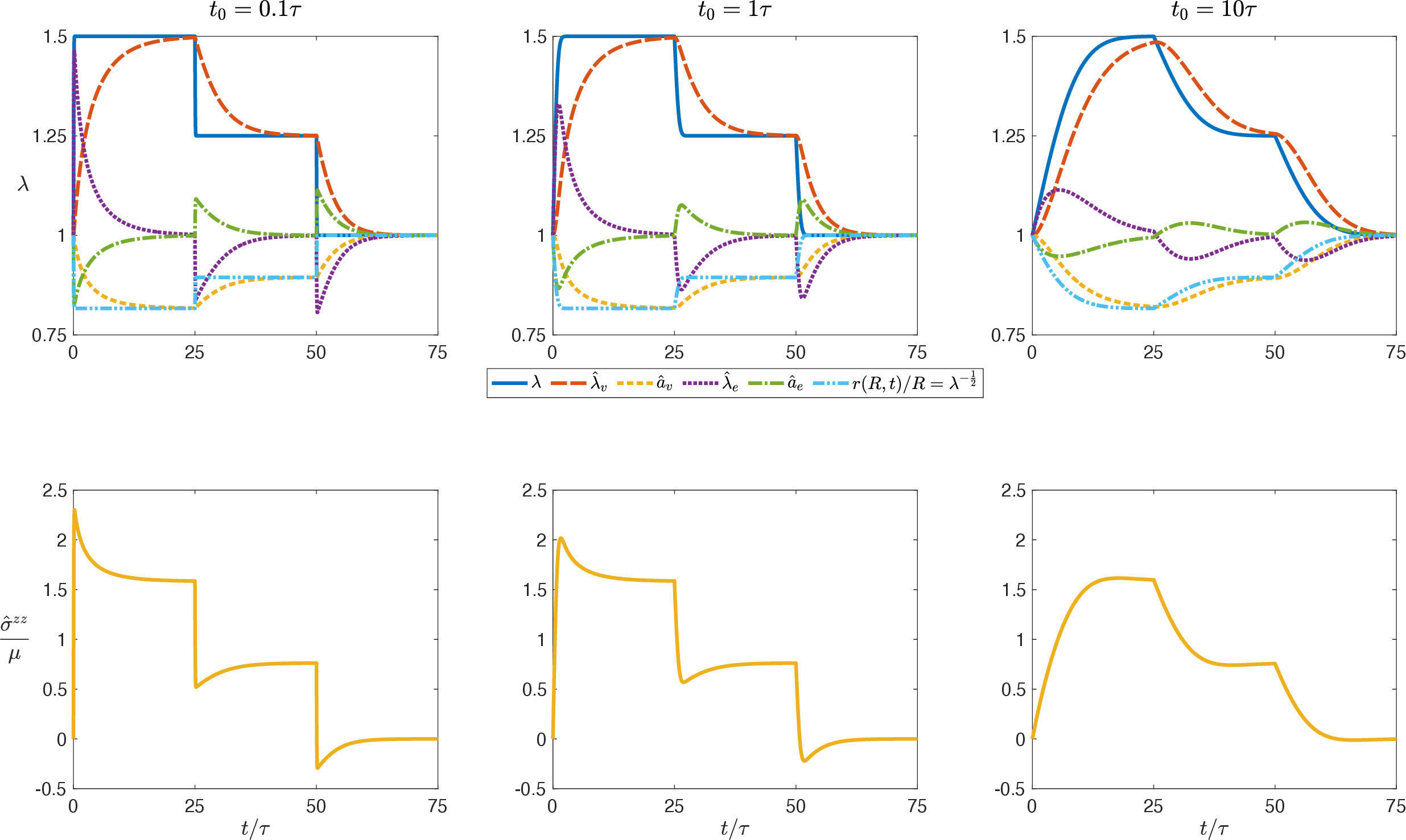}
\vspace*{-0.10in}
\caption{Numerical results for the time evolution of the strain and stress state of an isotropic neo-Hookean viscoelastic bar subject to displacement-control loading $\lambda(t)$ \eqref{disp_load} with different characteristic times $t_{0}$ versus the viscoelastic dissipation characteristic time $\tau=\frac{\eta_1}{\mu}$ of the system.}
\label{Fig:Ex1_D}
\end{figure}
%-----------------------------

%-----------------------------
\begin{figure}
\centering
\includegraphics[width=1\textwidth]{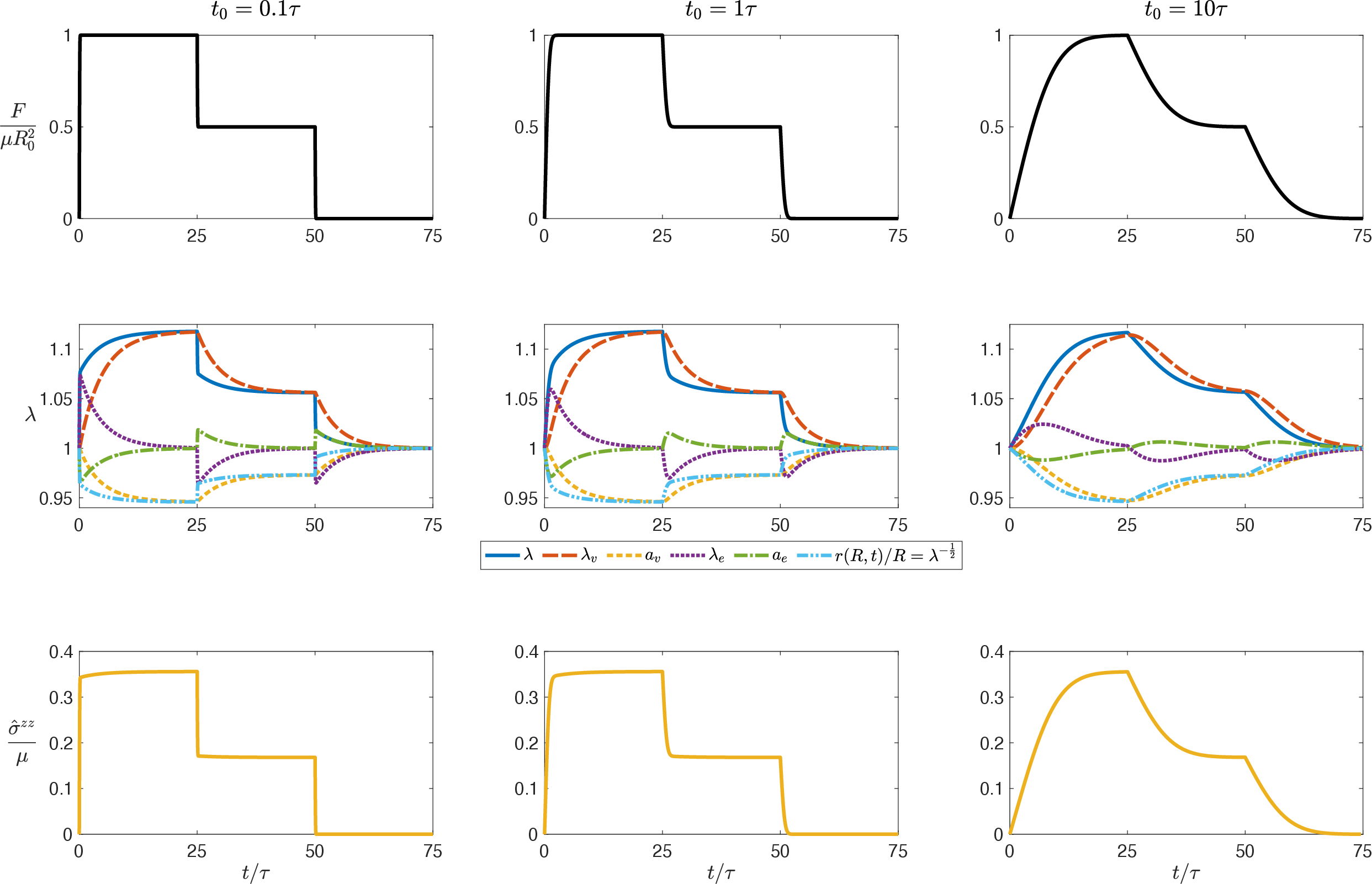}
\vspace*{-0.10in}
\caption{Numerical results for the time evolution of the strain and stress state of an isotropic neo-Hookean viscoelastic bar subject to force-control loading $F(t)$ \eqref{force_load} with different characteristic times $t_{0}$ versus the viscoelastic dissipation characteristic time $\tau=\frac{\eta_1}{\mu}$ of the system.}
\label{Fig:Ex1_F}
\end{figure}
%-----------------------------

\paragraph{Numerical results.}
Let us consider a solid cylinder made of an isotropic neo-Hookean viscoelastic solid with $\mu_e=\frac{1}{2}\mu$ and a dissipation potential of the form \eqref{Quadratic-Dissipation} such that $\eta_1=\eta_2=\frac{1}{2}\eta_3\,$. In this example, an explicit finite difference scheme has been used to numerically solve the governing equations.

Let us first look at the displacement-control loading case as we subject the structure to the longitudinal loading $\lambda(t)$ \eqref{disp_load} with $\lambda_{0}=1.5\,$. We numerically solve the displacement-control governing Eq.~\eqref{Ex1_D_ODEs} assuming different loading characteristic times $t_{0}\,$,  respectively smaller, equal, and larger than the characteristic time of the viscoelastic bar. 
In Fig.~\ref{Fig:Ex1_D}, we plot the profile of the given displacement-control loading $\lambda(t)$ and the resulting evolution of the kinematic quantities $\lambda_v\,$,  $a_v\,$,  $\lambda_e\,$,  $a_e\,$,  and the outer radius $r$ at $R=R_{0}$ as well as the longitudinal physical stress component.
As the cylindrical bar is subjected to a longitudinal stretch load in three stages\textemdash loading followed by partial unloading and then full unloading\textemdash we observe in each of the stages that the cylinder experiences stress relaxation: The bar first experiences an instantaneous fast elastic stress response followed by a slow stress relaxation (decrease in the loading stage and increase in the unloading stages) under a constant imposed displacement. Note, however, that this is not observed for $t_{0}=10\tau\,$,  since the loading does not reach a steady state of constant $\lambda\,$.

Next, we look at the force-control loading case as we subject the structure to the force loading $F(t)$ \eqref{force_load} with $F_{0}=\mu R_{0}^2\,$. We numerically solve the force-control governing Eq.~\eqref{eq:Force_Contr_t} assuming different loading characteristic times $t_{0}\,$,  respectively smaller, equal, and larger than the characteristic time of the viscoelastic bar. 
In  Fig.~\ref{Fig:Ex1_F}, we plot the profile of the given force-control loading $F(t)$ and the resulting evolution of the kinematic quantities $\lambda_v\,$,  $a_v\,$,  $\lambda_e\,$,  $a_e\,$,  and the outer radius $r$ at $R=R_{0}\,$, as well as the longitudinal physical stress component.
As the bar is subjected to an axial force in three stages\textemdash loading followed by partial unloading and then full unloading\textemdash we observe in each of the stages that the bar experiences creep: The bar first experiences a fast elastic deformation but continues to slowly deform even as the force reaches a steady state.

In both cases, we observe that the cross section shrinks as the bar is loaded and expands back again as the bar is unloaded.
Interestingly, and in both loading cases, the elastic deformation gradient features a behaviour akin to a strain relaxation as its physical components experience a fast elastic response followed by a slower relaxation under constant loading towards its initial value at the unloaded state. However, the viscous deformation gradient experiences what is akin to creep as it undergoes a fast response followed by a slow evolution towards matching the total deformation gradient $\hat{\mathbf F}$ at large times under constant loading\textemdash $\hat\lambda_v$ approaches $\lambda$ and $\hat a_v=\hat b_v$ approaches ${{\hat F}^r}_R={\hat F^\theta}_\Theta=\lambda^{-\frac{1}{2}}\,$. That is, at large times under constant loading in each of the three loading stages, we have $\widehat{\Fe}=\mathbf I$ and $\widehat{\Fv}=\mathbf F$ as previously discussed in Remark~\ref{rmrk:F_LargeTimes}.
\end{example}
%-----------------------------

%-----------------------------
%-----------------------------
\subsection{Example 2: Finite torsion of an incompressible transversely isotropic circular cylindrical bar}

\paragraph{Kinematics.}
Let us consider a solid circular cylindrical bar that, in its undeformed configuration, has radius $R_{0}$ and length $L_{0}\,$. 
In this example, we assume that the cylindrical bar is transversely isotropic with helical material preferred directions. One can think of this bar as being a homogeneous isotropic solid cylinder reinforced by helical fibers. More precisely, for fixed $R\in(0,R_{0}]\,$, it is assumed that fibers are along a family of helices tangent to the fiber direction $\mathbf{N}=\mathbf{N}(R,\Theta)\,$. 
Recall that in cylindrical coordinates, the tangent to a helix  has a vanishing radial coordinate.
Let us denote by $\gamma(R)$ the angle that $\mathbf{N}(R,\Theta)$ makes with $\mathbf{E}_{\Theta}(\Theta)=\frac{\partial}{\partial \Theta}\,$.
Thus, $\mathbf{N}(R,\Theta)={\cos\gamma(R)}/{R}\,\mathbf{E}_{\Theta}(\Theta)+\sin\gamma(R)\,\mathbf{E}_Z\,$,  where $\mathbf{E}_Z=\frac{\partial}{\partial Z}\,$.\footnote{Let us recall that $\mathbf{N}$ is a $\mathbf{G}$-unit vector, i.e.,~$N^A N^B G_{AB}=R^2(N^{\Theta})^2+(N^Z)^2=1$ and this is why the $\frac{1}{R}$ factor shows up in the $\Theta$-component.}

Given the helical symmetry of the problem, we assume the following deformation ansatz:
%-----------------------------
\begin{equation} \label{Deformation}
   r=r(R,t)\,,\qquad \theta=\Theta+\psi(t) Z\,,\qquad z=\lambda(t) Z\,,
\end{equation}
%-----------------------------
where $\psi(t)$ is the twist per unit length, and $\lambda(t)$ is the axial stretch. 
In a twist-force-control loading, the twist $\psi(t)$ and force $F(t)$ are prescribed, while $\lambda(t)$ is an unknown. In a torque-force-control loading, the torque $T(t)$ and force $F(t)$ are prescribed, and both $\psi(t)$ and $\lambda(t)$ are unknown functions.\footnote{The other two possible loadings are twist-displacement and torque-displacement-control loadings that we will not consider in our numerical examples.} The deformation gradient reads
%-----------------------------
\begin{equation}
   \mathbf{F}=\mathbf{F}(R,t)=\begin{bmatrix}
  r'(R,t) & 0  & 0  \\
  0 & 1  & \psi(t)  \\
  0 & 0  & \lambda(t)
\end{bmatrix}\,,
\end{equation}
%-----------------------------
where $r'(R,t)=\frac{\partial r(R,t)}{\partial R}\,$. The incompressibility implies that
%-----------------------------
\begin{equation}
	J=\sqrt{\frac{\det\mathbf{g}}{\det\mathbf{G}}}\det\mathbf{F}=\frac{\lambda(t)\,r(R,t)\,r'(R,t)}{R}=1\,.
\end{equation}
%-----------------------------
Assuming that $r(0,t)=0\,$, one obtains $r(R,t)=\frac{R}{\sqrt{\lambda(t)}}\,$.
In terms of its physical components, the deformation gradient reads
%-----------------------------
\begin{equation}
\hat{\mathbf F}=\hat{\mathbf F}(R,t)=\begin{bmatrix}
  \frac{1}{\sqrt{\lambda(t)}} & 0  & 0  \\
  0 & \frac{1}{\sqrt{\lambda(t)}}  & \frac{R \psi(t)}{\sqrt{\lambda(t)}}  \\
  0 & 0  & \lambda(t)
\end{bmatrix}\,.
\end{equation}
%-----------------------------
We again use a semi-inverse method and assume that the viscous deformation gradient has the following form
%-----------------------------
\begin{equation}
   \Fv=\Fv(R,t)=\begin{bmatrix}
  a_v(R,t) & 0  & 0  \\
  0 & b_v(R,t)  &  \psi_v(R,t)  \\
  0 & 0  & \lambda_v(R,t)
\end{bmatrix}\,.
\end{equation}
%-----------------------------
Incompressibility of the local viscous deformation implies that $a_v(R,t)\,b_v(R,t)\lambda_v(R,t)=1\,$.
The physical components of the viscous deformation gradient read
%-----------------------------
\begin{equation}
	\hat a_v(R,t)=a_v(R,t)\,,\quad
	\hat b_v(R,t)=b_v(R,t)=\frac{1}{a_v(R,t) \lambda_v(R,t)}\,, \quad
	\hat\lambda_v(R,t)=\lambda_v(R,t)\,, \quad
	\hat\psi_v(R,t)=\psi_v(R,t)
	\,.
\end{equation}
%-----------------------------
For a torque-force-control loading, the unknown fields of the problem are $\lambda(t)\,$, $\psi(t)\,$,  $a_v(R,t)\,$, $\psi_v(R,t)\,$, and $\lambda_v(R,t)\,$,  while for a twist-force-control loading, $\psi(t)$ is prescribed and the unknown fields are $\lambda(t)\,$,  $a_v(R,t)\,$,  $\psi_v(R,t)\,$, and $\lambda_v(R,t)\,$.
In this problem, the elastic deformation gradient has the following form
%-----------------------------
\begin{equation}
   \Fe=\Fe(R,t)=\begin{bmatrix}
  a_e(R,t) & 0  & 0  \\
  0 & b_e(R,t)  &  \psi_e(R,t)  \\
  0 & 0  & \lambda_e(R,t)
\end{bmatrix}\,.
\end{equation}
%-----------------------------
Knowing that $\mathbf{F}=\Fe\Fv$ implies that 
%-----------------------------
\begin{equation}
   \Fe=\Fe(R,t)=\begin{bmatrix}
	 \frac{1}{\sqrt{\lambda(t)}\, a_v(R,t)} & 0 & 0 \\
 	0 & a_v(R,t)\lambda_v(R,t) & \frac{\psi(t)}{\lambda_v(R,t)} - \psi_v(R,t) a_v(R,t) \\
 	0 & 0 & \frac{\lambda(t)}{\lambda_v(R,t)}
\end{bmatrix}\,.
\end{equation}
%-----------------------------
The physical components read
%-----------------------------
\begin{equation}
\begin{aligned}
	\hat a_e(R,t) &= \frac{1}{\lambda^{\frac{1}{2}}(t)a_v(R,t)}\,,&& \quad
	\hat b_e(R,t) = \frac{a_v(R,t)\lambda_v(R,t)}{\lambda^{\frac{1}{2}}(t)}\,, \\
	\hat \lambda_e(R,t) & = \frac{\lambda(t)}{\lambda_v(R,t)}\,, && \quad
	\hat \psi_e(R,t) = \frac{R\psi(t)}{\lambda^{\frac{1}{2}}(t)\lambda_v(R,t)}-\frac{R\psi_v(R,t)a_v(R,t)}{\lambda^{\frac{1}{2}}(t)} \,.
\end{aligned}
\end{equation}
%-----------------------------

%-----------------------------
\begin{remark}
Notice that $\mathbf{F}$ is homogeneous. However, {$\Fv$}, and consequently {$\Fe$}, are not compatible, in general, because
%-----------------------------
\begin{equation}
	\cFv^2{}_{2,1}-\cFv^2{}_{1,2}=\frac{\partial b_v(R,t)}{\partial R}\,,\qquad
	\cFv^2{}_{3,1}-\cFv^2{}_{1,3}=\frac{\partial \psi_v(R,t)}{\partial R}\,,\qquad
	\cFv^3{}_{3,1}-\cFv^3{}_{1,3}=\frac{\partial \lambda_v(R,t)}{\partial R}	\,.
\end{equation}
%-----------------------------
This implies that $\Fv$ is compatible if and only if 
%-----------------------------
\begin{equation}
	\frac{\partial b_v(R,t)}{\partial R}=\frac{\partial \psi_v(R,t)}{\partial R}
	=\frac{\partial \lambda_v(R,t)}{\partial R}=0	\,.
\end{equation}
%-----------------------------
\end{remark}
%-----------------------------

\paragraph{Stress and equilibrium equations.}
The principal invariants read
%-----------------------------
\begin{equation}
\begin{aligned}
	I_1& =\lambda^2(t)+\frac{2}{\lambda(t)}+\frac{R^2 \psi^2(t)}{\lambda(t)}\,, \qquad
	I_2=2\lambda(t)+\frac{1}{\lambda^2(t)}+\frac{R^2 \psi^2(t)}{\lambda^2(t)} \\
	I_4 &= \frac{\cos^2\gamma+\sin^2\gamma \left(\lambda^3+R^2 \psi^2\right)
	+R \psi \sin2 \gamma}{\lambda} \\
	I_5 &= \lambda^4 \sin^2\gamma+\frac{(\cos\gamma+R\, \psi \sin\gamma) 
	\left[R\, \psi \sin\gamma \left(2\lambda^3+R^2 \psi^2+1\right)
	+\cos\gamma \left(R^2\psi^2+1\right)\right]}{\lambda^2}
	\,.
\end{aligned}
\end{equation}
%-----------------------------
The non-zero physical components of the Cauchy stress are $\hat{\sigma}^{rr}=\sigma^{rr}\,$, $\hat{\sigma}^{\theta\theta}=r^2\sigma^{\theta\theta}\,$, $\hat{\sigma}^{zz}=\sigma^{zz}\,$, and $\hat{\sigma}^{\theta z}=r\sigma^{\theta z}\,$. The diagonal components read
%-----------------------------
\begin{equation}
\begin{aligned}
	\hat{\sigma}^{rr}(R,t) &=-p+\frac{2\overline{\Psi}_1}{\lambda}
	-2\lambda\,\overline{\Psi}_2+\frac{2\widetilde{\Psi}_1}{\lambda\,a_v^2}
	-2\widetilde{\Psi}_2\,\lambda\,a_v^2\,,
\end{aligned}
\end{equation}
%-----------------------------
%-----------------------------
\begin{equation}
\begin{aligned}
	\hat{\sigma}^{\theta\theta}(R,t) & = -p+2\frac{1+R^2 \psi^2}{\lambda} \overline{\Psi}_1
	-2\lambda\,\overline{\Psi}_2+\frac{2(\cos\gamma+R \,\psi  
	\sin\gamma)^2}{\lambda}\overline{\Psi}_4 \\
	& \quad +\frac{4(\cos\gamma+R \psi \sin\gamma) \left[R \psi
	\left(\lambda^3+R^2 \psi^2+1\right)\sin\gamma+ \left(R^2 \psi^2+1\right)\cos\gamma \right]}
	{\lambda^2}\overline{\Psi}_5
	 \,,\\
	 & \quad +\frac{2 \left[a_v \lambda_v \left(a_v \left(R^2 \lambda_v \psi_v^2+\lambda_v^3\right)
	 -2 R^2 \psi \psi_v\right)+R^2 \psi^2\right]}{\lambda \lambda_v^2}\widetilde{\Psi}_1 \\
	 &\quad -\frac{2 \lambda }{a_v^2 \lambda_v^2}\widetilde{\Psi}_2
	 +\frac{2\left[a_v \lambda_v \left(\lambda_v \cos\gamma -R \psi_v \sin\gamma \right)
	 +R \psi \sin\gamma\right]^2}	{\lambda\lambda_v^2}\widetilde{\Psi}_4 \\
	 & \quad +\frac{4}{\lambda^2 \lambda_v^4}
	 \left(a_v \lambda_v \left(\cos\gamma \lambda_v-R \sin\gamma \psi_v\right)
	 +R\psi \sin\gamma \right) \\
	 & \qquad \times \Big\{(R \sin\gamma \left(\psi -a_v \lambda_v \psi_v\right)   
	 \Big[a_v \lambda_v \left(a_v \left(R^2 \lambda_v \psi_v^2+\lambda_v^3\right)
	 -2 R^2 \psi \psi_v\right)
	 +\lambda^3+R^2 \psi^2\Big] \\
   	& \qquad\qquad +a_v \cos\gamma \lambda_v^2 \left(a_v \lambda_v  
	\left(a_v \left(R^2 \lambda_v \psi_v^2
	+\lambda_v^3\right)-2 R^2 \psi \psi_v\right)+R^2 \psi^2\right)\Big\}\widetilde{\Psi}_5 \,,
\end{aligned}
\end{equation}
%-----------------------------
and
%-----------------------------
\begin{equation}
\begin{aligned}
	\hat{\sigma}^{zz}(R,t) & =-p+2 \lambda^2 \,\overline{\Psi}_1
	-\frac{2\left(R^2 \psi^2+1\right)}{\lambda^2}\overline{\Psi}_2
	+2 \lambda ^2 \sin^2\gamma\,\overline{\Psi}_4\\
	&\quad +4 \lambda \sin\gamma \left[\lambda^3 \sin\gamma
	+R \psi(\cos\gamma+R\,\psi \sin\gamma)\right]
	\overline{\Psi}_5
	+\frac{2 \lambda^2}{\lambda_v^2}\widetilde{\Psi}_1 \\
	& \quad -\frac{2\left(a_v \lambda_v \left(a_v \left(R^2 \lambda_v \psi_v^2+\lambda_v^3\right)
	-2 R^2 \psi \psi_v\right)+R^2 \psi^2\right)}{\lambda^2 a_v^2 \lambda_v^2}\widetilde{\Psi}_2
	+\frac{2 \lambda^2 \sin^2\gamma}{\lambda_v^2}\widetilde{\Psi}_4	\\
	& \quad  +\frac{4 \lambda \sin\gamma}{\lambda_v^4} 
	\left\{ R a_v \lambda_v \left[R \sin\gamma \,\psi_v \left(a_v \lambda_v \psi_v-2 \psi \right)
	+\cos\gamma \,\lambda_v \left(\psi-a_v \lambda_v \psi_v\right)\right]+\sin\gamma 
	\left(\lambda^3+R^2 \psi^2\right) \right\} \widetilde{\Psi}_5
	\,.
\end{aligned}
\end{equation}
%-----------------------------
The only non-zero shear stress is
%-----------------------------
\begin{equation}
\begin{aligned}
	\hat{\sigma}^{\theta z}(R,t) & =2 R \sqrt{\lambda} \psi\,\overline{\Psi}_1
	+\frac{2R \psi}{\sqrt{\lambda}}\,\overline{\Psi}_2
	+2 \sqrt{\lambda} \sin\gamma\, (\cos\gamma+R \psi \sin\gamma)\,\overline{\Psi}_4 \\
	& \quad +\frac{\sin2\gamma \left(\lambda^3+3 R^2 \psi^2+1\right)-2 R \psi\cos 2\gamma 
	\left(\lambda^3+R^2\psi^2\right)+2R \psi \left(\lambda^3+R^2 \psi^2+1\right)}
	{\sqrt{\lambda}}\overline{\Psi}_5 \\
	& \quad +\frac{2R \sqrt{\lambda} \left(\psi -a_v \lambda_v \psi_v\right)}{\lambda_v^2}\widetilde{\Psi}_1
	+\frac{2R \left(\psi -a_v \lambda_v \psi_v\right)}{\sqrt{\lambda}\, a_v^2 \lambda_v^2}\widetilde{\Psi}_2
	+\frac{2 \sqrt{\lambda}\, \sin\gamma \left(a_v \lambda_v \left(\cos\gamma\, \lambda_v-R \sin\gamma 
	\,\psi _v\right) +R \psi \sin\gamma\right)}{\lambda_v^2}\widetilde{\Psi}_4 \\
	& \quad +\frac{1}{\sqrt{\lambda}\, \lambda _v^4} 
	\Big\{ 2 a_v \sin\gamma\, \lambda_v \left(\lambda^3+3 R^2 \psi^2\right) 
	\left(\cos\gamma\, \lambda_v-2 R \sin\gamma\, \psi _v\right) \\
	& \qquad\qquad +2 R \psi  a_v^2 \lambda _v^2 
	\left(6R^2 \sin^2\gamma\, \psi_v^2-3 R \sin 2\gamma\, \lambda_v \psi_v+\lambda_v^2\right) \\
	& \qquad\qquad 
	+a_v^3 \lambda_v^3 \left(-4 R^3 \sin^2\gamma\, \psi_v^3+3R^2 \sin 2\gamma\, \lambda_v \psi _v^2
	-2 R \lambda_v^2 \psi_v+\sin 2\gamma\, \lambda_v^3\right) \\
	& \qquad\qquad +4 R \psi \sin^2\gamma\, \left(\lambda^3+R^2 \psi^2\right) \Big\}  \widetilde{\Psi}_5
	\,.
\end{aligned}
\end{equation}
%-----------------------------

The only nontrivial equilibrium equation is $\sigma^{rr}{}_{,r}+\frac{1}{r}\sigma^{rr}-r\sigma^{\theta\theta}=0\,$. In terms of the referential coordinates, this reads 
%-----------------------------
\begin{equation}
	\frac{\partial}{\partial R}\sigma^{rr}(R,t) = f(R,t)	\,,
\end{equation}
%-----------------------------
where
%-----------------------------
\begin{equation}
\begin{aligned}
	f(R,t) & = \frac{2R\,\psi^2}{\lambda}\overline{\Psi}_1
	+\frac{2 (\cos\gamma+R \psi \sin\gamma)^2}{\lambda  R}\,\overline{\Psi}_4 \\
	& \quad +\frac{4(\cos\gamma+R \psi \sin\gamma) \left[R\,\psi \sin\gamma   
	\left(\lambda^3+R^2 \psi^2+1\right)
	+\cos\gamma \left(R^2 \psi^2+1\right)\right]}{\lambda^2R}\,\overline{\Psi}_5 \\
	& \quad +2\frac{a_v^2 \left[a_v \lambda_v \left(a_v \left(R^2 \lambda_v \psi_v^2
	+\lambda_v^3\right)
	-2 R^2 \psi \psi_v\right)+R^2 \psi^2\right]-\lambda_v^2}{\lambda  R a_v^2 \lambda_v^2}
	\, \widetilde{\Psi}_1 \\
	& \quad +\frac{2 \lambda \left(a_v^4 \lambda_v^2-1\right)}{R a_v^2 \lambda_v^2}\, 
	\widetilde{\Psi}_2
	+\frac{2 \left(a_v \lambda_v \left(\cos\gamma \lambda_v-R \sin\gamma \psi_v\right)
	+R \psi \sin\gamma\right)^2}{\lambda  R \lambda_v^2}\, \widetilde{\Psi}_4 \\
	& \quad \frac{4 \left[a_v \lambda_v \left(\cos\gamma \lambda _v-R \sin\gamma \psi_v\right)
	+R \psi  \sin\gamma \right] }{\lambda^2 R \lambda_v^4}
	\Bigg\{R \sin\gamma \left(\psi -a_v\lambda_v \psi_v\right)\\ 
	& \qquad \times\Big[a_v \lambda_v \left(a_v \left(R^2 \lambda_v \psi_v^2+\lambda_v^3\right)
	-2 R^2 \psi \psi_v\right) 
	+\lambda^3+R^2 \psi^2\Big]\\
	&\qquad +a_v\cos\gamma \lambda_v^2 
	\left(a_v \lambda_v \left(a_v \left(R^2 \lambda_v \psi_v^2+\lambda_v^3\right)
	-2 R^2 \psi  \psi_v\right)+R^2 \,\psi^2\right)\Bigg\}\, \widetilde{\Psi}_5
		\,.
\end{aligned}
\end{equation}
%-----------------------------
Assuming that the boundary cylinder is traction free, i.e., $\sigma^{rr}(R_0,t)=0\,$, one obtains
%-----------------------------
\begin{equation}
	\sigma^{rr}(R,t) = -\int_R^{R_0} f(\xi,t)\,d\xi	\,.
\end{equation}
%-----------------------------
This, in particular, implies that
%-----------------------------
\begin{equation}
	-p = -\int_R^{R_0} f(\xi,t)\,d\xi -\frac{2\overline{\Psi}_1}{\lambda}
	+2\lambda\,\overline{\Psi}_2-\frac{2\widetilde{\Psi}_1}{\lambda\,a_v^2}
	+2\widetilde{\Psi}_2\,\lambda\,a_v^2	\,.
\end{equation}
%-----------------------------
The normal stress components $\hat{\sigma}^{\theta\theta}$ and $\hat{\sigma}^{zz}$ are now simplified to read
%-----------------------------
\begin{equation}
\begin{aligned}
	\hat{\sigma}^{\theta\theta}(R,t) & = -\int_R^{R_0} f(\xi,t)\,d\xi 
	+\frac{2R^2 \psi^2}{\lambda} \overline{\Psi}_1	
	+\frac{2(\cos\gamma+R \,\psi  \sin\gamma)^2}{\lambda}\overline{\Psi}_4 \\
	& \quad +\frac{4(\cos\gamma+R \psi \sin\gamma) \left[R \psi\sin\gamma
	\left(\lambda^3+R^2 \psi^2+1\right)+\cos\gamma \left(R^2 \psi^2+1\right)\right]}
	{\lambda^2}\overline{\Psi}_5 \\
	 & \quad +\frac{2 \left(R^2 a_v^4 \lambda_v^2 \psi_v^2-2 R^2 \psi  a_v^3 \lambda_v \psi_v
	 +R^2 \psi^2 a_v^2+a_v^4 \lambda_v^4-\lambda_v^2\right)}
	 {\lambda  a_v^2 \lambda_v^2}\,\widetilde{\Psi}_1
	  \\
	 &\quad   +\frac{2\lambda \left(a_v^4 \lambda_v^2-1\right)}{a_v^2 \lambda _v^2}  \widetilde{\Psi}_2
	 +\frac{2\left[a_v \lambda_v \left(\cos\gamma \lambda_v-R \sin\gamma \psi_v\right)
	 +R \psi \sin\gamma\right]^2}	{\lambda\lambda_v^2}\widetilde{\Psi}_4 \\
	 & \quad +\frac{4}{\lambda^2 \lambda_v^4}
	 \left(a_v \lambda_v \left(\cos\gamma \lambda_v-R \sin\gamma \psi_v\right)
	 +R\psi \sin\gamma \right) \\
	 & \qquad \times \Big\{(R \sin\gamma \left(\psi -a_v \lambda_v \psi_v\right)   
	 \Big[a_v \lambda_v \left(a_v \left(R^2 \lambda_v \psi_v^2+\lambda_v^3\right)
	 -2 R^2 \psi \psi_v\right)
	 +\lambda^3+R^2 \psi^2\Big] \\
   	& \qquad\qquad +a_v \cos\gamma \lambda_v^2 \left(a_v \lambda_v  
	\left(a_v \left(R^2 \lambda_v \psi_v^2
	+\lambda_v^3\right)-2 R^2 \psi \psi_v\right)+R^2 \psi^2\right)\Big\}\widetilde{\Psi}_5 
	\,,
\end{aligned}
\end{equation}
%-----------------------------
and
%-----------------------------
\begin{equation}
\begin{aligned}
	\hat{\sigma}^{zz}(R,t) & =  -\int_R^{R_0} f(\xi,t)\,d\xi 
	+2\left(\lambda^2-\frac{1}{\lambda} \right)\overline{\Psi}_1
	+\frac{2}{\lambda^2}\left(\lambda^3-R^2 \psi^2-1\right)\,\overline{\Psi}_2  \\
	& \quad	+2 \lambda ^2 \sin^2\gamma\,\overline{\Psi}_4 
	+4 \lambda \sin\gamma \left[\lambda^3 \sin\gamma+R \psi(\cos\gamma+R\,\psi \sin\gamma)\right]
	\overline{\Psi}_5
	+2\frac{\lambda^3a_v^2-\lambda_v^2}{\lambda\,\lambda_v^2\,a_v}\,\widetilde{\Psi}_1  \\
	& \quad +2\frac{-R^2 a_v^2 \lambda_v^2 \psi_v^2+2 R^2 \psi  a_v \lambda_v \psi_v
	+\lambda^3 a_v^4 \lambda_v^2-a_v^2 \lambda_v^4-R^2 \psi^2}{\lambda^2 a_v^2 \lambda_v^2}
	\,\widetilde{\Psi}_2
	+\frac{2 \lambda^2 \sin^2\gamma}{\lambda_v^2}\widetilde{\Psi}_4	\\
	& \quad  +\frac{4 \lambda \sin\gamma}{\lambda_v^4} 
	\left[ R a_v \lambda_v \left(R \sin\gamma \,\psi_v \left(a_v \lambda_v \psi_v-2 \psi \right)
	+\cos\gamma \,\lambda_v \left(\psi-a_v \lambda_v \psi_v\right)\right)+\sin\gamma 
	\left(\lambda^3+R^2 \psi^2\right) \right]\widetilde{\Psi}_5
	\,.
\end{aligned}
\end{equation}
%-----------------------------

For a force-control loading at the two ends of the bar ($Z=0,L$), the axial force and torque required to maintain the deformation are calculated as
%-----------------------------
\begin{equation}
\begin{aligned}
	F(t) & =2\pi \int_{0}^{R_0}P^{zZ}(R,t)R\,dR=0\,,\\
	T(t) &=2\pi \int_{0}^{R_0}\bar{P}^{\theta Z}(R,t)R^2\,dR
	=2\pi \int_{0}^{R_0} P^{\theta Z}(R,t)\,r(R,t)R^2\,dR\,,
\end{aligned}
\end{equation}
%-----------------------------
where $\bar{P}^{zZ}=P^{zZ}$ is the $zZ$-component of the first Piola-Kirchhoff stress and $\bar{P}^{\theta Z}=rP^{\theta Z}$ is the physical $\theta Z$ component of the first Piola-Kirchhoff stress. Recall the relation $\mathbf{P}=J\boldsymbol{\sigma}\mathbf{F}^{-\star}\,$,  or in components $P^{aA}=J\sigma^{ab}F^{-A}\mbox{}_b\,$. Thus, $P^{zZ}=\lambda^{-1}\sigma^{zz}$ and $P^{\theta Z}=\lambda^{-1}\sigma^{\theta z}\,$,  and hence
%-----------------------------
\begin{equation}
\begin{aligned}
	\bar{P}^{zZ}(R,t) &= -\frac{1}{\lambda} \int_R^{R_0} f(\xi,t)\,d\xi 
	+2\left(\lambda-\frac{1}{\lambda^2} \right)\overline{\Psi}_1
	+\frac{2}{\lambda^3}\left(\lambda^3-R^2 \psi^2-1\right)\,\overline{\Psi}_2  \\
	& \quad	+2 \lambda \sin^2\gamma\,\overline{\Psi}_4 
	+4 \sin\gamma \left[\lambda^3 \sin\gamma+R \psi(\cos\gamma+R\,\psi \sin\gamma)\right]
	\overline{\Psi}_5
	+2\frac{\lambda^3a_v^2-\lambda_v^2}{\lambda^2\,\lambda_v^2\,a_v^2}\,\widetilde{\Psi}_1  \\
	& \quad +2\frac{-R^2 a_v^2 \lambda_v^2 \psi_v^2+2 R^2 \psi  a_v \lambda_v \psi_v
	+\lambda^3 a_v^4 \lambda_v^2-a_v^2 \lambda_v^4-R^2 \psi^2}{\lambda^3 a_v^2 \lambda_v^2}
	\,\widetilde{\Psi}_2
	+\frac{2 \lambda \sin^2\gamma}{\lambda_v^2}\widetilde{\Psi}_4	\\
	& \quad  +\frac{4 \sin\gamma}{\lambda_v^4} 
	\Big\{ R a_v \lambda_v \left[R \sin\gamma \,\psi_v \left(a_v \lambda_v \psi_v-2 \psi \right)
	+\cos\gamma \,\lambda_v \left(\psi-a_v \lambda_v \psi_v\right)\right]+\sin\gamma 
	\left(\lambda^3+R^2 \psi^2\right) \Big\} \widetilde{\Psi}_5\,, 
\end{aligned}
\end{equation}
%-----------------------------
and 
%-----------------------------
\begin{equation}
\begin{aligned}
	\bar{P}^{\theta Z}(R,t) &= \frac{2R\, \psi}{\lambda^{\frac{1}{2}}} \overline{\Psi}_1	
	+\frac{2R\, \psi}{\lambda^{\frac{3}{2}}} \overline{\Psi}_2	
	+\frac{2 \sin\gamma (\cos \gamma+R \psi  \sin\gamma)}{\lambda^{\frac{1}{2}}} \overline{\Psi}_4	\\
	& +\frac{\sin 2 \gamma \left(\lambda^3+3 R^2 \psi ^2+1\right)-2 R\, \psi \cos 2\gamma
  	\left(\lambda^3+R^2\,\psi^2\right)+2 R\,\psi \left(\lambda^3+R^2 \psi^2+1\right)}
	{\lambda^{\frac{3}{2}}}\overline{\Psi}_5 \\
	& +\frac{2 R \left(\psi -a_v \lambda_v \psi_v\right)}{\lambda^{\frac{1}{2}} \lambda_v^2}
	\,\widetilde{\Psi}_1
	+\frac{2 R \left(\psi -a_v \lambda_v \psi_v\right)}{\lambda^{\frac{3}{2}} a_v^2 \lambda_v^2}
	\,\widetilde{\Psi}_2 \\
	& +\frac{2 \sin\gamma \left[a_v \lambda _v \left( \lambda _v \cos\gamma-R  \psi_v \sin\gamma \right)
	 +R \psi \sin\gamma\right]}{\lambda^{\frac{1}{2}} \lambda_v^2}\,\widetilde{\Psi}_4 \\
	 & +\frac{1}{\lambda^{\frac{3}{2}} \lambda_v^4}
	\Big\{ 2 a_v  \lambda_v \left(\lambda^3+3 R^2 \psi^2\right) 
	\left( \lambda_v\cos\gamma-2 R  \,\psi_v \sin\gamma \right)\sin\gamma \\
	& \qquad\qquad +2 R \psi  a_v^2 \lambda_v^2 
	\left(6R^2 \psi_v^2 \sin^2\gamma-3 R \lambda_v \psi_v \sin2\gamma+\lambda_v^2\right) \\
	& \qquad\qquad +a_v^3 \lambda_v^3 \left(-4 R^3 \psi_v^3 \sin^2\gamma
	+3 R^2 \lambda_v \psi_v^2 \sin2\gamma
	-2 R \lambda_v^2 \psi_v+ \lambda_v^3 \sin2\gamma \right) \\
	& \qquad\qquad +4 R \psi \left(\lambda^3+R^2 \psi^2\right) \sin^2\gamma \Big\}\, \widetilde{\Psi}_5
	 \,.
\end{aligned}
\end{equation}
%-----------------------------

\paragraph{Kinetic equations.}
The three kinetic equations for $\dot{a}_v\,$, $\dot{\lambda}_v\,$, and $\dot{\psi}_v$ are written as
%---------------------
\begin{equation}\label{eq:Ex2_KinEq}
\begin{dcases}
	 \frac{\partial \phi}{\partial \dot{a}_v}+ \frac{\partial \Ie_1}{\partial a_v}\,\widetilde{\Psi}_1
	+ \frac{\partial \Ie_2}{\partial a_v}\,\widetilde{\Psi}_2 + \frac{\partial \Ie_4}{\partial a_v}\,\widetilde{\Psi}_4
	+ \frac{\partial \Ie_5}{\partial a_v}\,\widetilde{\Psi}_5=0\,, \\
	 \frac{\partial \phi}{\partial \dot{\lambda}_v}+ \frac{\partial \Ie_1}{\partial \lambda_v}\,\widetilde{\Psi}_1
	+ \frac{\partial \Ie_2}{\partial \lambda_v}\,\widetilde{\Psi}_2 
	+ \frac{\partial \Ie_4}{\partial \lambda_v}\,\widetilde{\Psi}_4
	+ \frac{\partial \Ie_5}{\partial \lambda_v}\,\widetilde{\Psi}_5=0\,, \\
	 \frac{\partial \phi}{\partial \dot{\psi}_v}+ \frac{\partial \Ie_1}{\partial \psi_v}\,\widetilde{\Psi}_1
	+ \frac{\partial \Ie_2}{\partial \psi_v}\,\widetilde{\Psi}_2 
	+ \frac{\partial \Ie_4}{\partial \psi_v}\,\widetilde{\Psi}_4
	+ \frac{\partial \Ie_5}{\partial \psi_v}\,\widetilde{\Psi}_5=0\,.
\end{dcases}
\end{equation}
%---------------------

%------------------------------	
\begin{example}
For the numerical examples, we consider the following incompressible neo-Hookean reinforced model: 
%------------------------------	
\begin{equation} \label{neo-Hookean-reinforced}
	\Psi_{\text{EQ}} = \overline{\Psi}(I_1,I_4)=\frac{\mu}{2}(I_1-3)
	+\frac{\mu_1}{2}\left(I_4-1\right)^2   \,,\qquad
	\Psi_{\text{NEQ}}  = \widetilde{\Psi}(\Ie_1,\Ie_4)=\frac{\mu_{e}}{2}(\Ie_1-3)
	+\frac{\mu_{e1}}{2}\big(\Ie_4-1\big)^2 
	\,,
\end{equation}	
%------------------------------
where $\mu_{1}$ and $\mu_{e1}$ are positive constants.
Thus, $\overline{\Psi}_1=\frac{\mu}{2}\,$,  $\overline{\Psi}_2=0\,$,  $\overline{\Psi}_4=\mu_{1}(I_4-1)\,$,  $\overline{\Psi}_5=0\,$,  $ \widetilde{\Psi}_1=\frac{\mu_e}{2}\,$,  $ \widetilde{\Psi}_2=0\,$,  $ \widetilde{\Psi}_4=\mu_{e1}(\Ie_4-1)\,$, and $ \widetilde{\Psi}_5=0$\,.
For this material
%-----------------------------
\begin{equation}
\begin{aligned}
	f(R,t) & = \frac{R\,\psi^2}{\lambda}\,\mu
	+\frac{2 (\cos\gamma+R \psi \sin\gamma)^2}{\lambda  R}\,\mu_{1}(I_4-1) \\
	& \quad +\frac{ a_v^2 \left(a_v \lambda_v \left(a_v \left(R^2 \lambda_v \psi_v^2
	+\lambda_v^3\right)
	-2 R^2 \psi \psi_v\right)+R^2 \psi^2\right)-\lambda_v^2 }{\lambda  R a_v^2 \lambda_v^2}
	\, \mu_e \\
	& \quad 
	+\frac{2 \left(a_v \lambda_v \left(\cos\gamma \lambda_v-R \sin\gamma \psi_v\right)
	+R \psi \sin\gamma\right)^2}{\lambda  R \lambda_v^2}\,\mu_{e1}(\Ie_4-1)   \,.
\end{aligned}
\end{equation}
%-----------------------------
Thus, the axial force and torque are written as 
%-----------------------------
\begin{subequations}\label{eq:Ex2_IntEq}
\begin{align}
	\begin{split}\label{eq:Ex2_IntEq1}
	\frac{F(t)}{2\pi} & = -\frac{1}{\lambda}\int_{0}^{R_0} R \int_R^{R_0} f(\xi,t)\,d\xi \,dR
	+\left(\lambda-\frac{1}{\lambda^2} \right)\frac{\mu}{2}R_0^2
	+2\mu_{1}\lambda  \int_0^{R_0} R (I_4-1) \sin^2\gamma\,dR \\
	& \quad 
	+ \mu_e\int_0^{R_0} \frac{\lambda^3a_v^2-\lambda_v^2}{\lambda^2\,\lambda_v^2\,a_v^2} R\,dR
	+2\mu_{e1}\,\lambda \int_0^{R_0} (\Ie_4-1)\frac{\sin^2\gamma}{\lambda_v^2}R\,dR \,,
	\end{split}
	\\%
	\begin{split}\label{eq:Ex2_IntEq2}
	\frac{T(t)}{2\pi} &=  \frac{R_0^4\, \psi}{4\lambda^{\frac{1}{2}}} \mu
	+2\mu_1 \int_{0}^{R_0} (I_4-1)\,\frac{\sin\gamma (\cos \gamma+R\psi \sin\gamma)}
	{\lambda^{\frac{1}{2}}} R^2\,dR
	+\mu_e \int_{0}^{R_0} \frac{R \left(\psi -a_v \lambda_v \psi_v\right)}
	{\lambda^{\frac{1}{2}} \lambda_v^2}  R^2\,dR \\
	& \quad +2\mu_{e1} \int_{0}^{R_0}(\Ie_4-1)\frac{\sin\gamma 
	\left[a_v \lambda _v \left(\cos\gamma \lambda _v-R \sin\gamma \psi_v\right)
	 +R \psi \sin\gamma\right]}{\lambda^{\frac{1}{2}} \lambda_v^2} R^2\,dR\,.
	\end{split}
\end{align}
\end{subequations}
%-----------------------------

For the neo-Hookean material \eqref{neo-Hookean-reinforced}, the kinetic equations \eqref{eq:Ex2_KinEq} simplify to read
%-----------------------------
\begin{equation}\label{eq:Ex2_nH_KinEq1}
\begin{aligned}
	& \frac{a_v^4 \lambda ^2 \lambda_v^3\left(\eta _2+\eta _3\right)
	+a_v^2 \lambda ^2 \lambda_v^2 \left(a_v^2   \lambda_v-1\right)\eta _1
	- \lambda ^2 \lambda_v \left(a_v^2  \lambda_v-1\right)\eta _1
	+ \lambda ^2 \lambda_v \left(\eta _2+\eta _3\right)}{a_v^4 \lambda ^2 \lambda_v^3}\,\dot{a}_v \\
	& \quad +\frac{a_v^2 \lambda ^2 \lambda_v^2 \left(a_v^2 \lambda_v-1\right) \eta _1
	-a_v \lambda ^2 \left(a_v^2 \lambda_v-1\right)\eta _1+a_v \lambda^2 \left(\eta _2+\eta _3\right) }
	{a_v^4 \lambda^2 \lambda_v^3}\,\dot{\lambda}_v \\
	& \quad +\frac{a_v^4 \lambda_v^3+\lambda_v \left(a_v^4 R^2  \psi_v^2-1\right)
	-a_v^3 R^2 \psi \psi_v}{a_v^3 \lambda  \lambda_v}\,\mu_e \\
	& \quad +\frac{2 (\lambda_v \cos\gamma-R \psi_v \sin\gamma)
	\left[a_v \lambda_v^2 \cos\gamma+R (\psi -a_v \lambda_v \psi_v)\sin\gamma \right]}
	{\lambda^2  \lambda_v^3}  \\
	&  \qquad \times \left[a_v^2 \lambda_v^4\cos^2\gamma+
	\left(R^2 (\psi -a_v \lambda_v \psi_v)^2+\lambda^3\right)\sin^2\gamma
	-\lambda_v^2 (a_v R (a_v \lambda_v \psi_v-\psi )\sin 2\gamma+\lambda )\right]\,\mu_{e1} =0\,, 
\end{aligned}
\end{equation}
%-----------------------------
%-----------------------------
\begin{equation}\label{eq:Ex2_nH_KinEq2}
\begin{aligned}
	& \frac{a_v^2 \lambda^2 \lambda_v^3 \left(a_v \lambda_v^2-1\right)\eta_1
	- \lambda^2 \lambda_v^2 \left(a_v \lambda_v^2-1\right)\eta_1
	+ \lambda ^2 \lambda_v^2 \left(\eta _2+\eta _3\right) }{a_v^3 \lambda ^2 \lambda_v^5}\,\dot{a}_v \\
	& \quad + \frac{a_v^3 \lambda ^2 \lambda_v^5\left(\eta _2+\eta _3\right)
	+a_v^2 \lambda ^2 \lambda_v^3 \left(a_v   \lambda_v^2-1\right)\eta_1
	-a_v \lambda ^2 \lambda_v   \left(a_v \lambda_v^2-1\right)\eta_1
	+a_v \lambda ^2 \lambda_v \left(\eta _2+\eta _3\right) }{a_v^3 \lambda ^2 \lambda_v^5}\,
	\dot{\lambda}_v \\
	& \quad +\frac{a_v^2 \lambda_v^4+a_v \lambda_v R^2 \psi \psi_v-\lambda^3-R^2 \psi ^2}
	{\lambda  \lambda_v^3}\,\mu_e  \\
	& \quad  +\frac{\left[a_v^2 \lambda_v^4+ \left(a_v^2 \lambda_v^4
	-a_v \lambda_v R^2 \psi  \psi_v+\lambda^3+R^2 \psi ^2\right)\cos2\gamma
	-a_v^2 \lambda_v^3 R \psi_v \sin2\gamma +a_v \lambda_v R^2 \psi \psi_v-\lambda ^3-R^2 \psi^2\right]}
	{\lambda^2 \lambda_v^5} \\
	& \qquad \times \left\{a_v^2 \lambda_v^4 \cos^2\gamma+
	 \left[R^2 (\psi -a_v \lambda_v \psi_v)^2+\lambda^3\right] \sin^2\gamma
	 -\lambda_v^2 \left[a_v R (a_v  \lambda_v \psi_v-\psi )\sin2\gamma+\lambda \right] \right\}\,\mu_{e1}=0\,,
\end{aligned}
\end{equation}
%-----------------------------
and
%-----------------------------
\begin{equation}\label{eq:Ex2_nH_KinEq3}
\begin{aligned}
	& \eta _3 R \,\dot{\psi}_v+\frac{a_v R (a_v \lambda_v \psi_v-\psi )}{\lambda  \lambda_v}
	\,\mu_e\\
	& \quad -\frac{2 a_v \left[a_v \lambda_v^2 \cos\gamma
	+R (\psi -a_v \lambda_v \psi_v) \sin\gamma\right] \sin\gamma}{\lambda ^2 \lambda_v^3} \\
	& \qquad  \times\left\{a_v^2 \lambda_v^4 \cos^2\gamma
	+ \left[R^2(\psi-a_v \lambda_v \psi_v)^2+\lambda^3\right] \sin^2\gamma
	-\lambda_v^2 \left[a_v R (a_v \lambda_v \psi_v-\psi)\sin2\gamma+\lambda \right] \right\}\,\mu_{e1}=0
	 \,.
\end{aligned}
\end{equation}
%-----------------------------

%-----------------------------
\begin{figure}
\centering
\includegraphics[width=1\textwidth]{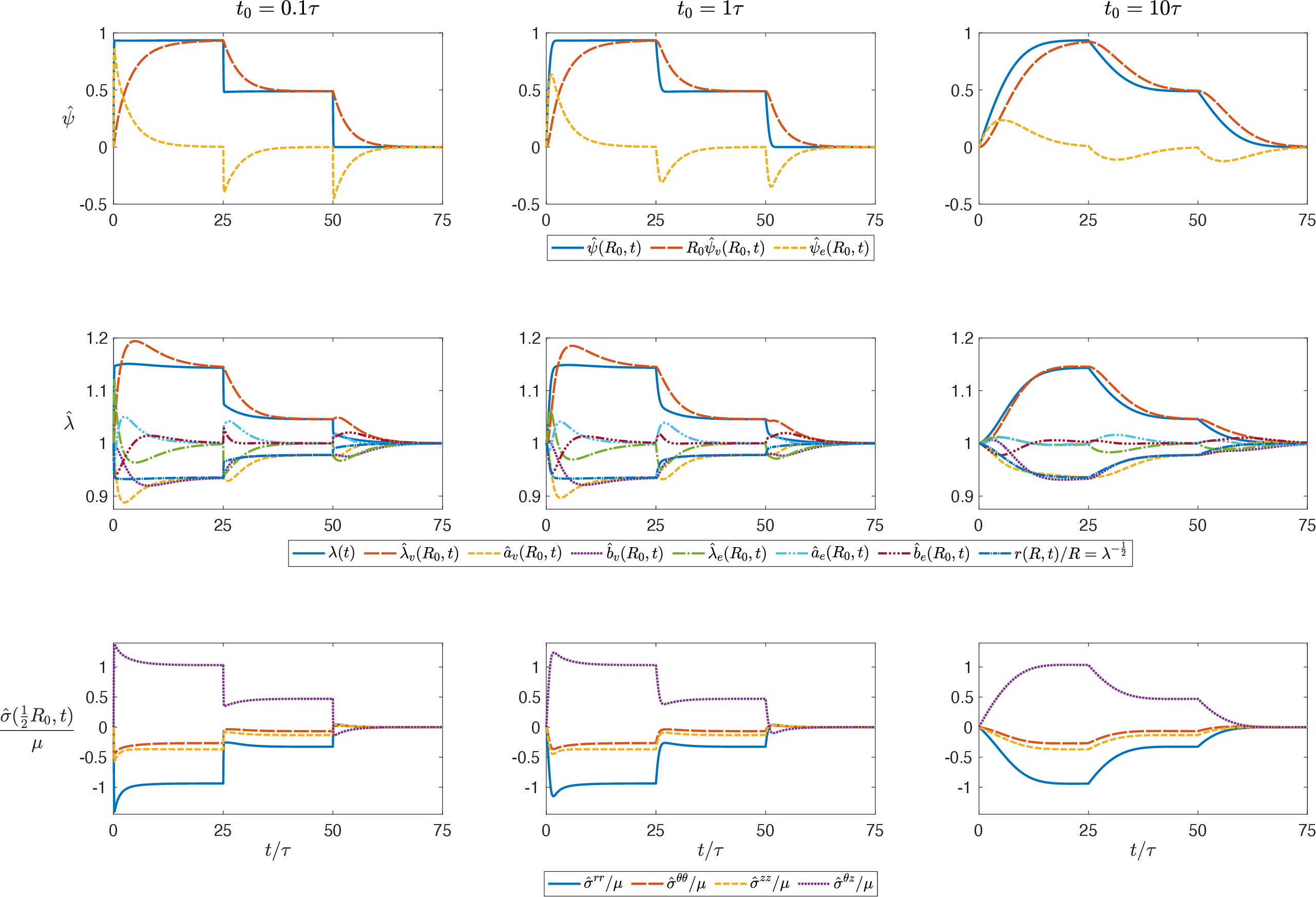}
\vspace*{-0.10in}
\caption{Numerical results for the time evolution of the strain and stress state of a transversely isotropic neo-Hookean bar subject to twist-force-control loading $F(t)=0$ and \eqref{twist_load} at the two ends of the bar with different characteristic times $t_{0}$ versus the viscoelastic dissipation characteristic time $\tau=\frac{\eta_1}{\mu}$ of the system.}
\label{Fig:Ex2Tw_t}
\end{figure}
%-----------------------------

\vspace*{0.0in}
%-----------------------------
\begin{figure}
\centering
\includegraphics[width=1\textwidth]{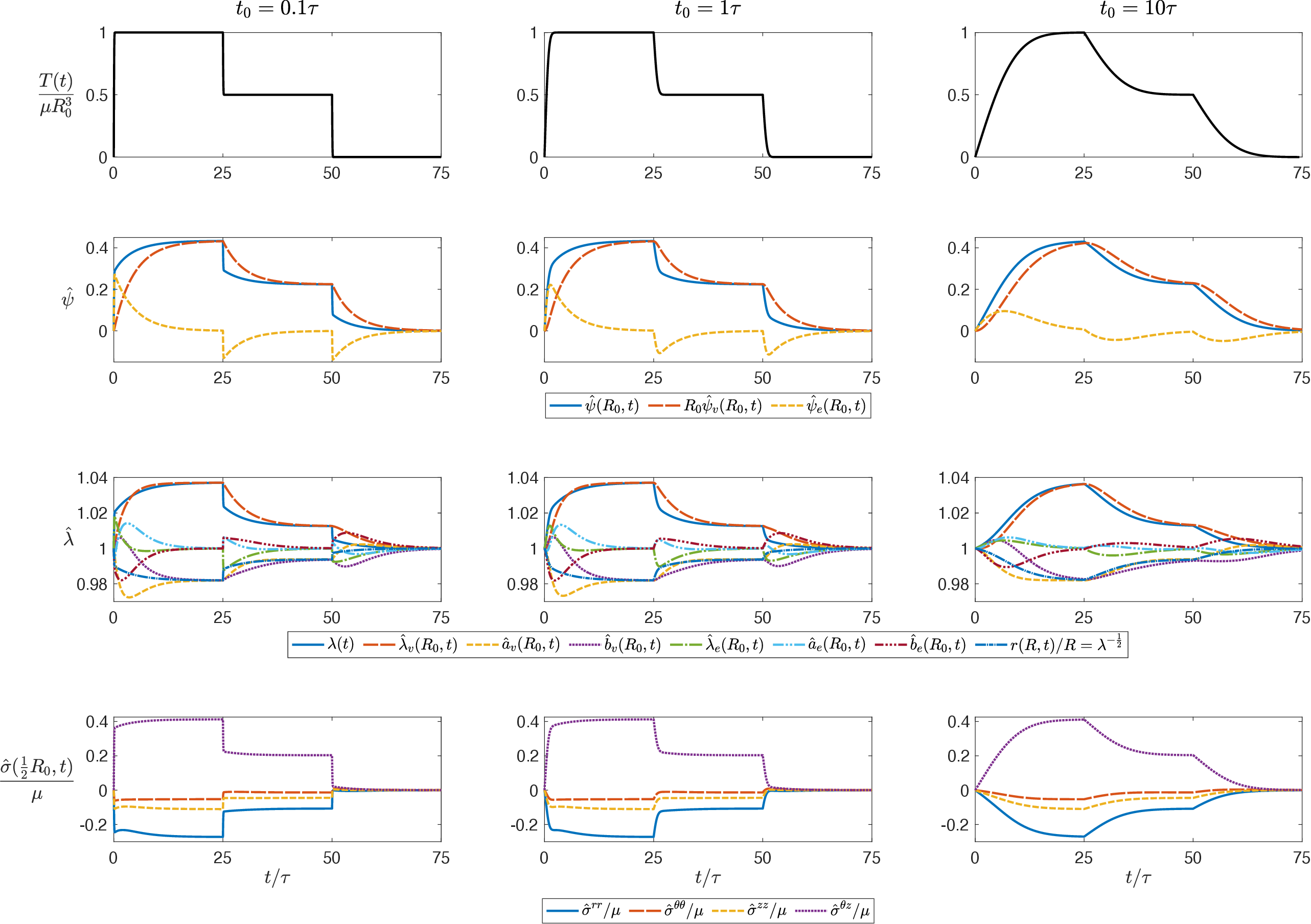}
\vspace*{-0.10in}
\caption{Numerical results for the time evolution of the strain and stress state of a transversely isotropic neo-Hookean bar subject to torque-force-control loading $F(t)=0$ and \eqref{torque_load} at the ends of the bar with different characteristic times $t_{0}$ versus the viscoelastic dissipation characteristic time $\tau=\frac{\eta_1}{\mu}$ of the system.}
\label{Fig:Ex2Tq_t}
\end{figure}
%-----------------------------
\end{example}
%------------------------------	

\paragraph{Numerical results.}
We consider a solid cylinder made of a transversely isotropic neo-Hookean viscoelastic solid such that $\mu_e=\frac{1}{2}\mu\,$,  $\mu_1=\frac{1}{2}\mu\,$,  and $\mu_{e1}=\frac{1}{2}\mu$ with helically symmetric preferred directions along $\Theta=\frac{1}{6}\pi\,$,  and a dissipation potential of the form \eqref{Quadratic-Dissipation} such that $\eta_1=\eta_2=\frac{1}{2}\eta_3\,$. In this example, an implicit finite difference scheme has been used to numerically solve the governing equations.

Let us first look at the twist-force-control loading case and let the bar be free to deform in the longitudinal direction, i.e.,~$F(t)=0\,$, and subject it to the following twist loading
%-----------------------------
\begin{equation}\label{twist_load}
\psi(t)=
\begin{dcases}
\psi_0\erf\left(\frac{t}{t_0}\right)\,,\quad & 0\leq t\leq t_1\,,\\
\psi_0+\frac{\psi(t_1)}{2}\erf\left(\frac{t-t_1}{t_0}\right)\,,\quad & t_1 \leq t\leq t_2\,,\\
\frac{\psi_0}{2}-\psi(t_2)\erf\left(\frac{t-t_2}{t_0}\right)\,,\quad & t_2 \leq t\leq t_f\,,
\end{dcases}
\end{equation}
%----------------------------- 
where $t_{0}$ is the loading characteristic time, $t_1=25t_{0}\,$,  $t_2=50t_{0}\,$,  $t_f=75t_{0}\,$,  and $\psi_{0}=\frac{1}{R_{0}}$ is the angle of twist per unit length at large times ${t_{0}\ll t<t_1}\,$.
In this case, we need to solve the governing PDEs (kinetic equations) \eqref{eq:Ex2_nH_KinEq1}--\eqref{eq:Ex2_nH_KinEq3} coupled with the integral Eq.~\eqref{eq:Ex2_IntEq1} with $F=0$ and prescribing a twist loading $\psi(t)$ given by \eqref{twist_load}. We numerically solve this system of equations assuming different loading characteristic times $t_{0}\,$,  respectively smaller than, equal to, and larger than the characteristic time of the viscoelastic cylinder $\tau=\frac{\eta_1}{\mu}\,$.
In Fig.~\ref{Fig:Ex2Tw_t}, given the twist loading $\psi (t)\,$,  we plot the profile of the corresponding physical component $\hat \psi(R_{0},t) = \frac{R_{0} \psi(t)}{\sqrt{\lambda(t)}} $ as well as the resulting time evolution of the kinematic quantities $\lambda$ $\hat \psi_v\,$,  $\hat \lambda_v\,$,  $\hat a_v\,$,  $\hat b_v\,$,  $\hat \psi_e\,$,  $\hat \lambda_e\,$,  $\hat a_e\,$,  $\hat b_e\,$,  and $r$ at $R=R_{0}\,$, as well as the non-zero stress physical components at $R=\frac{1}{2}R_{0}\,$.
As the bar is twist-loaded, we observe stress relaxation on all the non-zero stress components: First, the bar experiences a fast elastic stress response followed by a slow relaxation towards a steady state of stress under a constant twist angle in each of the loading stages. We also observe that the bar elongation follows the trend of the imposed twist while its outer radius follows an inverse trend. However, the physical components of both the viscous and elastic deformation gradients experience what may be described as strain relaxation. Ultimately, we see that the elastic deformation gradient relaxes towards its unloaded state, and the viscous deformation gradient approaches the total deformation gradient, i.e.,~$\widehat{\Fe}=\mathbf I$ and $\widehat{\Fv} = \hat{\mathbf F}\,$, at large times under constant twist for each of the loading stages, as was previously discussed in Remark~\ref{rmrk:F_LargeTimes}.

Next, we look at the torque-force-control loading case and let the cylinder be free to deform in the longitudinal direction, i.e.,~$F(t)=0\,$, and subject it to the following end torque
%-----------------------------
\begin{equation}\label{torque_load}
T(t)=
\begin{dcases}
T_0\erf\left(\frac{t}{t_0}\right)\,,\quad & 0\leq t\leq t_1\,,\\
T_0+\frac{T(t_1)}{2}\erf\left(\frac{t-t_1}{t_0}\right)\,,\quad & t_1 \leq t\leq t_2\,,\\
\frac{T_0}{2}-T(t_2)\erf\left(\frac{t-t_2}{t_0}\right)\,,\quad & t_2 \leq t\leq t_f\,,
\end{dcases}
\end{equation}
%----------------------------- 
where $t_{0}$ is the loading characteristic time, $t_1=25t_{0}\,$,  $t_2=50t_{0}\,$,  $t_f=75t_{0}\,$,  and $T_{0}=\mu R_{0}^3$ is the end torque at large times ${t_{0}\ll t<t_1}\,$.
In this case, we need to solve the governing PDEs (kinetic equations) \eqref{eq:Ex2_nH_KinEq1}--\eqref{eq:Ex2_nH_KinEq3} coupled with the integral Eqs.~\eqref{eq:Ex2_IntEq} with $F=0$ and $T(t)$ as given by \eqref{torque_load}. We numerically solve this system of equations assuming different loading characteristic times $t_{0}\,$,  respectively smaller than, equal to, and larger than the characteristic time of the viscoelastic cylinder $\tau=\frac{\eta_1}{\mu}\,$.
In Fig.~\ref{Fig:Ex2Tq_t}, we plot the profile of the torque loading $T$ as well as the resulting time evolution of the kinematic quantities $\hat \psi\,$,  $\lambda$ $\hat \psi_v\,$,  $\hat \lambda_v\,$,  $\hat a_v\,$,  $\hat b_v\,$,  $\hat \psi_e\,$,  $\hat \lambda_e\,$,  $\hat a_e\,$,  $\hat b_e\,$,  and $r$ at $R=R_{0}\,$, as well as the non-zero stress physical components at $R=\frac{1}{2}R_{0}\,$.
As the cylinder is torque-loaded, we observe that it experiences creep: First, it elastically deforms instantaneously then continues to slowly deform even as the torque reaches a constant steady state. This can be seen in the evolution of the elongation $\lambda(t)$ and the outer radius $r(R_{0},t)\,$. Except for the viscous elongation $\hat\lambda_v\,$,  which experiences an evolution akin to creep, all the other physical components of the viscous and the elastic deformation gradients experience a strain relaxation. However, here again, and in accordance with  Remark~\ref{rmrk:F_LargeTimes}, the elastic deformation gradient relaxes towards its unloaded state, while the viscous deformation gradient approaches the total deformation gradient, i.e.,~$\widehat {\Fe}=\mathbf I$ and $\widehat{\Fv} = \widehat{\mathbf F}\,$, at large times under constant torque for each of the loading stages.

%-----------------------------
%-----------------------------
\subsection{Example 3: Inflation of an incompressible isotropic viscoelastic thick spherical shell}

\paragraph{Kinematics.}
Let us consider a thick spherical shell subject to a uniform time-dependent inner pressure $p_i(t)\,$. In its undeformed configuration, it has inner and outer radii $R_1$ and $R_2\,$,  respectively.
Let $(R,\Theta,\Phi)$ and $(r,\theta,\phi)$ be spherical coordinate systems in the reference and current configurations, respectively, with their origins at the centers of the respective configurations of the spherical shell.
Following the spherical symmetry of the problem, we consider a radial deformation ansatz:
%-----------------------------
\begin{equation} 
	r=r(R,t)\,,\qquad \theta= \Theta\,,\qquad 	\phi=\Phi	\,.
\end{equation}
%-----------------------------
Therefore, the material and spatial metrics have the following representations:
%-----------------------------
\begin{equation}
   \mathbf{G}=\begin{bmatrix}
  1 & 0  & 0  \\
  0 & R^2  & 0  \\
  0 & 0  & R^2\sin^2\Theta
\end{bmatrix}\,,\qquad
 \mathbf{g}=\begin{bmatrix}
  1 & 0  & 0  \\
  0 & r^2  & 0  \\
  0 & 0  & r^2\sin^2\Theta
\end{bmatrix}\,,
\end{equation}
%-----------------------------
and the deformation gradient reads
%-----------------------------
\begin{equation}
   \mathbf{F}=\mathbf{F}(R,t)=\begin{bmatrix}
  r'(R,t) & 0  & 0  \\
  0 & 1  & 0  \\
  0 & 0  & 1
\end{bmatrix}\,,
\end{equation}
%-----------------------------
where $r'(R,t)=\partial r(R,t)/\partial R\,$.
Incompressibility $J=1$ implies that $r^2(R,t)\,r'(R,t)=R^2\,$. Thus, $r(R,t)=\left[R^3+C^3(t)\right]^{\frac{1}{3}}$\, for some time-dependent function $C(t)\,$. At $t=0\,$,  the thick shell is in its undeformed configuration, i.e.,~$r(R,0)=R$\,; hence, the unknown function $C(t)$ satisfies the initial condition $C(0)=0\,$.
In terms of its physical components, the deformation gradient reads
%-----------------------------
\begin{equation}
  \hat{\mathbf F}=\hat{\mathbf F}(R,t)=\begin{bmatrix}
  \frac{R^2}{r^2(R,t)} & 0  & 0  \\
  0 & \frac{r(R,t)}{R} & 0  \\
  0 & 0  & \frac{r(R,t)}{R}
\end{bmatrix}\,.
\end{equation}
%-----------------------------
In order to be consistent with spherically-symmetric universal eigenstrains~\citep{Goodbrake2021}, we assume the following form for the viscous deformation gradient
%-----------------------------
\begin{equation}
   \Fv=\Fv(R,t)=\begin{bmatrix}
  a_v(R,t) & 0  & 0  \\
  0 & b_v(R,t)  & 0  \\
  0 & 0  & b_v(R,t)
\end{bmatrix}\,.
\end{equation}
%-----------------------------
At $t=0\,$, in the initial unloaded state, $\Fv(R,0)=\mathbf I\,$, i.e., $a_v(R,0)=b_v(R,0)=1\,$. Incompressibility $\Jv=1$ implies that $a_v(R,t)\,b_v^2(R,t)=1\,$.
The physical components of the viscous deformation gradient read
%-----------------------------
\begin{equation}
	\hat a_v(R,t)=a_v(R,t)\,,\qquad
	\hat b_v(R,t)=b_v(R,t)=\frac{1}{\sqrt{a_v(R,t)}}\,.
\end{equation}
%-----------------------------
Since $\mathbf{F}=\Fe\Fv\,$, it follows that the elastic deformation gradient has the following form
%-----------------------------
\begin{equation}
   \Fe=\Fe(R,t)=\begin{bmatrix}
  a_e(R,t) & 0  & 0  \\
  0 & b_e(R,t)   & 0  \\
  0 & 0  & b_e(R,t)
\end{bmatrix}\,,
\end{equation}
%-----------------------------
where
%-----------------------------
\begin{equation}
	a_e(R,t) = \frac{R^2}{\left[R^3+C^3(t)\right]^{\frac{2}{3}}} a_v^{-1}(R,t)\,, \qquad 
	b_e(R,t) = b_v^{-1}(R,t) = \sqrt{a_v(R,t)}\,.\end{equation}
%-----------------------------
Its physical components read
%-----------------------------
\begin{equation}
	\hat a_e(R,t) = \frac{R^2}{\left[R^3+C^3(t)\right]^{\frac{2}{3}}} a_v^{-1}(R,t)\,, \qquad
	\hat b_e(R,t) = \frac{\left[R^3+C^3(t)\right]^{\frac{1}{3}}}{R} b_v^{-1}(R,t) = \frac{\left[R^3+C^3(t)\right]^{\frac{1}{3}}}{R}\sqrt{a_v(R,t)}\,.
	\end{equation}
%-----------------------------

%-----------------------------
\begin{remark} 
$\Fv$ is compatible if and only if $\frac{\partial b_v(R,t)}{\partial R}=0\,$.
\end{remark}
%-----------------------------

\paragraph{Stress and equilibrium equations.}
Following \eqref{eq:sigma_iso}, the non-zero components of the Cauchy stress are
%-----------------------------
\begin{subequations} \label{sigmarr1}
\begin{align}
	\label{sigmarr1-a}
	\sigma^{rr}(R,t) &= -p+\frac{2 R^4}{\left(R^3+C^3\right)^{\frac{4}{3}}}\,\overline{\Psi}_1
	-\frac{2\left(R^3+C^3\right)^{\frac{4}{3}}}{R^4}\,\overline{\Psi}_2
	+\frac{2 R^4}{a_v^2 \left(R^3+C^3\right)^{\frac{4}{3}}}\,\widetilde{\Psi}_1
	-\frac{2 a_v^2 \left(R^3+C^3\right)^{\frac{4}{3}}}{R^4}\,\widetilde{\Psi}_2 \,,\\ 
	\label{sigmarr1-b}
	\sigma^{\theta\theta}(R,t) & = -\frac{p}{\left(R^3+C^3\right)^{\frac{2}{3}}}
	+\frac{2}{R^2}\,\overline{\Psi}_1-\frac{2 R^2}{\left(R^3+C^3\right)^{\frac{4}{3}}}\,\overline{\Psi}_2
	+\frac{2}{R^2 b_v^2}\,\widetilde{\Psi}_1
	-\frac{2 R^2 b_v^2}{\left(R^3+C^3\right)^{\frac{4}{3}}}\,\widetilde{\Psi}_2\,,\\
	\label{sigmarr1-c}
	\sigma^{\phi\phi}(R,t) & = \frac{1}{\sin^2\Theta} \left[-\frac{p}{\left(R^3+C^3\right)^{\frac{2}{3}}}
	+\frac{2}{R^2}\,\overline{\Psi}_1-\frac{2 R^2}{\left(R^3+C^3\right)^{\frac{4}{3}}}\,\overline{\Psi}_2
	+\frac{2}{R^2 b_v^2}\,\widetilde{\Psi}_1-\frac{2R^2 b_v^2}{\left(R^3+C^3\right)^{\frac{4}{3}}}\,\widetilde{\Psi}_2
	\right]
	\,,
\end{align}
\end{subequations}
%-----------------------------
where $p=p(R,t)$ is the Lagrange multiplier enforcing incompressibility, i.e., $J=1\,$.
The only non-trivial equilibrium is 
%-----------------------------
\begin{equation}\label{Equilibrium-Spherical}
	\sigma^{rr}{}_{,r}+\frac{2}{r}\sigma^{rr}-r\sigma^{\theta\theta}-r\sin^2\theta~\sigma^{\phi\phi}=0.
\end{equation}
%-----------------------------
Or
%-----------------------------
\begin{equation}
	\sigma^{rr}{}_{,R} = -r'\left[\frac{2}{r}\sigma^{rr}-2r\sigma^{\theta\theta}\right]\,.
\end{equation}
%-----------------------------
Thus, recalling that the inner boundary is under a time-dependent pressure $p_i(t)\,$, i.e., $\sigma^{rr}(R_1,t)=-p_i(t)\,$, we find
%-----------------------------
\begin{equation} \label{sigmarr2}
	\sigma^{rr}(R,t)=-p_i(t)+\int_{R_1}^{R} f(\xi,t)\,d\xi\,,
\end{equation}
%-----------------------------
where
%-----------------------------
\begin{equation}
\begin{aligned}
	f(R,t) &=-r'\left[\frac{2}{r}\sigma^{rr}-2r\sigma^{\theta\theta}\right]
	=\frac{4 C^3 \left(2 R^3+C^3\right)}{\left(R^3+C^3\right)^{\frac{7}{3}}}\,\overline{\Psi}_1
	+\frac{4C^3 \left(2 R^3+C^3\right)}{R^2 \left(R^3+C^3\right)^{\frac{5}{3}}}\,
	\overline{\Psi}_2 \\
	& \quad+\frac{4\left[a_v^3 \left(R^3+C^3\right)^2-R^6 \right]}
	{a_v^2 \left(R^3+C^3\right)^{\frac{7}{3}}}\,\widetilde{\Psi}_1
	+\frac{4\left[a_v^3 \left(R^3+C^3\right)^2-R^6 \right]}
	{a_v R^2\left(R^3+C^3\right)^{\frac{5}{3}}}\,\widetilde{\Psi}_2
	\,.
\end{aligned}
\end{equation}
%-----------------------------
From \eqref{sigmarr1-a} and \eqref{sigmarr2}, the pressure field is calculated as
%-----------------------------
\begin{equation} 
\begin{aligned}
	 p(R,t) &=p_i(t)-\int_{R_1}^{R} f(\xi,t)\,d\xi
	 +\frac{2 R^4}{\left(R^3+C^3\right)^{\frac{4}{3}}}\,\overline{\Psi}_1
	-\frac{2\left(R^3+C^3\right)^{\frac{4}{3}}}{R^4}\,\overline{\Psi}_2 \\
	& \quad +\frac{2 R^4}{a_v^2 \left(R^3+C^3\right)^{\frac{4}{3}}}\,\widetilde{\Psi}_1
	-\frac{2 a_v^2 \left(R^3+C^3\right)^{\frac{4}{3}}}{R^4}\,\widetilde{\Psi}_2 
	\,.
\end{aligned}
\end{equation}
%-----------------------------
We assume that the outer boundary is traction-free, i.e., $\sigma^{rr}(R_2,t)=0\,$. Thus
%-----------------------------
\begin{equation} \label{pi-conditiion}
	\int_{R_1}^{R_2} f(\xi,t)\,d\xi=p_i(t)
	\,.
\end{equation}
%-----------------------------
At this point, the unknown fields of the problem are $C(t)$ and $a_v(R,t)\,$. The boundary condition at $R_2$ above needs to be supplemented by the kinetic equation to solve the problem herein.

\paragraph{Kinetic equation.}
Assuming the quadratic dissipation potential introduced in \S\ref{SubSec:Quadr-Diss-Pot}, the kinetic equation\textemdash following \eqref{eq:Quadr-Diss_Kin-Eq}\textemdash reads as the following system of equations
%-----------------------------
\begin{subequations} \label{eq:Ex3-KinEq2}
\begin{align}
	& \dot a_v \left[(\eta_1+\eta_2+\eta_3) a_v-\frac{\eta_1}{\sqrt{a_v}}\right]
	-\frac{2 R^4 \widetilde{\Psi}_1}{\left(C^3+R^3\right)^{\frac{4}{3}} a_v^2}
	-\frac{4 R^2 \widetilde{\Psi}_2}{\left(C^3+R^3\right)^{\frac{2}{3}} a_v}=q\,,\\
	& \dot a_v \left(\frac{\eta_1}{\sqrt{a_v}}-\frac{2\eta_1+\eta_2+\eta_3}{2 a_v^2}\right)
	-\frac{2 \widetilde{\Psi}_2 \left(C^3+R^3\right)^{\frac{4}{3}} a_v^2}{R^4}
	-\frac{2 \widetilde{\Psi}_1 \left(C^3+R^3\right)^{\frac{2}{3}} a_v}{R^2}
	-\frac{2 R^2 \widetilde{\Psi}_2}{\left(C^3+R^3\right)^{\frac{2}{3}} a_v}=q\,,
\end{align}
\end{subequations}
%-----------------------------
where we recall that $q=q(R,t)$ is the Lagrange multiplier corresponding to viscous incompressibility, i.e.,~$\Jv=1\,$. We eliminate $q$ from \eqref{eq:Ex3-KinEq2} and end up with a single ordinary differential equation for $a_v$ as the kinetic equation:
%-----------------------------
\begin{equation}\label{eq:Ex3-KinEq1}
\begin{aligned}
	\left[2 \eta_1 \left(a_v^{\frac{3}{2}}-1\right)^2+\left(\eta _2+\eta _3\right)\left(2 a_v^3+1\right)\right]\dot{a}_v
	+\frac{4\left[a_v^3 \left(R^3+C^3\right)^2-R^6\right]}{R^2\left(R^3+C^3\right)^{\frac{4}{3}}}\widetilde{\Psi}_1+\frac{4\left[a_v^4 \left(R^3+C^3\right)^2-R^6 a_v\right]}{R^4\left(R^3+C^3\right)^{\frac{2}{3}}}\widetilde{\Psi}_2=0
	\,.
\end{aligned}
\end{equation}
%-----------------------------

%-----------------------------
\begin{example}
Let us consider a neo-Hookean viscoelastic  solid, i.e., $\overline{\Psi}_1=\frac{1}{2}\mu\,$, $\overline{\Psi}_2=0\,$, $\widetilde{\Psi}_1=\frac{1}{2}\mu_e\,$, and $\widetilde{\Psi}_2=0\,$. 
In this case, the kinetic equation \eqref{eq:Ex3-KinEq1} is simplified to read
%-----------------------------
\begin{equation} 
\begin{aligned}
	\left[2 \eta_1 \left(a_v^{\frac{3}{2}}-1\right)^2+\left(\eta_2+\eta_3\right) \left(2a_v^3+1\right)\right]\dot{a}_v(R,t)
	+2\mu_e\frac{\left(C^3(t)+R^3\right)^{\frac{2}{3}}}{R^2}a_v^3=  2\mu_e \frac{R^4}{\left(C^3(t)+R^3\right)^{\frac{4}{3}}}
	\,.
\end{aligned}
\end{equation}
%-----------------------------
For this model
%-----------------------------
\begin{equation}
	f(R,t)=2\mu\frac{C^3 \left(2 R^3+C^3\right)}{\left(R^3+C^3\right)^{\frac{7}{3}}}
	 +2\mu_e\frac{\left[a_v^3 \left(R^3+C^3\right)^2-R^6 \right]}
	 	{a_v^2 \left(R^3+C^3\right)^{\frac{7}{3}}}
	\,.
\end{equation}
%-----------------------------
The boundary condition \eqref{pi-conditiion} is simplified to read
%-----------------------------
\begin{equation}
	\mu\left[\frac{4R_2 C^3+5R_2^4}{2\left(C^3+R_2^3\right)^{\frac{4}{3}}}
	-\frac{4R_1 C^3+5R_1^4}{2\left(C^3+R_1^3\right)^{\frac{4}{3}}}\right]
	+2 \mu_e \int_{R_1}^{R_2} \frac{a_v^3 \left(R^3+C^3\right)^2-R^6}{a_v^2 \left(R^3+C^3\right)^{\frac{7}{3}}}\,dR
	=p_i(t)\,.
\end{equation}
%-----------------------------
The following system of ODE-integral equation governs the unknowns $a_v(R,t)$ and $C(t)$:\footnote{Assuming that $a_v(R,0)=1\,$,  the integral equation at $t=0$ is simplified to read
%-----------------------------
\begin{equation} 
	(\mu+\mu_e)\left[\frac{4R_2 C^3+5R_2^4}{4\left(C^3+R_2^3\right)^{\frac{4}{3}}}
	-\frac{4R_1 C^3+5R_1^4}{4\left(C^3+R_1^3\right)^{\frac{4}{3}}}\right]
	=0\,,
\end{equation}
%-----------------------------
which implies that $C(0)=0\,$. Thus, only one initial condition is needed.}
%-----------------------------
\begin{equation}\label{Ex3_Gov_Eq}
\begin{dcases}
	\left[2 \eta_1 \left(a_v^{\frac{3}{2}}-1\right)^2+\left(\eta_2+\eta_3\right) \left(2a_v^3+1\right)\right]\dot{a}_v(R,t)
	+2\mu_e \frac{\left(C^3(t)+R^3\right)^{\frac{2}{3}}}{R^{2}}a_v^3= 
	2\mu_e \frac{R^4}{\left(C^3(t)+R^3\right)^{\frac{4}{3}}}\,, \\
	2\mu\left[\frac{4R_2 C^3+5R_2^4}{4\left(C^3+R_2^3\right)^{\frac{4}{3}}}
	-\frac{4R_1 C^3+5R_1^4}{4\left(C^3+R_1^3\right)^{\frac{4}{3}}}\right]
	+2\mu_e \int_{R_1}^{R_2} \frac{a_v^3 \left(R^3+C^3\right)^2-R^6}{a_v^2
	\left(R^3+C^3\right)^{\frac{7}{3}}}\,dR=p_i(t)\,, \\
	a_v(R,0)=1\,,	
\end{dcases}
\end{equation}
%-----------------------------
where the inner pressure loading is given by
%-----------------------------
\begin{equation}\label{pressure_load}
p_i(t)=
\begin{dcases}
p_f\erf\left(\frac{t}{t_0}\right)\,,\quad & 0\leq t\leq t_1\,,\\
p_f+\frac{p_i(t_1)}{2}\erf\left(\frac{t-t_1}{t_0}\right)\,,\quad & t_1 \leq t\leq t_2\,,\\
\frac{p_f}{2}-p_i(t_2)\erf\left(\frac{t-t_2}{t_0}\right)\,,\quad & t_2 \leq t\leq t_f\,,
\end{dcases}
\end{equation}
%-----------------------------
where $t_0$ is the loading characteristic time, $t_1=25t_0\,$, $t_2=50t_0\,$, $t_f=75t_0\,$, and $p_f$ is the force at large times ${t_0\ll t<t_1}\,$.

%-----------------------------
\begin{figure}
\centering
\includegraphics[width=1\textwidth]{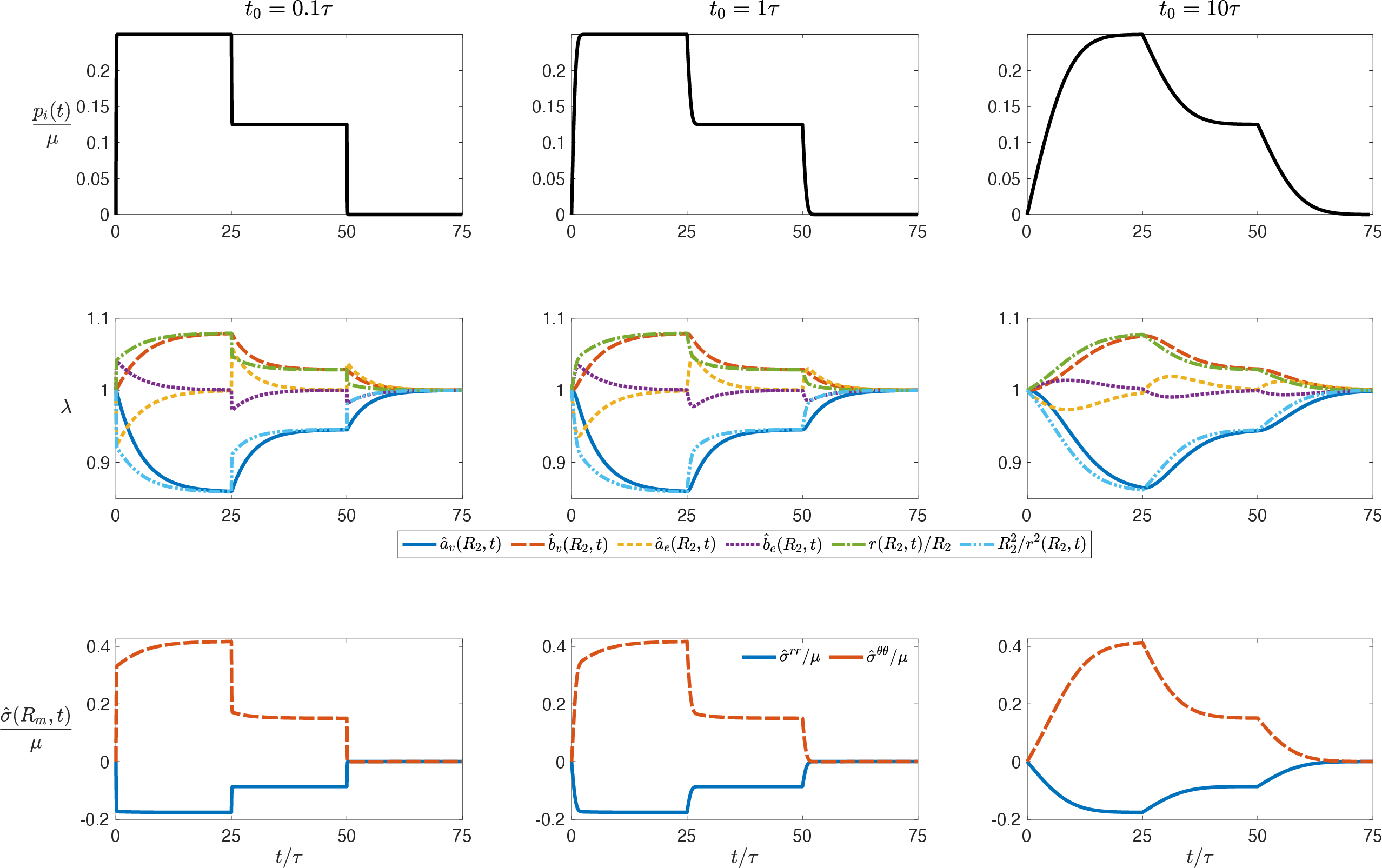}
\vspace*{-0.10in}
\caption{Numerical results for the time evolution of strain and stress state of an isotropic neo-Hookean viscoelastic thick shell of outer radius $R_2$ and inner radius $R_1=0.75R_2$ subject to pressure-control loading $p_i(t)$ \eqref{pressure_load} at its inner wall with different characteristic times $t_{0}$ versus the viscoelastic dissipation characteristic time $\tau=\frac{\eta_1}{\mu}$ of the system.}
\label{Fig:Ex3_t}
\end{figure}
%-----------------------------

\paragraph{Numerical results.}
Let us consider a thick spherical shell made of a neo-Hookean viscoelastic solid such that $\mu_e=\frac{1}{2}\mu\,$,  with a dissipation potential of the form \eqref{Quadratic-Dissipation} such that $\eta_1=\eta_2=\frac{1}{2}\eta_3\,$,  and subject it to the inner pressure $p_i(t)$ \eqref{pressure_load} with $p_f=0.25 \mu$ at $R_1 = 0.75 R_2\,$. In this example, an explicit finite difference scheme has been used to numerically solve the governing equations.

We numerically solve the governing Eq.~\eqref{Ex3_Gov_Eq} assuming different characteristic times $t_{0}\,$,  respectively smaller, equal, and larger than the characteristic time of the viscoelastic shell.
In Fig.~\ref{Fig:Ex3_t}, we show the profile of the inner pressure loading $p_i(t)$ and the resulting evolution of the kinematic quantities $a_v\,$,  $b_v\,$,  $a_e\,$,  $b_e\,$,  and $r$ at $R=R_2\,$,  as well as the non-zero physical stress components at $R=R_m=\frac{1}{2}(R_1+R_2)\,$.
As pressure is applied on the inner wall of the spherical shell, we observe that in each of the three loading then unloading stages, the thick shell experiences creep as its outer radius first experiences a fast elastic response followed by a slow deformation even as the inner pressure reaches a constant steady state.
However, the elastic deformation gradient experiences a strain relaxation as its physical components first elastically increase at a fast rate and then slowly relax into their initial unloaded state as the loading reaches a constant steady state, i.e.,~$\widehat{\Fe}=\mathbf I\,$, as previously discussed in  Remark~\ref{rmrk:F_LargeTimes}. We also see that the viscous deformation gradient experiences a phenomenon akin to creep which is in accordance with  Remark~\ref{rmrk:F_LargeTimes} since its physical components tend to match those of the total deformation gradient as the loading reaches a constant steady state\textemdash $\hat a_v(R_2,t)$ approaches ${F^r}_R(R_2,t)=\frac{R_2^2}{r^2(R_2,t)}$ and $\hat b_v(R_2,t)$ approaches ${F^{\theta}}_\Theta(R_2,t)=\frac{r(R_2,t)}{R_2}$ so that $\widehat{\Fv}=\widehat{\mathbf F}$ at large times in each of the three loading stages.

\end{example}
%-----------------------------

%-----------------------------
%-----------------------------
\section{Conclusions} \label{Sec:Conclusions}

In this paper, we first revisited the multiplicative decomposition of the deformation gradient {$\mathbf{F}=\Fe\Fv$} in nonlinear viscoelasticity from a geometric point of view. We showed, based on invariance and physical arguments, that the viscous deformation gradient has to be a material tensor while the elastic deformation gradient is a two-point tensor. We assumed an additive split of the free energy density into equilibrium and non-equilibrium parts. The equilibrium free energy depends on the total deformation gradient, while the non-equilibrium part depends only on the elastic deformation gradient. We also assumed the existence of a dissipation potential that depends on the total deformation gradient, the viscous deformation gradient, and its rate, and that it is convex in the rate of the viscous deformation gradient. 
It was concluded that there is a subtle but crucial difference between anelasticity and viscoelasticity; the intermediate configuration is stress-free in anelasticity but it is stressed in viscoelasticity.

We derived the balance laws using a two-potential approach and the Lagrange--d'Alembert principle. More specifically, the variational principle gives us the balance of linear momentum and a kinetic equation for the viscous deformation gradient.
We also discussed thermodynamics of viscoelasticity. Next, material symmetry was discussed and it was emphasized that the equilibrium and non-equilibrium free energies and the dissipation potential are all invariant under the same symmetry group. We derived the representations of the Cauchy stress in terms of the principal and structural invariants for isotropic, transversely isotropic, orthotropic, and monoclinic solids. The explicit form of the kinetic equation was also derived for these four classes of solids.

Three examples of universal deformations were studied for both isotropic and transversely isotropic solids. These were extension and torsion of a solid circular bar, and inflation of a spherical shell. Assuming incompressible solids, the kinematics in each case is reduced to depend on one or two unknown time-dependent functions. The viscous deformation gradient has one or two unknown functions of a radial coordinate and time. The governing equations were reduced to an initial-value problem for a coupled system of partial differential and integral equations. For specific examples of solids, these problems were solved numerically.

In a future communication, a theory of small-on-large viscoelasticity will be formulated, a particular case of which is linearized viscoelasticity.
In another future communication, a nonlinear theory of visco-anelasticity will be developed in order to study the coupling of anelasticity and viscoelasticity.

%---------------------------
%------------------------------
\section*{Acknowledgement}

AY benefited from discussions with A. Kumar and O. Lopez-Pamies.
SS benefited from discussions with L. De Cunto on the numerical implementation of the examples. The authors are grateful to S. Govindjee for his valuable comments on an earlier draft of the paper. This work was partially supported by IFD -- Eurostars III Grant No. 2103-00007B and NSF -- Grant No. CMMI~1939901.

%-------------------------
%----------------------------
\bibliographystyle{abbrvnat}
\bibliography{Visco_ref}

\begin{thebibliography}{109}
\providecommand{\natexlab}[1]{#1}
\providecommand{\url}[1]{\texttt{#1}}
\expandafter\ifx\csname urlstyle\endcsname\relax
  \providecommand{\doi}[1]{doi: #1}\else
  \providecommand{\doi}{doi: \begingroup \urlstyle{rm}\Url}\fi

\bibitem[Badal et~al.(2023)Badal, Friedrich, and Kru{\v{z}}{\'\i}k]{Badal2023}
R.~Badal, M.~Friedrich, and M.~Kru{\v{z}}{\'\i}k.
\newblock Nonlinear and linearized models in thermoviscoelasticity.
\newblock \emph{Archive for Rational Mechanics and Analysis}, 247\penalty0
  (1):\penalty0 5, 2023.

\bibitem[Bahreman et~al.(2022)Bahreman, Darijani, and Narooei]{Bahreman2022}
M.~Bahreman, H.~Darijani, and K.~Narooei.
\newblock Investigation of multiplicative decompositions in the form of
  $\mathbf{F}_e\mathbf{F}_v$ and $\mathbf{F}_v\mathbf{F}_e$ to extend
  viscoelasticity laws from small to finite deformations.
\newblock \emph{Mechanics of Materials}, 167:\penalty0 104235, 2022.

\bibitem[Banks et~al.(2011)Banks, Hu, and Kenz]{Banks2011}
H.~T. Banks, S.~Hu, and Z.~R. Kenz.
\newblock A brief review of elasticity and viscoelasticity for solids.
\newblock \emph{Advances in Applied Mathematics and Mechanics}, 3\penalty0
  (1):\penalty0 1--51, 2011.

\bibitem[Berdichevskii and Sedov(1967)]{BerdiSedov1967}
V.~Berdichevskii and L.~Sedov.
\newblock Dynamic theory of continuously distributed dislocations. its relation
  to plasticity theory: {PMM} vol. 31, no. 6, 1967, pp. 981--1000.
\newblock \emph{Journal of Applied Mathematics and Mechanics}, 31\penalty0
  (6):\penalty0 989--1006, 1967.

\bibitem[Bilby et~al.(1955)Bilby, Bullough, and Smith]{bilby1955}
B.~A. Bilby, R.~Bullough, and E.~Smith.
\newblock Continuous distributions of dislocations: a new application of the
  methods of non-riemannian geometry.
\newblock \emph{Proceedings of the Royal Society of London. Series A.
  Mathematical and Physical Sciences}, 231\penalty0 (1185):\penalty0 263--273,
  1955.

\bibitem[Biot(1954)]{biot1954theory}
M.~A. Biot.
\newblock Theory of stress-strain relations in anisotropic viscoelasticity and
  relaxation phenomena.
\newblock \emph{Journal of Applied Physics}, 25\penalty0 (11):\penalty0
  1385--1391, 1954.

\bibitem[Boehler(1979)]{Boehler1979}
J.-P. Boehler.
\newblock A simple derivation of representations for non-polynomial
  constitutive equations in some cases of anisotropy.
\newblock \emph{Zeitschrift f{\"u}r Angewandte Mathematik und Mechanik},
  59\penalty0 (4):\penalty0 157--167, 1979.

\bibitem[Boehler(1987)]{boehler1987}
J.-P. Boehler.
\newblock \emph{Applications of Tensor Functions in Solid Mechanics}, volume
  292.
\newblock Springer, 1987.

\bibitem[Boltzmann(1874)]{Boltzmann1874}
L.~Boltzmann.
\newblock Zur theorie der elastischen nachwirkung.
\newblock \emph{Sitzungsberichte der Mathematisch-Naturwissenschaftlichen
  Classe der Kaiseflichen Akademie der Wissenschaffen}, 70:\penalty0 275--300,
  1874.

\bibitem[Carroll(1967)]{Carroll1967}
M.~M. Carroll.
\newblock Controllable deformations of incompressible simple materials.
\newblock \emph{International Journal of Engineering Science}, 5\penalty0
  (6):\penalty0 515--525, 1967.

\bibitem[Ciambella and Nardinocchi(2021)]{Ciambella2021}
J.~Ciambella and P.~Nardinocchi.
\newblock A structurally frame-indifferent model for anisotropic
  visco-hyperelastic materials.
\newblock \emph{Journal of the Mechanics and Physics of Solids}, 147:\penalty0
  104247, 2021.

\bibitem[Coleman and Gurtin(1967)]{ColeGurt1967internal}
B.~D. Coleman and M.~E. Gurtin.
\newblock Thermodynamics with internal state variables.
\newblock \emph{The journal of chemical physics}, 47\penalty0 (2):\penalty0
  597--613, 1967.

\bibitem[Coleman and Noll(1961)]{ColemanNoll1961}
B.~D. Coleman and W.~Noll.
\newblock Foundations of linear viscoelasticity.
\newblock \emph{Reviews of Modern Physics}, 33\penalty0 (2):\penalty0 239,
  1961.

\bibitem[Doyle and Ericksen(1956)]{Doyle1956}
T.~C. Doyle and J.~L. Ericksen.
\newblock Nonlinear elasticity.
\newblock \emph{Advances in Applied Mechanics}, 4:\penalty0 53--115, 1956.

\bibitem[Drapaca et~al.(2007)Drapaca, Sivaloganathan, and Tenti]{Drapaca2007}
C.~S. Drapaca, S.~Sivaloganathan, and G.~Tenti.
\newblock Nonlinear constitutive laws in viscoelasticity.
\newblock \emph{Mathematics and mechanics of solids}, 12\penalty0 (5):\penalty0
  475--501, 2007.

\bibitem[Ericksen(1954)]{Ericksen1954}
J.~L. Ericksen.
\newblock Deformations possible in every isotropic, incompressible, perfectly
  elastic body.
\newblock \emph{Zeitschrift f{\"u}r Angewandte Mathematik und Physik},
  5\penalty0 (6):\penalty0 466--489, 1954.

\bibitem[Ericksen(1955)]{Ericksen1955}
J.~L. Ericksen.
\newblock Deformations possible in every compressible, isotropic, perfectly
  elastic material.
\newblock \emph{Studies in Applied Mathematics}, 34\penalty0 (1-4):\penalty0
  126--128, 1955.

\bibitem[Germain et~al.(1983)Germain, Suquet, and Nguyen]{Germain1983}
P.~Germain, P.~Suquet, and Q.~S. Nguyen.
\newblock Continuum thermodynamics.
\newblock \emph{Journal of Applied Mechanics}, 50:\penalty0 1010--1020, 1983.

\bibitem[Goldstein et~al.(2002)Goldstein, Poole, and Safko]{Goldstein2002}
H.~Goldstein, C.~Poole, and J.~Safko.
\newblock \emph{Classical Mechanics}.
\newblock American Association of Physics Teachers, 2002.

\bibitem[Goodbrake et~al.(2020)Goodbrake, Yavari, and Goriely]{Goodbrake2020}
C.~Goodbrake, A.~Yavari, and A.~Goriely.
\newblock The anelastic {E}ricksen problem: {U}niversal deformations and
  universal eigenstrains in incompressible nonlinear anelasticity.
\newblock \emph{Journal of Elasticity}, 142\penalty0 (2):\penalty0 291--381,
  2020.

\bibitem[Goodbrake et~al.(2021)Goodbrake, Goriely, and Yavari]{Goodbrake2021}
C.~Goodbrake, A.~Goriely, and A.~Yavari.
\newblock The mathematical foundations of anelasticity: {E}xistence of smooth
  global intermediate configurations.
\newblock \emph{Proceedings of the Royal Society A}, 477\penalty0
  (2245):\penalty0 20200462, 2021.

\bibitem[Green and Naghdi(1971)]{GreenNaghdi1971}
A.~E. Green and P.~M. Naghdi.
\newblock Some remarks on elastic-plastic deformation at finite strain.
\newblock \emph{International Journal of Engineering Science}, 9\penalty0
  (12):\penalty0 1219--1229, 1971.

\bibitem[Green and Rivlin(1957)]{GreenRivlin1957}
A.~E. Green and R.~S. Rivlin.
\newblock The mechanics of non-linear materials with memory {I}.
\newblock \emph{Archive for Rational Mechanics and Analysis}, 1\penalty0
  (1):\penalty0 1--21, 1957.

\bibitem[Green and Rivlin(1959)]{GreenRivlin1959}
A.~E. Green and R.~S. Rivlin.
\newblock The mechanics of non-linear materials with memory {III}.
\newblock \emph{Archive for Rational Mechanics and Analysis}, 4:\penalty0
  387--404, 1959.

\bibitem[Green et~al.(1959)Green, Rivlin, and Spencer]{GreenRivlinSpencer1959}
A.~E. Green, R.~S. Rivlin, and A.~J.~M. Spencer.
\newblock The mechanics of non-linear materials with memory {II}.
\newblock \emph{Archive for Rational Mechanics and Analysis}, 3:\penalty0
  82--90, 1959.

\bibitem[Green and Tobolsky(1946)]{green1946new}
M.~S. Green and A.~V. Tobolsky.
\newblock A new approach to the theory of relaxing polymeric media.
\newblock \emph{The Journal of chemical physics}, 14\penalty0 (2):\penalty0
  80--92, 1946.

\bibitem[Gurtin(1972)]{Gurtin1972}
M.~E. Gurtin.
\newblock The linear theory of elasticity.
\newblock In \emph{Handbuch der Physik, Band VIa/2.} Springer-Verlag, Berlin,
  1972.

\bibitem[Gurtin(1974)]{gurtin1974modern}
M.~E. Gurtin.
\newblock Modern continuum thermodynamics.
\newblock \emph{Mechanics Today}, 1:\penalty0 168--213, 1974.

\bibitem[Gurtin and Anand(2005)]{Gurtin2005}
M.~E. Gurtin and L.~Anand.
\newblock The decomposition $\mathbf{F}=\mathbf{ F}^e\mathbf{F}^p$, material
  symmetry, and plastic irrotationality for solids that are
  isotropic-viscoplastic or amorphous.
\newblock \emph{International Journal of Plasticity}, 21\penalty0 (9):\penalty0
  1686--1719, 2005.

\bibitem[Gurtin and Sternberg(1962)]{GurtinSternberg1962}
M.~E. Gurtin and E.~Sternberg.
\newblock On the linear theory of viscoelasticity.
\newblock \emph{Archive for Rational Mechanics and Analysis}, 11:\penalty0
  291--356, 1962.

\bibitem[Hilbert(1993)]{Hilbert1993}
D.~Hilbert.
\newblock \emph{Theory of Algebraic Invariants}.
\newblock Cambridge University Press, 1993.

\bibitem[Holzapfel(1996)]{Holzapfel1996b}
G.~A. Holzapfel.
\newblock On large strain viscoelasticity: continuum formulation and finite
  element applications to elastomeric structures.
\newblock \emph{International Journal for Numerical Methods in Engineering},
  39\penalty0 (22):\penalty0 3903--3926, 1996.

\bibitem[Holzapfel and Simo(1996)]{holzapfel1996a}
G.~A. Holzapfel and J.~C. Simo.
\newblock A new viscoelastic constitutive model for continuous media at finite
  thermomechanical changes.
\newblock \emph{International Journal of Solids and Structures}, 33\penalty0
  (20-22):\penalty0 3019--3034, 1996.

\bibitem[Jog(2006)]{Jog2006}
C.~Jog.
\newblock A concise proof of the representation theorem for fourth-order
  isotropic tensors.
\newblock \emph{Journal of Elasticity}, 85:\penalty0 119--124, 2006.

\bibitem[Khatyr et~al.(2004)Khatyr, Imberdis, Vescovo, Varchon, and
  Lagarde]{khatyr2004model}
F.~Khatyr, C.~Imberdis, P.~Vescovo, D.~Varchon, and J.-M. Lagarde.
\newblock Model of the viscoelastic behaviour of skin in vivo and study of
  anisotropy.
\newblock \emph{Skin research and technology}, 10\penalty0 (2):\penalty0
  96--103, 2004.

\bibitem[Klingbeil and Shield(1966)]{KlingbeilShield1966}
W.~W. Klingbeil and R.~T. Shield.
\newblock On a class of solutions in plane finite elasticity.
\newblock \emph{Zeitschrift f{\"u}r angewandte Mathematik und Physik},
  17\penalty0 (4):\penalty0 489--511, 1966.

\bibitem[Kosmann-Schwarzbach et~al.(2011)Kosmann-Schwarzbach, Schwarzbach, and
  Kosmann-Schwarzbach]{Kosmann2011}
Y.~Kosmann-Schwarzbach, B.~E. Schwarzbach, and Y.~Kosmann-Schwarzbach.
\newblock \emph{The {N}oether Theorems}.
\newblock Springer, 2011.

\bibitem[Kr{\"o}ner(1959)]{kroner1959}
E.~Kr{\"o}ner.
\newblock Allgemeine kontinuumstheorie der versetzungen und eigenspannungen.
\newblock \emph{Archive for Rational Mechanics and Analysis}, 4\penalty0
  (1):\penalty0 273--334, 1959.

\bibitem[Kumar and Lopez-Pamies(2016)]{Kumar2016}
A.~Kumar and O.~Lopez-Pamies.
\newblock On the two-potential constitutive modeling of rubber viscoelastic
  materials.
\newblock \emph{Comptes Rendus Mecanique}, 344\penalty0 (2):\penalty0 102--112,
  2016.

\bibitem[Latorre and Mont{\'a}ns(2016)]{Latorre2016}
M.~Latorre and F.~J. Mont{\'a}ns.
\newblock Fully anisotropic finite strain viscoelasticity based on a reverse
  multiplicative decomposition and logarithmic strains.
\newblock \emph{Computers \& Structures}, 163:\penalty0 56--70, 2016.

\bibitem[Le~Tallec et~al.(1993)Le~Tallec, Rahier, and Kaiss]{LeTallec1993}
P.~Le~Tallec, C.~Rahier, and A.~Kaiss.
\newblock Three-dimensional incompressible viscoelasticity in large strains:
  formulation and numerical approximation.
\newblock \emph{Computer methods in applied mechanics and engineering},
  109\penalty0 (3-4):\penalty0 233--258, 1993.

\bibitem[Lee and Liu(1967)]{leeliu1967}
E.~Lee and D.~Liu.
\newblock Finite-strain elastic---plastic theory with application to plane-wave
  analysis.
\newblock \emph{Journal of Applied Physics}, 38\penalty0 (1):\penalty0 19--27,
  1967.

\bibitem[Lee(1969)]{lee1969ElastoPlast}
E.~H. Lee.
\newblock Elastic-plastic deformation at finite strains.
\newblock \emph{Journal of Applied Mechanics}, 36\penalty0 (1):\penalty0 1--6,
  1969.

\bibitem[Leonov(1976)]{Leonov1976}
A.~I. Leonov.
\newblock Nonequilibrium thermodynamics and rheology of viscoelastic polymer
  media.
\newblock \emph{Rheologica Acta}, 15\penalty0 (2):\penalty0 85--98, 1976.

\bibitem[Liu et~al.(2019)Liu, Holzapfel, Skallerud, and Prot]{Liu2019}
H.~Liu, G.~A. Holzapfel, B.~H. Skallerud, and V.~Prot.
\newblock Anisotropic finite strain viscoelasticity: {C}onstitutive modeling
  and finite element implementation.
\newblock \emph{Journal of the Mechanics and Physics of Solids}, 124:\penalty0
  172--188, 2019.

\bibitem[Liu(1982)]{liu1982}
I.~Liu.
\newblock On representations of anisotropic invariants.
\newblock \emph{International Journal of Engineering Science}, 20\penalty0
  (10):\penalty0 1099--1109, 1982.

\bibitem[Liu et~al.(2021)Liu, Latorre, and Marsden]{Liu2021}
J.~Liu, M.~Latorre, and A.~L. Marsden.
\newblock A continuum and computational framework for viscoelastodynamics: {I}.
  {F}inite deformation linear models.
\newblock \emph{Computer Methods in Applied Mechanics and Engineering},
  385:\penalty0 114059, 2021.

\bibitem[Lu(2012)]{Lu2012}
J.~Lu.
\newblock A covariant constitutive theory for anisotropic hyperelastic solids
  with initial strains.
\newblock \emph{Mathematics and Mechanics of Solids}, 17\penalty0 (2):\penalty0
  104--119, 2012.

\bibitem[Lu and Papadopoulos(2000)]{lu2000covariant}
J.~Lu and P.~Papadopoulos.
\newblock A covariant constitutive description of anisotropic non-linear
  elasticity.
\newblock \emph{Zeitschrift f{\"u}r Angewandte Mathematik und Physik},
  51\penalty0 (2):\penalty0 204--217, 2000.

\bibitem[Lubliner(1985)]{Lubliner1985}
J.~Lubliner.
\newblock A model of rubber viscoelasticity.
\newblock \emph{Mechanics Research Communications}, 12\penalty0 (2):\penalty0
  93--99, 1985.

\bibitem[Marsden and Ratiu(2013)]{MarsRat2013MechSym}
J.~Marsden and T.~Ratiu.
\newblock \emph{Introduction to Mechanics and Symmetry: A Basic Exposition of
  Classical Mechanical Systems}.
\newblock Texts in Applied Mathematics. Springer New York, 2013.

\bibitem[Marsden and Hughes(1983)]{MarsdenHughes1983}
J.~E. Marsden and T.~J.~R. Hughes.
\newblock \emph{Mathematical Foundations of Elasticity}.
\newblock Prentice-Hall, 1983.

\bibitem[Mazzucato and Rachele(2006)]{MazzucatoRachele2006}
A.~L. Mazzucato and L.~V. Rachele.
\newblock Partial uniqueness and obstruction to uniqueness in inverse problems
  for anisotropic elastic media.
\newblock \emph{Journal of Elasticity}, 83\penalty0 (3):\penalty0 205--245,
  2006.

\bibitem[Merodio(2006)]{Merodio2006}
J.~Merodio.
\newblock On constitutive equations for fiber-reinforced nonlinearly
  viscoelastic solids.
\newblock \emph{Mechanics Research Communications}, 33\penalty0 (6):\penalty0
  764--770, 2006.

\bibitem[Merodio and Ogden(2020)]{merodio2020finite}
J.~Merodio and R.~W. Ogden.
\newblock Finite deformation elasticity theory.
\newblock In \emph{Constitutive Modelling of Solid Continua}, pages 17--52.
  Springer, 2020.

\bibitem[Mielke and Roub{\'\i}{\v{c}}ek(2020)]{Mielke2020}
A.~Mielke and T.~Roub{\'\i}{\v{c}}ek.
\newblock Thermoviscoelasticity in {K}elvin--{V}oigt rheology at large strains.
\newblock \emph{Archive for Rational Mechanics and Analysis}, 238\penalty0
  (1):\penalty0 1--45, 2020.

\bibitem[Nedjar(2007)]{Nedjar2007}
B.~Nedjar.
\newblock An anisotropic viscoelastic fibre--matrix model at finite strains:
  {C}ontinuum formulation and computational aspects.
\newblock \emph{Computer Methods in Applied Mechanics and Engineering},
  196\penalty0 (9-12):\penalty0 1745--1756, 2007.

\bibitem[Nguyen et~al.(2007)Nguyen, Jones, and Boyce]{Nguyen2007}
T.~D. Nguyen, R.~E. Jones, and B.~L. Boyce.
\newblock Modeling the anisotropic finite-deformation viscoelastic behavior of
  soft fiber-reinforced composites.
\newblock \emph{International Journal of Solids and Structures}, 44\penalty0
  (25-26):\penalty0 8366--8389, 2007.

\bibitem[Nishikawa(2002)]{nishikawa2002variation}
S.~Nishikawa.
\newblock \emph{Variational Problems in Geometry}, volume 205.
\newblock American Mathematical Society, 2002.

\bibitem[Noll(1958)]{noll1958}
W.~Noll.
\newblock A mathematical theory of the mechanical behavior of continuous media.
\newblock \emph{Archive for Rational Mechanics and Analysis}, 2\penalty0
  (1):\penalty0 197--226, 1958.

\bibitem[Olive et~al.(2017)Olive, Kolev, and Auffray]{Olive2017}
M.~Olive, B.~Kolev, and N.~Auffray.
\newblock A minimal integrity basis for the elasticity tensor.
\newblock \emph{Archive for Rational Mechanics and Analysis}, 226:\penalty0
  1--31, 2017.

\bibitem[Onsager(1931)]{onsager1931reciprocal}
L.~Onsager.
\newblock Reciprocal relations in irreversible processes. i.
\newblock \emph{Physical review}, 37\penalty0 (4):\penalty0 405, 1931.

\bibitem[Pipkin(1964)]{Pipkin1964}
A.~Pipkin.
\newblock Small finite deformations of viscoelastic solids.
\newblock \emph{Reviews of Modern Physics}, 36\penalty0 (4):\penalty0 1034,
  1964.

\bibitem[Pipkin and Rogers(1968)]{PipkinRogers1968}
A.~Pipkin and T.~Rogers.
\newblock A non-linear integral representation for viscoelastic behaviour.
\newblock \emph{Journal of the Mechanics and Physics of Solids}, 16\penalty0
  (1):\penalty0 59--72, 1968.

\bibitem[Reese and Govindjee(1998{\natexlab{a}})]{Reese1998}
S.~Reese and S.~Govindjee.
\newblock A theory of finite viscoelasticity and numerical aspects.
\newblock \emph{International Journal of Solids and Structures}, 35\penalty0
  (26-27):\penalty0 3455--3482, 1998{\natexlab{a}}.

\bibitem[Reese and Govindjee(1998{\natexlab{b}})]{ReeseGovindjee1998}
S.~Reese and S.~Govindjee.
\newblock Theoretical and numerical aspects in the thermo-viscoelastic material
  behaviour of rubber-like polymers.
\newblock \emph{Mechanics of Time-Dependent Materials}, 1:\penalty0 357--396,
  1998{\natexlab{b}}.

\bibitem[Rivlin(1965)]{Rivlin1965}
R.~S. Rivlin.
\newblock Nonlinear viscoelastic solids.
\newblock \emph{SIAM Review}, 7\penalty0 (3):\penalty0 323--340, 1965.

\bibitem[Rivlin and Ericksen(1955)]{RivlinEricksen1955}
R.~S. Rivlin and J.~L. Ericksen.
\newblock Stress-deformation relations for isotropic materials.
\newblock \emph{Journal of Rational Mechanics and Analysis}, 4\penalty0
  (6):\penalty0 323--425, 1955.

\bibitem[Sadik and Yavari(2017{\natexlab{a}})]{Sadik2017}
S.~Sadik and A.~Yavari.
\newblock On the origins of the idea of the multiplicative decomposition of the
  deformation gradient.
\newblock \emph{Mathematics and Mechanics of Solids}, 22\penalty0 (4):\penalty0
  771--772, 2017{\natexlab{a}}.

\bibitem[Sadik and Yavari(2017{\natexlab{b}})]{Sadik2017Thermoelasticity}
S.~Sadik and A.~Yavari.
\newblock Geometric nonlinear thermoelasticity and the time evolution of
  thermal stresses.
\newblock \emph{Mathematics and Mechanics of Solids}, 22\penalty0 (7):\penalty0
  1546--1587, 2017{\natexlab{b}}.

\bibitem[Schapery(2000)]{Schapery2000}
R.~Schapery.
\newblock Nonlinear viscoelastic solids.
\newblock \emph{International Journal of Solids and Structures}, 37\penalty0
  (1-2):\penalty0 359--366, 2000.

\bibitem[Segers{\"a}ll et~al.(2014)Segers{\"a}ll, Moverare, Leidermark, and
  Simonsson]{segersall2014creep}
M.~Segers{\"a}ll, J.~J. Moverare, D.~Leidermark, and K.~Simonsson.
\newblock Creep and stress relaxation anisotropy of a single-crystal
  superalloy.
\newblock \emph{Metallurgical and Materials Transactions A}, 45\penalty0
  (5):\penalty0 2532--2544, 2014.

\bibitem[{\c{S}}eng{\"u}l(2021)]{Csengul2021}
Y.~{\c{S}}eng{\"u}l.
\newblock Nonlinear viscoelasticity of strain rate type: {A}n overview.
\newblock \emph{Proceedings of the Royal Society A}, 477\penalty0
  (2245):\penalty0 20200715, 2021.

\bibitem[Shariff(2022)]{Shariff2022}
M.~Shariff.
\newblock On the smallest number of functions representing isotropic functions
  of scalars, vectors and tensors.
\newblock \emph{arXiv preprint arXiv:2207.09617}, 2022.

\bibitem[Sidoroff(1973)]{sidoroff1973geometrical}
F.~Sidoroff.
\newblock The geometrical concept of intermediate configuration and
  elastic-plastic finite strain.
\newblock \emph{Arch. Mech}, 25\penalty0 (2):\penalty0 299--308, 1973.

\bibitem[Sidoroff(1974)]{sidoroff1974}
F.~Sidoroff.
\newblock Un mod{\`e}le visco{\'e}lastique non lin{\'e}aire avec configuration
  interm{\'e}diaire.
\newblock \emph{Journal de M{\'e}canique}, 13\penalty0 (4):\penalty0 679--713,
  1974.

\bibitem[Simo(1987)]{simo1987fully}
J.~C. Simo.
\newblock On a fully three-dimensional finite-strain viscoelastic damage model:
  {F}ormulation and computational aspects.
\newblock \emph{Computer Methods in Applied Mechanics and Engineering},
  60\penalty0 (2):\penalty0 153--173, 1987.

\bibitem[Simo(1988)]{Simo1988}
J.~C. Simo.
\newblock A framework for finite strain elastoplasticity based on maximum
  plastic dissipation and the multiplicative decomposition: {P}art {I}.
  {C}ontinuum formulation.
\newblock \emph{Computer Methods in Applied Mechanics and Engineering},
  66\penalty0 (2):\penalty0 199--219, 1988.

\bibitem[Simo(1998)]{Simo1998Handbook}
J.~C. Simo.
\newblock Numerical analysis and simulation of plasticity.
\newblock In P.~J. Ciarlet and J.~L. Lions, editors, \emph{Handbook of
  Numerical Analysis}, volume~VI, pages 183--499. Elsevier, Berlin, 1998.

\bibitem[Simo and Hughes(2006)]{SimoHughes2006}
J.~C. Simo and T.~J. Hughes.
\newblock \emph{Computational Inelasticity}, volume~7.
\newblock Springer Science \& Business Media, 2006.

\bibitem[Singh and Pipkin(1965)]{SinghPipkin1965}
M.~Singh and A.~C. Pipkin.
\newblock Note on {E}ricksen's problem.
\newblock \emph{Zeitschrift f{\"u}r angewandte Mathematik und Physik},
  16\penalty0 (5):\penalty0 706--709, 1965.

\bibitem[Sivasithamparam et~al.(2015)Sivasithamparam, Karstunen, and
  Bonnier]{sivasithamparam2015modelling}
N.~Sivasithamparam, M.~Karstunen, and P.~Bonnier.
\newblock Modelling creep behaviour of anisotropic soft soils.
\newblock \emph{Computers and Geotechnics}, 69:\penalty0 46--57, 2015.

\bibitem[Spencer(1982)]{spencer1982formulation}
A.~Spencer.
\newblock The formulation of constitutive equation for anisotropic solids.
\newblock In \emph{Mechanical Behavior of Anisotropic Solids/Comportment
  M{\'e}chanique des Solides Anisotropes}, pages 3--26. Springer, 1982.

\bibitem[Spencer(1971)]{Spencer1971}
A.~J.~M. Spencer.
\newblock Part {III}. {T}heory of {I}nvariants.
\newblock \emph{Continuum Physics}, 1:\penalty0 239--353, 1971.

\bibitem[Spencer(1986)]{Spencer1986}
A.~J.~M. Spencer.
\newblock Modelling of finite deformations of anisotropic materials.
\newblock In \emph{Large Deformations of Solids: Physical Basis and
  Mathematical Modelling}, pages 41--52. Springer, 1986.

\bibitem[Truesdell(1952)]{truesdell1952mechanical}
C.~Truesdell.
\newblock The mechanical foundations of elasticity and fluid dynamics.
\newblock \emph{Journal of Rational Mechanics and Analysis}, 1\penalty0
  (1):\penalty0 125?300, 1952.

\bibitem[Truesdell(1953)]{Truesdell1953physical}
C.~Truesdell.
\newblock The physical components of vectors and tensors.
\newblock \emph{Zeitschrift f{\"u}r Angewandte Mathematik und Mechanik},
  33\penalty0 (10-11):\penalty0 345--356, 1953.

\bibitem[Truesdell(1966)]{Truesdell1966}
C.~Truesdell.
\newblock \emph{The Elements of Continuum Mechanics}.
\newblock Springer-Verlag, 1966.

\bibitem[Valanis(1972)]{valanis1972irreversible}
K.~Valanis.
\newblock \emph{Irreversible Thermodynamics of Continuous Media: Internal
  Variable Theory}.
\newblock CISM Series. Springer, 1972.

\bibitem[Volterra(1909)]{Volterra1909}
V.~Volterra.
\newblock Sulle equazioni integro-differenziali della theoria dell'elasticita.
\newblock \emph{Atti Reale Accad. naz. Lincei. Rend. Cl. sci. fis., mat. e
  natur.}, 18:\penalty0 295--300, 1909.

\bibitem[Wang(1965)]{Wang1965}
C.~C. Wang.
\newblock The principle of fading memory.
\newblock \emph{Archive for Rational Mechanics and Analysis}, 18:\penalty0
  343--366, 1965.

\bibitem[Wang et~al.(2022)Wang, Chehade, Govindjee, and Nguyen]{Wang2022}
Z.~Wang, A.~E.~H. Chehade, S.~Govindjee, and T.~D. Nguyen.
\newblock A nonlinear viscoelasticity theory for nematic liquid crystal
  elastomers.
\newblock \emph{Journal of the Mechanics and Physics of Solids}, 163:\penalty0
  104829, 2022.

\bibitem[Wineman(2020)]{Wineman2020}
A.~Wineman.
\newblock \emph{Viscoelastic Solids}, pages 81--123.
\newblock Springer International Publishing, 2020.

\bibitem[Yavari(2013)]{Yavari2013}
A.~Yavari.
\newblock Compatibility equations of nonlinear elasticity for
  non-simply-connected bodies.
\newblock \emph{Archive for Rational Mechanics and Analysis}, 209:\penalty0
  237--253, 2013.

\bibitem[Yavari(2021)]{Yavari2021a}
A.~Yavari.
\newblock Universal deformations in inhomogeneous isotropic nonlinear elastic
  solids.
\newblock \emph{Proceedings of the Royal Society A}, 477\penalty0
  (2253):\penalty0 20210547, 2021.

\bibitem[Yavari(2023)]{Yavari2023Fibers}
A.~Yavari.
\newblock Universal displacements in inextensible fiber-reinforced linear
  elastic solids.
\newblock 2023.

\bibitem[Yavari and Goriely(2016)]{YavariGoriely2016}
A.~Yavari and A.~Goriely.
\newblock The anelastic {E}ricksen problem: {U}niversal eigenstrains and
  deformations in compressible isotropic elastic solids.
\newblock \emph{Proceedings of the Royal Society A}, 472\penalty0
  (2196):\penalty0 20160690, 2016.

\bibitem[Yavari and Goriely(2021)]{YavariGoriely2021}
A.~Yavari and A.~Goriely.
\newblock Universal deformations in anisotropic nonlinear elastic solids.
\newblock \emph{Journal of the Mechanics and Physics of Solids}, 156:\penalty0
  104598, 2021.

\bibitem[Yavari and Goriely(2022)]{Yavari2022Anelastic-Universality}
A.~Yavari and A.~Goriely.
\newblock Universality in anisotropic linear anelasticity.
\newblock \emph{Journal of Elasticity}, 150\penalty0 (2):\penalty0 241--259,
  2022.

\bibitem[Yavari and Goriely(2023)]{yavari2023universal}
A.~Yavari and A.~Goriely.
\newblock The universal program of linear elasticity.
\newblock \emph{Mathematics and Mechanics of Solids}, 28\penalty0 (1):\penalty0
  251--268, 2023.

\bibitem[Yavari and Sozio(2023)]{YavariSozio2023}
A.~Yavari and F.~Sozio.
\newblock On the direct and reverse multiplicative decompositions of
  deformation gradient in nonlinear anisotropic anelasticity.
\newblock \emph{Journal of the Mechanics and Physics of Solids}, 170:\penalty0
  105101, 2023.

\bibitem[Yavari et~al.(2006)Yavari, Marsden, and Ortiz]{yavari2006spatial}
A.~Yavari, J.~E. Marsden, and M.~Ortiz.
\newblock On spatial and material covariant balance laws in elasticity.
\newblock \emph{Journal of Mathematical Physics}, 47:\penalty0 042903, 2006.

\bibitem[Yavari et~al.(2016)Yavari, Ozakin, and Sadik]{Yavari2016}
A.~Yavari, A.~Ozakin, and S.~Sadik.
\newblock Nonlinear elasticity in a deforming ambient space.
\newblock \emph{Journal of Nonlinear Science}, 26:\penalty0 1651--1692, 2016.

\bibitem[Yavari et~al.(2020)Yavari, Goodbrake, and Goriely]{Yavari2020}
A.~Yavari, C.~Goodbrake, and A.~Goriely.
\newblock Universal displacements in linear elasticity.
\newblock \emph{Journal of the Mechanics and Physics of Solids}, 135:\penalty0
  103782, 2020.

\bibitem[Zener(1948)]{Zener1948}
C.~M. Zener.
\newblock \emph{Elasticity and Anelasticity of Metals}.
\newblock University of Chicago Press, 1948.

\bibitem[Zheng(1994)]{zheng1994theory}
Q.~S. Zheng.
\newblock Theory of representations for tensor functions.
\newblock \emph{Applied Mechanics Reviews}, 47\penalty0 (11):\penalty0
  545--587, 1994.

\bibitem[Zheng and Spencer(1993)]{zheng1993}
Q.-S. Zheng and A.~J.~M. Spencer.
\newblock Tensors which characterize anisotropies.
\newblock \emph{International Journal of Engineering Science}, 31\penalty0
  (5):\penalty0 679--693, 1993.

\bibitem[Ziegler(1958)]{ziegler1958attempt}
H.~Ziegler.
\newblock An attempt to generalize {O}nsager's principle, and its significance
  for rheological problems.
\newblock \emph{Zeitschrift f{\"u}r angewandte Mathematik und Physik},
  9\penalty0 (5-6):\penalty0 748--763, 1958.

\bibitem[Ziegler and Wehrli(1987)]{ziegler1987derivation}
H.~Ziegler and C.~Wehrli.
\newblock The derivation of constitutive relations from the free energy and the
  dissipation function.
\newblock In \emph{Advances in applied mechanics}, volume~25, pages 183--238.
  Elsevier, 1987.

\end{thebibliography}

\appendix

\section{Variations for the Lagrange-d'Alembert Principle}
\label{eq:LD_deets}

The variation of the velocity vector, $\delta \mathbf V\,$, is computed as
%---------------------
\begin{equation}
\delta \mathbf V
	= D^{\mathbf g}_\epsilon \mathbf V_\epsilon|_{\epsilon=0}
	= D^{\mathbf g}_\epsilon \left(\left.\frac{\partial \varphi_{\epsilon}(X,t)}{\partial t}\right)\right|_{\epsilon=0}
	= D^{\mathbf g}_t \left.\left(\frac{\partial \varphi_{\epsilon}(X,t)}{\partial \epsilon}\right)\right|_{\epsilon=0}
	= D^{\mathbf g}_t\delta\varphi\,,
\end{equation}
%---------------------
where $D^{\mathbf g}_\epsilon$ denotes the covariant derivative along the curve
$\epsilon\mapsto\varphi_{t,\epsilon}(X)$ for fixed $t$ and $X\,$,  $D^{\mathbf g}_t$ denotes
the covariant derivative along the curve $t\mapsto\varphi_{t,\epsilon}(X)$ for fixed $\epsilon$
and $X\,$,  and use was made of the symmetry lemma for covariant
derivatives~\citep{nishikawa2002variation}: $D^{\mathbf g}_\epsilon \frac{\partial}{\partial t} = D^{\mathbf g}_t \frac{\partial}{\partial \epsilon}\,$. Hence, it follows
that
%---------------------
\begin{equation}\label{eq:LD_V}
\begin{split}
	\int_{t_0}^{t_1}\!\int_{\mathcal B} \frac{\partial \hat{\mathscr L}}{\partial \mathbf V} \delta \mathbf V  \, dV dt
	&= \int_{t_0}^{t_1}\!\int_{\mathcal B} \frac{\partial \hat{\mathscr L}}{\partial \mathbf V} 
	D^{\mathbf g}_t\delta\varphi  \, dV dt\\
	&= \int_{t_0}^{t_1}\!\int_{\mathcal B} \left[\frac{d}{dt}\left(\frac{\partial \hat{\mathscr L}}{\partial \mathbf V} 
	\delta\varphi\right) - D^{\mathbf g}_t\left(\frac{\partial \hat{\mathscr L}}{\partial \mathbf V}\right) 
	\delta\varphi \right] \, dV dt\\
	&=  \int_{\mathcal B} \left[\frac{\partial \hat{\mathscr L}}{\partial \mathbf V} \delta\varphi\right]_{t_0}^{t_1}\,dV 
	- \int_{t_0}^{t_1}\!\int_{\mathcal B} D^{\mathbf g}_t\left(\frac{\partial \hat{\mathscr L}}{\partial \mathbf V}\right) 
	\delta\varphi  \, dV dt \,.
\end{split}
\end{equation}
%---------------------
Following \eqref{cd_variations3,cd_variations4}, one may see that $\delta\varphi_{t_{0}}=\delta\varphi_{t_1}=\mathbf 0\,$.
From \eqref{eq:Lagrangian}, one finds that ${\partial \hat{\mathscr L}}/{\partial \mathbf V}=\rho_o \mathbf g \mathbf V$ and $D^{\mathbf g}_t\left({\partial \hat{\mathscr L}}/{\partial \mathbf V}\right)=\rho_o \mathbf g \mathbf A\,$.\footnote{Note that $D^{\mathbf g}_t\mathbf g=\nabla^{\mathbf g}_{\mathbf V}\mathbf g=\mathbf 0$ per compatibility of the Levi-Civita connection $\nabla^{\mathbf g}\,$.} Therefore, it follows that \eqref{eq:LD_V} yields
%---------------------
\begin{equation}\label{eq:LD_Vf}
	\int_{t_0}^{t_1}\!\int_{\mathcal B} \frac{\partial \hat{\mathscr L}}{\partial \mathbf V} 
	\delta \mathbf V  \, dV dt
	=  - \int_{t_0}^{t_1}\!\int_{\mathcal B} \llangle \rho_o \mathbf A , 
	\delta\varphi \rrangle_{\mathbf g}  \, dV dt \,.
\end{equation}
%---------------------

The variation of the Cauchy-Green deformation tensor, $\delta \mathbf C^\flat\,$, is computed as
%---------------------
\begin{equation}\label{eq:del-C}
\begin{split}
\delta \mathbf C^\flat
	&= \left.\frac{d \mathbf C^\flat_\epsilon}{d\epsilon}\right|_{\epsilon=0}
	= \left.\frac{d}{d\epsilon}(\varphi_\epsilon^*\mathbf g)\right|_{\epsilon=0}
	= \left.\frac{d}{d\epsilon}(\varphi^*\varphi_*\varphi_\epsilon^*\mathbf g)\right|_{\epsilon=0}
	= \varphi^*\left.\frac{d}{d\epsilon}(\varphi_*\varphi_\epsilon^*\mathbf g)\right|_{\epsilon=0}
	= \varphi^*(L_{\delta\varphi}\mathbf g)
	= \mathbf F^\star (L_{\delta\varphi}\mathbf g) \mathbf F\\
	&= \mathbf F^\star \left(\mathbf g (\nabla^{\mathbf g}\delta\varphi) + (\nabla^{\mathbf g}\delta\varphi)^\star \mathbf g \right)\mathbf F \,,
\end{split}
\end{equation}
%---------------------
where $L$ denotes the total Lie derivative operator\footnote{Note that in terms of the autonomous Lie derivative $\mathfrak{L}\,$,  one has $L_{\delta \varphi}=\frac{\partial }{\partial \epsilon} + \mathfrak{L}_{\delta \varphi}$} and use was made of the compatibility of the Levi-Civita connection to write $L_{\delta\varphi}\mathbf g=\mathbf g (\nabla^{\mathbf g}\delta\varphi) + (\nabla^{\mathbf g}\delta\varphi)^\star \mathbf g\,$. It follows that
%---------------------
\begin{equation}\label{eq:LD_Cf}
\begin{split}
	\int_{t_0}^{t_1}\!\int_{\mathcal B} \frac{\partial \hat{\mathscr L}}{\partial \mathbf C^\flat}\!:\!
	\delta \mathbf C^\flat  \, dV dt
	&= -\int_{t_0}^{t_1}\!\int_{\mathcal B} \frac{\partial \hat{\Psi}}{\partial \mathbf C^\flat}\!:\!
	\mathbf F^\star\left[ \mathbf g (\nabla^{\mathbf g}\delta\varphi) + (\nabla^{\mathbf g}\delta\varphi)^\star 
	\mathbf g \right]\mathbf F  \, dV dt\\
	&= -\int_{t_0}^{t_1}\!\int_{\mathcal B} 2\frac{\partial \hat{\Psi}}{\partial \mathbf C^\flat}\!:\!\mathbf 
	F^\star \mathbf g (\nabla^{\mathbf g}\delta\varphi) \mathbf F  \, dV dt\\
	&= -\int_{t_0}^{t_1}\!\int_{\mathcal B} 2\frac{\partial \hat{\Psi}}{\partial \mathbf C^\flat}\!:\!\mathbf 
	F^\star \mathbf g [(\varphi^*\nabla^{\mathbf g})\delta\varphi]  \, dV dt\\
	&= -\int_{t_0}^{t_1}\!\int_{\mathcal B} 2\mathbf g \mathbf F \frac{\partial \hat{\Psi}}{\partial \mathbf 
	C^\flat}\!:\![(\varphi^*\nabla^{\mathbf g})\delta\varphi]  \, dV dt\\
	&= -\int_{t_0}^{t_1}\!\int_{\mathcal B} \operatorname{Div}\left(2\delta\varphi \mathbf g \mathbf F 
	\frac{\partial \hat{\Psi}}{\partial \mathbf C^\flat} \right)  \, dA\, dt + \int_{t_0}^{t_1}\!
	\int_{\mathcal B} \delta\varphi \operatorname{Div} \left(2\mathbf g \mathbf F \frac{\partial \hat{\Psi}}
	{\partial \mathbf C^\flat}\right)  \, dV dt\\
	&= -\int_{t_0}^{t_1}\!\int_{\partial\mathcal B} 2\delta\varphi\mathbf g \mathbf F 
	\frac{\partial \hat{\Psi}}{\partial \mathbf C^\flat} \mathbf N \, dA\, dt + \int_{t_0}^{t_1}\!
	\int_{\mathcal B} \mathbf g \operatorname{Div} \left(2\mathbf F \frac{\partial \hat{\Psi}}
	{\partial \mathbf C^\flat}\right) \delta\varphi  \, dV dt\\
	&= -\int_{t_0}^{t_1}\!\int_{\partial\mathcal B} \llangle 2\mathbf F \frac{\partial \hat{\Psi}}{\partial 
	\mathbf C^\flat} \mathbf N, \delta\varphi \rrangle_{\mathbf g} \, dA\, dt + \int_{t_0}^{t_1}\!
	\int_{\mathcal B} \llangle \operatorname{Div}\left(2\mathbf F \frac{\partial \hat{\Psi}}{\partial 
	\mathbf C^\flat}\right) , \delta\varphi \rrangle_{\mathbf g}  \, dV dt\,,
\end{split}
\end{equation}
%---------------------
where $(\varphi^{*}\nabla^{\mathbf g})\delta\varphi=(\nabla^{\mathbf g}\delta\varphi) \mathbf F\,$,  $\operatorname{Div}$ denotes the material Levi-Civita divergence operator, and use was made of Stokes' theorem with $\mathbf N$ being the $\mathbf G$-unit normal to $\partial \mathcal B\,$.

The variation of the elastic Cauchy-Green deformation tensor, $\delta \Ce\,$, is calculated as
%---------------------
\begin{equation}\label{eq:del-Ce}
\begin{split}
\delta \Ce^\flat
	&= \delta (\Fv_*\mathbf C^\flat)
	= \delta (\Fv^{-\star}\mathbf C^\flat\Fv^{-1})\\
	&= \Fv^{-\star}\delta\mathbf C^\flat\Fv^{-1}
	+ \delta \Fv^{-\star} \mathbf C^\flat\Fv^{-1}
	+ \Fv^{-\star}\mathbf C^\flat\delta \Fv^{-1}\\
	&= \Fv^{-\star}\mathbf F^\star (L_{\delta\varphi}\mathbf g) \mathbf F \Fv^{-1}
	-\Fv^{-\star}\delta \Fv^\star \Fv^{-\star} \mathbf C^\flat\Fv^{-1}
	- \Fv^{-\star}\mathbf C^\flat\Fv^{-1} \delta \Fv\Fv^{-1}\\
	&= \Fe^\star (L_{\delta\varphi}\mathbf g)\Fe
	- \Fv^{-\star}(\delta \Fv)^\star\Ce^\flat
	- \Ce^\flat(\delta\Fv)\Fv^{-1}\\
	&=  \Fe^\star \left(\mathbf g (\nabla^{\mathbf g}\delta\varphi) + (\nabla^{\mathbf g}\delta\varphi)^\star \mathbf g \right)\Fe
	- \Ce^\flat(\delta\Fv)\Fv^{-1}
	- \Fv^{-\star}(\delta \Fv)^\star\Ce^\flat\,.
\end{split}
\end{equation}
%---------------------
It follows that
%---------------------
\begin{equation}\label{eq:LD_Ce}
\begin{split}
	\int_{t_0}^{t_1}\!\int_{\mathcal B} \frac{\partial \hat{\mathscr L}}{\partial \Ce^\flat}\!:\!\delta \Ce^\flat  \, dV dt 
	&= \int_{t_0}^{t_1}\!\int_{\mathcal B} \frac{\partial \hat{\mathscr L}}{\partial \Ce^\flat} \!:\!
	\left[\Fe^\star \left(\mathbf g (\nabla^{\mathbf g}\delta\varphi) 
	+ (\nabla^{\mathbf g}\delta\varphi)^\star \mathbf g \right)\Fe
	- \Ce^\flat(\delta\Fv)\Fv^{-1} - \Fv^{-\star}(\delta \Fv)^\star\Ce^\flat\right] \, dV dt\\
	&= -\int_{t_0}^{t_1}\!\int_{\mathcal B} \left[2\frac{\partial \hat{\Psi}}{\partial \Ce^\flat} 
	: \Fe^\star \mathbf g (\nabla^{\mathbf g}\delta\varphi) \Fe 
	- 2\frac{\partial \hat{\Psi}}{\partial \Ce^\flat}\!:\! \Ce^\flat(\delta\Fv)\Fv^{-1} \right]  \, dV dt \\
	&= -\int_{t_0}^{t_1}\!\int_{\mathcal B} \left[2\frac{\partial \hat{\Psi}}{\partial \Ce^\flat} 
	: \Fe^\star \mathbf g [(\varphi^*\nabla^{\mathbf g})\delta\varphi] \Fv^{-1} 
	- 2\Ce^\flat\frac{\partial \hat{\Psi}}{\partial \Ce^\flat}\Fv^{-\star}\!:\! \delta\Fv \right]  \, dV dt\\
	&= -\int_{t_0}^{t_1}\!\int_{\mathcal B} \left[2 \mathbf g \Fe\frac{\partial \hat{\Psi}}{\partial \Ce^\flat} 
	\Fv^{-\star}\!:\![(\varphi^*\nabla^{\mathbf g})\delta\varphi] 
	- 2\Ce^\flat\frac{\partial \hat{\Psi}}{\partial \Ce^\flat}\Fv^{-\star}\!:\! \delta\Fv \right]  \, dV dt\,.
\end{split}
\end{equation}
%---------------------
Isolating the first term, one may write
%---------------------
\begin{equation}\label{eq:LD_Ceb}
\begin{split}
	& \int_{t_0}^{t_1}\!\int_{\mathcal B} \left\{2 \mathbf g \Fe\frac{\partial \hat{\Psi}}{\partial \Ce^\flat} \Fv^{-\star}\!:\![(\varphi^*\nabla^{\mathbf g})\delta\varphi] \right\}  \, dV dt \\
	& \quad = \int_{t_0}^{t_1}\!\int_{\mathcal B} \left[\operatorname{Div}
	\left(2 \delta\varphi \mathbf g \Fe\frac{\partial \hat{\Psi}}{\partial \Ce^\flat} \Fv^{-\star}\right) - \delta\varphi \operatorname{Div} 
	\left( 2 \mathbf g \Fe\frac{\partial \hat{\Psi}}{\partial \Ce^\flat} \Fv^{-\star} \right) \right]  \, dV dt\\
	& \quad = \int_{t_0}^{t_1} \int_{\partial \mathcal B} 2 \delta\varphi \mathbf g \Fe\frac{\partial \hat{\Psi}}{\partial \Ce^\flat} \Fv^{-\star} \mathbf N  \, dA \, dt
	- \int_{t_0}^{t_1}\!\int_{\mathcal B} \delta\varphi \mathbf g \operatorname{Div} 
	\left(2 \Fe\frac{\partial \hat{\Psi}}{\partial \Ce^\flat} \Fv^{-\star} \right)  \, dV dt\\
	& \quad = \int_{t_0}^{t_1} \int_{\partial \mathcal B} \llangle 2 \Fe\frac{\partial \hat{\Psi}}{\partial \Ce^\flat} \Fv^{-\star} \mathbf N, \delta\varphi \rrangle_{\mathbf g} \, dA \, dt
	- \int_{t_0}^{t_1}\!\int_{\mathcal B}  \llangle \operatorname{Div} 
	\left( 2 \Fe\frac{\partial \hat{\Psi}}{\partial \Ce^\flat} \Fv^{-\star}  \right), \delta\varphi \rrangle_{\mathbf g}  \, dV dt\,.
\end{split}
\end{equation}
%---------------------
Therefore, it follows from \eqref{eq:LD_Ce} and \eqref{eq:LD_Ceb} that
%---------------------
\begin{equation}\label{eq:LD_Cef}
\begin{split}
	\int_{t_0}^{t_1}\!\int_{\mathcal B} \frac{\partial \hat{\mathscr L}}{\partial \Ce^\flat}\!:\!\delta \Ce^\flat  \, dV dt
	&= \int_{t_0}^{t_1}\!\int_{\mathcal B} \left[\llangle \operatorname{Div} 
	\left( 2 \Fe\frac{\partial \hat{\Psi}}{\partial \Ce^\flat} \Fv^{-\star} \right), \delta\varphi \rrangle_{\mathbf g} 
	+ 2\Ce^\flat\frac{\partial \hat{\Psi}}{\partial \Ce^\flat}\Fv^{-\star}\!:\! \delta\Fv \right]  \, dV dt \\
	&\quad- \int_{t_0}^{t_1} \int_{\partial \mathcal B} \llangle 2 
	\Fe\frac{\partial \hat{\Psi}}{\partial \Ce^\flat} \Fv^{-\star} \mathbf N, \delta\varphi \rrangle_{\mathbf g} \, dA 
	\, dt \,.
\end{split}
\end{equation}
%---------------------

The variation of the material metric, $\delta\mathbf G\,$, is identically zero since the material metric $\mathbf G$ remains unaltered by the one-parameter family $\epsilon\mapsto(\varphi_\epsilon,\Fv_\epsilon)$:
%---------------------
\begin{equation}\label{eq:LD_G}
	\delta \mathbf G = \mathbf 0\,.
\end{equation}
%---------------------
The variation of the spatial metric, $\delta\mathbf g\,$, is also identically zero following from the compatibility of the Levi-Civita connection $\nabla^{\mathbf g}\,$:
%---------------------
\begin{equation}\label{eq:LD_g}
	\delta \mathbf g = D^{\mathbf g}_\epsilon(\mathbf g \circ\varphi_\epsilon)
	=\delta\varphi \nabla^{\mathbf g}\mathbf g=\mathbf 0\,.
\end{equation}
%---------------------

The variation of the Jacobian of the total deformation, $\delta J\,$, reads
%---------------------
\begin{equation}\label{eq:del_J}
\begin{split}
\delta J
	&= \frac{d}{d\epsilon}\left(\sqrt{\det \mathbf C_\epsilon}\right)\\
	&= \frac{1}{2\sqrt{\det \mathbf C}}\frac{d}{d\epsilon}\left(\det \mathbf C_\epsilon\right)\\
	&= \frac{1}{2\sqrt{\det \mathbf C}}\frac{d}{d\mathbf C}\left(\det \mathbf C\right)\!:\!\delta\mathbf C\\
	&= \frac{1}{2}\sqrt{\det \mathbf C}\,\mathbf C^{-1}\!:\!\delta\mathbf C \quad\text{\textemdash using \eqref{Determinant-Variation}}\\
	&= \frac{1}{2}J\mathbf C^{-1}\!:\!\mathbf G^\sharp\mathbf F^\star \left(\mathbf g (\nabla^{\mathbf g}\delta\varphi) + (\nabla^{\mathbf g}\delta\varphi)^\star \mathbf g \right)\mathbf F \quad\text{\textemdash using \eqref{eq:del-C}}\\
	&= J\mathbf C^{-1}\!:\!\mathbf G^\sharp\mathbf F^\star \mathbf g (\nabla^{\mathbf g}\delta\varphi) \mathbf F\quad\text{\textemdash recalling that $\mathbf C^\star = \mathbf C$}\\
	&= J\mathbf C^{-\star}\!:\!\mathbf F^{\mathsf T} (\nabla^{\mathbf g}\delta\varphi) \mathbf F  \quad\text{\textemdash recalling that $\mathbf F^{\mathsf T} = \mathbf G^\sharp\mathbf F^\star \mathbf g$}\\
	&= J {\mathbf F^{\mathsf T}}^\star\mathbf C^{-\star}\!:\! (\nabla^{\mathbf g}\delta\varphi)\mathbf F\\
	&= J \mathbf F ^{-\star} \!:\! (\varphi^*\nabla^{\mathbf g})\delta\varphi \quad\text{\textemdash since ${\mathbf F^{\mathsf T}}^\star\mathbf C^{-\star} = \mathbf F ^{-\star} $ and recalling  $(\varphi^*\nabla^{\mathbf g})\delta\varphi=(\nabla^{\mathbf g}\delta\varphi) \mathbf F$}\,.
\end{split}
\end{equation}
%---------------------
Therefore
%---------------------
\begin{equation}\label{eq:LD_J}
\begin{split}
\int_{t_0}^{t_1}\!\int_{\mathcal B} p\,\delta J \, dV dt
	& = \int_{t_0}^{t_1}\!\int_{\mathcal B} pJ \mathbf F^{-\star} \!:\! (\varphi^*\nabla^{\mathbf g})\delta\varphi \, dV dt\\
	& = \int_{t_0}^{t_1}\!\int_{\mathcal B} \left[\operatorname{Div}(pJ \delta\varphi \mathbf F^{-\star}) - \delta\varphi \operatorname{Div}(pJ \mathbf F^{-\star})\right] \, dV dt \\
	& = \int_{t_0}^{t_1}\!\int_{\partial \mathcal B} pJ \delta\varphi \mathbf F^{-\star} \mathbf N \, dA \, dt - \int_{t_0}^{t_1}\!\int_{\mathcal B} \delta\varphi \operatorname{Div}(p J \mathbf F^{-\star}) \, dV dt \\
	& = \int_{t_0}^{t_1}\!\int_{\partial \mathcal B} \llangle pJ \mathbf g^\sharp\mathbf F^{-\star} \mathbf N, \delta\varphi\rrangle_{\mathbf g} \, dA \, dt - \int_{t_0}^{t_1}\!\int_{\mathcal B} \llangle \operatorname{Div}(p J \mathbf g^\sharp\mathbf F^{-\star}), \delta\varphi\rrangle_{\mathbf g} \, dV dt\,.
\end{split}
\end{equation}
%---------------------

The variation of the Jacobian of the viscous contribution to the deformation, $\delta \Jv\,$, reads
%---------------------
\begin{equation}\label{eq:del_Jv}
\begin{split}
\delta\Jv
	&= \frac{d}{d\epsilon}\left(\det \Fv_\epsilon\right)\\
	&= \frac{d}{d\Fv}(\det \Fv)\!:\!\delta\Fv\\
	&= (\det \Fv) \Fv^{-\star}\!:\!\delta\Fv \quad\text{\textemdash similarly to \eqref{Determinant-Variation}}\\
	&= \Jv \Fv^{-\star}\!:\!\delta\Fv\,.
\end{split}
\end{equation}
%---------------------
And it follows that
%---------------------
\begin{equation}\label{eq:LD_Jv}
\begin{split}
\int_{t_0}^{t_1}\!\int_{\mathcal B} q\,\delta \Jv \, dV dt
	& = \int_{t_0}^{t_1}\!\int_{\mathcal B} q \Jv \Fv^{-\star}\!:\!\delta\Fv \, dV dt\,.
	\end{split}
\end{equation}
%---------------------

%---------------------
%---------------------
\section{Derivatives of the principal invariants} \label{appendix:derivatives}

Suppose $f(\mathbf{C}^{\flat})$ is a scalar-valued functional. For an arbitrary second-order covariant tensor $\mathbf{H}\,$, which has the coordinate representation ${\mathbf{H}=\mathrm H_{AB}\,dX^A \otimes dX^B}\,$, one writes
%---------------------------------
\begin{equation}
	f(\mathbf{C}^{\flat}+\epsilon\,\mathbf{H})=f(\mathbf{C}^{\flat})
	+\frac{\partial f}{\partial \mathbf{C}^{\flat}}\!:\!\mathbf{H}\,\epsilon+o(\epsilon)
	\,.
\end{equation}
%---------------------------------
Note that $I_1=\operatorname{tr}_{\mathbf{G}}\mathbf{C}^{\flat}=\mathbf{C}^{\flat}\!:\!\mathbf{G}=\mathrm{C}_{AB}\mathrm{G}^{AB}=\mathrm{C}^A{}_A\,$. This implies that
%---------------------
\begin{equation}
	\frac{\partial I_1}{\partial\mathbf{C}^\flat}=\mathbf{G}^{\sharp}\,.
\end{equation}
%---------------------
For $I_3=\det \mathbf{C}\,$, note that
%---------------------
\begin{equation} \label{Determinant-Variation}
\begin{aligned}
	\det(\mathrm{C}^A{}_B+\epsilon\,\mathrm{H}^A{}_B) 
	&=\det\left[\mathrm{C}^A{}_D\left(\delta^D_B
	+(\mathrm{C}^{-1})^D{}_M\,\epsilon\,\mathrm{H}^M{}_B\right)
	\right] 
	 =\det \mathbf{C}\,\det\left(\mathbf{I}+\epsilon\,\mathbf{C}^{-1}\mathbf{H}\right) \\
	& =\det \mathbf{C}\left[1+\epsilon\,\operatorname{tr}
	\left(\mathbf{C}^{-1}\mathbf{H}\right)+o(\epsilon)\right]
	 =\det \mathbf{C} + (\det \mathbf{C})\,\mathbf{C}^{-1}\!:\mathbf{H}\,\epsilon+o(\epsilon)\,.
\end{aligned}
\end{equation}
%---------------------
This implies that
%---------------------
\begin{equation}
	\frac{\partial I_3}{\partial\mathbf{C}}=I_3\mathbf{C}^{-1}\,,\qquad \frac{\partial I_3}{\partial \mathrm{C}^A{}_B}=I_3(\mathrm{C}^{-1})^B{}_A\,.
\end{equation}
%---------------------
Therefore
%---------------------
\begin{equation}
	\frac{\partial I_3}{\partial\mathbf{C}^{\flat}}=I_3\mathbf{C}^{-\sharp}\,,\qquad 
	\frac{\partial I_3}{\partial \mathrm{C}_{AB}}=I_3(\mathrm{C}^{-1})^{AB}\,,
\end{equation}
%---------------------
where $(\mathrm{C}^{-1})^{AB}=(\mathrm{C}^{-1})^A{}_M\,\mathrm{G}^{MB}\,$.
For $I_2=(\det\mathbf{C})\operatorname{tr}\mathbf{C}^{-1}=I_3\,\operatorname{tr}\mathbf{C}^{-1}\,$,
%---------------------
\begin{equation}
	\frac{\partial I_2}{\partial\mathbf{C}}=\frac{\partial I_3}{\partial\mathbf{C}}\operatorname{tr}\mathbf{C}^{-1}
	+I_3\,\frac{\partial}{\partial\mathbf{C}}\operatorname{tr}\mathbf{C}^{-1}
	=I_3\,\mathbf{C}^{-1}\operatorname{tr}\mathbf{C}^{-1}
	+I_3\frac{\partial}{\partial\mathbf{C}}\operatorname{tr}\mathbf{C}^{-1}
	=I_2\,\mathbf{C}^{-1}+I_3\frac{\partial}{\partial\mathbf{C}}\operatorname{tr}\mathbf{C}^{-1}\,.
\end{equation}
%---------------------
The second term on the right-hand side is calculated as
%---------------------
\begin{equation}
	\left(\mathbf{C}+\epsilon\,\mathbf{H}\right)^{-1}
	=\left[\mathbf{C}\left(\mathbf{I}+\epsilon\,\mathbf{C}^{-1}\mathbf{H}\right)\right]^{-1}
	=\left(\mathbf{I}+\epsilon\,\mathbf{C}^{-1}\mathbf{H}\right)^{-1}\mathbf{C}^{-1}\,.
\end{equation}
%---------------------
Note that for a small enough $\epsilon\,$, one has
%---------------------
\begin{equation}
\begin{aligned}
	\left(\mathbf{I}+\epsilon\,\mathbf{C}^{-1}\mathbf{H}\right)^{-1}
	&=\left[\mathbf{I}-\left(-\epsilon\,\mathbf{C}^{-1}\mathbf{H}\right)\right]^{-1}
	=\mathbf{I}+\left(-\epsilon\,\mathbf{C}^{-1}\mathbf{H}\right)
	+\left(-\epsilon\,\mathbf{C}^{-1}\mathbf{H}\right)^2+o(\epsilon^2) \\
	&=\mathbf{I}-\epsilon\left(\mathbf{C}^{-1}\mathbf{H}\right)
	+\epsilon^2\left(\mathbf{C}^{-1}\mathbf{H}\right)^2+o(\epsilon^2)
	\,.
\end{aligned}
\end{equation}
%---------------------
Thus
%---------------------
\begin{equation}
	\left(\mathbf{C}+\epsilon\,\mathbf{H}\right)^{-1}
	=\mathbf{C}^{-1}-\epsilon\left(\mathbf{C}^{-1}\mathbf{H}\mathbf{C}^{-1}\right)
	+\epsilon^2\left(\mathbf{C}^{-1}\mathbf{H}\right)^2\mathbf{C}^{-1}+o(\epsilon^2)
	\,.
\end{equation}
%---------------------
This implies that
%---------------------
\begin{equation}
	\operatorname{tr}\left(\mathbf{C}+\epsilon\,\mathbf{H}\right)^{-1}
	=\operatorname{tr}\mathbf{C}^{-1}-\epsilon\operatorname{tr}\left(\mathbf{C}^{-1}\mathbf{H}\mathbf{C}^{-1}\right)
	+\epsilon^2\operatorname{tr}\left[\left(\mathbf{C}^{-1}\mathbf{H}\right)^2\mathbf{C}^{-1}\right]+o(\epsilon^2)\,.
\end{equation}
%---------------------
Note that
%---------------------
\begin{equation}
	\operatorname{tr}\left(\mathbf{C}^{-1}\mathbf{H}\mathbf{C}^{-1}\right)
	=(\mathrm{C}^{-1})^A{}_B\,\mathrm{H}^B{}_D\,(\mathrm{C}^{-1})^D{}_A=(\mathrm{C}^{-1})^D{}_A\,(\mathrm{C}^{-1})^A{}_B\,\mathrm{H}^B{}_D
	=(\mathrm{C}^{-2})^D{}_B\,\mathrm{H}^B{}_D=\mathbf{C}^{-2}\!:\mathbf{H}\,.
\end{equation}
%---------------------
Therefore
 %---------------------
\begin{equation}
	\frac{\partial}{\partial\mathbf{C}}\operatorname{tr}\mathbf{C}^{-1}=\mathbf{C}^{-2}\,.
\end{equation}
%---------------------
Finally
%---------------------
\begin{equation}
	\frac{\partial I_2}{\partial\mathbf{C}}=I_2\,\mathbf{C}^{-1}+I_3\,\mathbf{C}^{-2}\,,\qquad
	\frac{\partial I_2}{\partial\mathbf{C}^{\flat}}=I_2\,(\mathbf{C}^{-1})^{-\sharp}+I_3\,(\mathbf{C}^{-2})^{\sharp}
	=I_2\,\mathbf{C}^{-\sharp}+I_3\,\mathbf{C}^{-2\sharp}\,.
\end{equation}
%---------------------
Similarly,
%---------------------
\begin{equation}\label{eq:part_Ie}
	\frac{\partial \Ie_1}{\partial\Ce^\flat}=\mathbf{G}^{\sharp}\,,\qquad 
	\frac{\partial \Ie_2}{\partial\Ce^\flat}=\Ie_2\Ce^{-\sharp}-\Ie_3\Ce^{-2\sharp}\,,\qquad
	\frac{\partial \Ie_3}{\partial\Ce^\flat}=\Ie_3\Ce^{-\sharp}\,.
\end{equation}
%---------------------

\end{document}